\documentclass[12pt,twoside]{article}
\usepackage{openbib,epsf,rotating,epsfig}
\usepackage{amsmath,amssymb,pifont,eurosym}
\usepackage{breakurl}
\usepackage{color}
\usepackage{graphicx}
\usepackage{setspace}
\usepackage{enumitem}
\usepackage{scrlayer-scrpage}
\usepackage{natbib}
\usepackage{tocloft}
\usepackage{chemformula}
\let\ce\ch
\usepackage[latin1]{inputenc}
\usepackage[german,english]{babel}
\usepackage[colorlinks=true,urlcolor=black,citecolor=black,linkcolor=black]
           {hyperref}

\definecolor{light}{gray}{0.95}
\definecolor{heavy}{gray}{0.35}
\def\bff{\sffamily\bfseries}
\def\itt{\sffamily\itshape}
\def\3{\ss}
\definecolor{dblue}{rgb}{0.0,0.0,0.6}
\definecolor{pale}{rgb}{0.98,0.98,0.98}
\newenvironment{itemize*}%
  {\vspace*{-1mm}\begin{itemize}%
    \setlength{\itemsep}{3pt}%
    \setlength{\parskip}{0.5pt}%
    \setlength{\topsep}{0.5pt}%
    \setlength{\parsep}{0.5pt}%
    \setlength{\partopsep}{-30pt}}%
  {\end{itemize}\vspace*{-1mm}}
\newenvironment{enumerate*}%
  {\vspace*{-1mm}\begin{enumerate}%
    \setlength{\itemsep}{3pt}%
    \setlength{\parskip}{0.5pt}%
    \setlength{\topsep}{0.5pt}%
    \setlength{\parsep}{0.5pt}%
    \setlength{\partopsep}{-30pt}}%
  {\end{enumerate}\vspace*{-1mm}}

\def\myparagraph#1{\vspace{2mm}\noindent{\bf #1}\ }

\def\minvalue(#1,#2){\ifnum #1<#2 #1\else #2\fi}

\def\la{\mathrel{\mathchoice {\vcenter{\offinterlineskip\halign{\hfil
$\displaystyle##$\hfil\cr<\cr\sim\cr}}}
{\vcenter{\offinterlineskip\halign{\hfil$\textstyle##$\hfil\cr
<\cr\sim\cr}}}
{\vcenter{\offinterlineskip\halign{\hfil$\scriptstyle##$\hfil\cr
<\cr\sim\cr}}}
{\vcenter{\offinterlineskip\halign{\hfil$\scriptscriptstyle##$\hfil\cr
<\cr\sim\cr}}}}}
\def\ga{\mathrel{\mathchoice {\vcenter{\offinterlineskip\halign{\hfil
$\displaystyle##$\hfil\cr>\cr\sim\cr}}}
{\vcenter{\offinterlineskip\halign{\hfil$\textstyle##$\hfil\cr
>\cr\sim\cr}}}
{\vcenter{\offinterlineskip\halign{\hfil$\scriptstyle##$\hfil\cr
>\cr\sim\cr}}}
{\vcenter{\offinterlineskip\halign{\hfil$\scriptscriptstyle##$\hfil\cr
>\cr\sim\cr}}}}}

\def\ProDiMo{{\sf ProDiMo}}

\def\etal{${\rm \hspace*{0.8ex}et\hspace*{0.7ex}al.\hspace*{0.5ex}}$}

\def\pabl#1#2{\frac{{\rm\partial} #1}{{\rm\partial} #2}}
\def\amin{a_{\rm min}}
\def\amax{a_{\rm max}}
\def\apow{a_{\rm pow}}

\def\Td{T_{\rm d}}
\def\Tg{T_{\rm g}}
\def\nH{n_{\rm\langle H\rangle}}
\def\Pe{P^{\rm \,esc}}
\def\Pp{P^{\rm \,pump}}
\def\Jcont{J^{\rm cont}}

\def\apj{ApJ}
\def\pasp{PASP}

\def\aap{A\&A}
\def\aaps{A\&AS}
\def\apjl{ApJL}

\def\mnras{MNRAS}

\def\psj{The Planetary Science Journal}

\def\Euro{\EUR}
      
\begin{document} 
\sloppy

\pagestyle{empty}
\centerline{ }
\vspace{5mm}
{\center \large \sf

{Kumulative Habilitationsschrift} \\[3mm]
{zur Erlangung der Lehrbefugnis im Fach} \\[3mm]
{\itt Astrophysik} \\[22mm]

{\huge\bff  Physico-chemical Processes} \\[6mm]
{\huge\bff  in Planet-forming Discs} \\[22mm]

vorgelegt von \\[3mm]
{Dr.\ rer.\ nat.\ Peter Woitke} \\[3mm]
{aus Berlin} \\[22mm] 

{\large \sf  Institut f\"ur Theoretische Physik} \\[3mm]
Technische Universit\"at Graz \\[3mm]
Petersgasse 16 \\[3mm] 
8010 Graz \\[22mm]
Graz, im November 2023\\
}

\cleardoublepage

\cftaddtitleline{toc}{section}{\bff List of submitted publications}{}

{\sf

\noindent\hspace*{-3mm}  
\begin{tabular}{cc}
  \begin{minipage}[t]{74mm}
    {\bff Dr.~Peter Woitke}\\
    Space Research Institute \\
    (Institut f\"ur Weltraumforschung, IWF) \\
    Austrian Academy of Science \\
    Schmiedlstra\3e 6 \\
    8042 Graz, Austria 
  \end{minipage}
  &
  \begin{minipage}[t]{75mm}
    {\ } \\[-0.5mm]
    Tel: +43~316~4120320 \\
    peter.woitke@oeaw.ac.at \\
    \resizebox{84mm}{!}{\tt
    https://www.oeaw.ac.at/en/iwf/institut/das-team/}\\[-2mm]
    \resizebox{84mm}{!}{\tt
    protoplanetare-scheiben-und-astrochemie/peter-woitke}
  \end{minipage}
\end{tabular}\\[5mm]

\noindent  
Nachstehende Publikationen werden als Habilitationsschrift
eingereicht:

\begin{itemize}
\item[\bff P1] \underline{Woitke, P.}, Kamp, I., Thi, W.-F.:
  {\bff Radiation thermo-chemical models of protoplanetary
    disks.  I.~Hydrostatic disk structure and inner rim},\\
  Astronomy \& Astrophysics 501, 383--406,
  \href{https://doi.org/10.1051/0004-6361/200911821}{(2009)} 

  \item[\bff P2] \underline{Woitke, P.}, {Pinte}, C., {Tilling}, I.,
  {M{\'e}nard}, F., {Kamp}, I., {Thi}, W.-F., {Duch{\^e}ne}, G.,
  {Augereau}, J.-C.:
  {\bff Continuum and line modelling of discs
  around young stars -- 300000 disc models for HERSCHEL/GASPS},\\
  Monthly Notices of the Royal Astronomical Society 405, L26--L30,
  \href{https://doi.org/10.1111/j.1745-3933.2010.00852.x}{(2010)}

  \item[\bff P3] \underline{Woitke, P.}, {Riaz}, B.,
  {Duch{\^e}ne}, G., {Pascucci}, I., {Lyo}, A. -R., {Dent},
  W.~R.~F., {Phillips}, N., {Thi}, W.-F., {M{\'e}nard},
  F., {Herczeg}, G.~J., {Bergin}, E., {Brown}, A., {Mora},
  A., {Kamp}, I., {Aresu}, G., {Brittain}, S., {de
  Gregorio-Monsalvo}, I., {Sandell}, G.:
  {\bff The unusual protoplanetary disk around the T\,Tauri star
  ET Chamaeleontis},\\
  Astronomy \& Astrophysics 534, A44,
  \href{https://doi.org/10.1051/0004-6361/201116684}{(2011)}

  \item[\bff P4] \underline{Woitke, P.}, {Min}, M., {Pinte}, C.,
    {Thi}, W.-F., {Kamp}, I., {Rab}, C., {Anthonioz}, F.,
    {Antonellini}, S., {Baldovin-Saavedra}, C., {Carmona}, A.,
    {Dominik}, C., {Dionatos}, O., {Greaves}, J., {G{\"u}del}, M.,
    {Ilee}, J.~D., {Liebhart}, A., {M{\'e}nard}, F., {Rigon}, L.,
    {Waters}, L.B.F.M., {Aresu}, G., {Meijerink}, R., {Spaans}, M.:
    {\bff Consistent dust and gas models for protoplanetary disks.
    I.~Disk shape, dust settling, opacities, and PAHs},\\
    Astronomy \& Astrophysics 586, A103,
    \href{https://doi.org/10.1051/0004-6361/201526538}{(2016)}

  \item[\bff P5] \underline{Woitke, P.}, {Min}, M., {Thi}, W.-F.,
  {Roberts}, C., {Carmona}, A., {Kamp}, I., {M{\'e}nard}, F., {Pinte},
  C.:
  {\bff Modelling mid-infrared molecular emission lines from T
    Tauri stars},\\
  Astronomy \& Astrophysics 618, A57,
  \href{https://doi.org/10.1051/0004-6361/201731460}{(2018)}


  \item[\bff P6] \underline{Woitke, P.}, {Kamp}, I., {Antonellini},
    S., {Anthonioz}, F., {Baldovin-Saveedra}, C., {Carmona}, A.,
    {Dionatos}, O., {Dominik}, C., {Greaves}, J., {G{\"u}del}, M.,
    {Ilee}, J.~D., {Liebhardt}, A., {Menard}, F., {Min}, M., {Pinte},
    C., {Rab}, C., {Rigon}, L., {Thi}, W.~F., {Thureau}, N., {Waters},
    L.~B.~F.~M.:
    {\bff Consistent Dust and Gas Models for Protoplanetary
    Disks.  III.~Models for Selected Objects from the FP7 DIANA
    Project}\\
    Publications of the Astronomical Society of the Pacific
    131, 064301,
    \href{https://doi.org/10.1088/1538-3873/aaf4e5}{(2019)}

  \item[\bff P7] Dionatos, O., \underline{Woitke, P.}, {G{\"u}del}, M.,
    {Degroote}, P., {Liebhart}, A., {Anthonioz}, F.,
    {Antonellini}, S., {Baldovin-Saavedra}, C., {Carmona},
    A., {Dominik}, C., {Greaves}, J., {Ilee}, J.~D.,
    {Kamp}, I., {M{\'e}nard}, F., {Min}, M., {Pinte}, C.,
    {Rab}, C., {Rigon}, L., {Thi}, W.~F., {Waters},
    L.~B.~F.~M.:
    {\bff Consistent dust and gas models for protoplanetary disks.
      IV.~A panchromatic view of protoplanetary disks},\\
    Astronomy \& Astrophysics 625, A66,
    \href{https://doi.org/10.1051/0004-6361/201832860}{(2019)}

\end{itemize}

}

\cleardoublepage

\selectlanguage{\english}
\cftaddtitleline{toc}{section}{\bff Abstract}{}

{\sf

\noindent  
{\Large\bff Physico-chemical Processes in Planet-forming Discs}\\[4mm]
{\large Peter Woitke} \\[2mm]
Space Research Institute (IWF), Austrian Academy of Science, Graz,
Austria\\[10mm] 
{\large\bff Abstract}\\[0mm]

\noindent Planets form in clouds of gas and dust rotating around new-born
stars, as a common byproduct of star formation.  The physical and
chemical conditions in these planet-forming discs decide upon when,
where and how many planets will be formed, how massive they are, and
which initial solid and gas composition they obtain.  Recent exoplanet
statistics have shown that the process of planet formation must be
very robust indeed, as on average, every star is observed to have at
least one planet.

This thesis summarises my scientific works in the research field of
thermo-chemical modelling of planet-forming discs since 2009, in
particular the development and applications of the \underline{Pro}toplanetary
\underline{Di}sc \underline{Mo}del (\ProDiMo).  By combining chemical
rate networks with continuum \& line radiative transfer, and the
calculation of all relevant dust and gas heating \& cooling rates in
an axisymmetric disc structure, these models make detailed predictions
about the molecular composition of the disc, the discs' internal gas
and dust temperature structure, and the stability and composition of
ice layers which form on top of the refractory dust grain
surfaces. The development of \ProDiMo\ was a process that took about
15\,years, is still ongoing, and involved a team of international
researchers, in particular Inga Kamp (Groningen University),
Wing-Fai Thi and Christian Rab (both now in Garching), besides some of
mine and some of their PhD-students.  I have started the \ProDiMo\
project, and am considered as the main developer, but without this
team-work, \ProDiMo\ would not have its capabilities and would not be
at the level of international recognition that is has achieved to
date.

Using formal solutions of the line \& continuum radiative transfer
equation along a bundle of parallel rays toward the observer, we can
predict the spectral appearance of these discs from optical to
millimetre wavelengths, for example the continuum and line fluxes,
monochromatic images, radial intensity profiles, high-resolution line
profiles that probe the disc dynamics, visibilities and channel
maps. A large part of this thesis describes the publications that
compared these predictions to disc observations that have been
obtained by various space-borne and ground-based astronomical
instruments, in particular Herschel/PACS, Spitzer/IRS, VLT/CRIRES,
JWST/MIRI and ALMA, which observe at wavelengths between a few micron
to a few millimetres.  These observations probe the gas and the dust
in different radial disc regions and in different layers above the
midplane.  Therefore, a recurring theme in my scientific work was
is to combine these multi-wavelength line and continuum observations,
and to develop disc models that can predict all observations
simultaneously, as good as possible, even at the expense of fitting
certain observations less convincingly.  Only if we succeed in combining
the observational data from different instruments and different
wavelength regions, a true holistic understanding of planet-forming discs
can be achieved.

This thesis summarises the conclusions drawn from the \ProDiMo\
thermo-chemical disc models about the chemical and physical state of
protoplanetary discs as the birth places of exoplanets.

}

\vspace*{7cm}

\noindent\parbox{12cm}{
{\bff Acknowledgements}
\vspace*{3mm}
{\small\sf
  
I would like to thank my wife Christiane Helling for her continuous
support and scientific inspiration. Between 2007 and 2011, when most
of \ProDiMo\ was developed, she had to continue life together with our
daughter Johanna, while I was commuting to Edinburgh and later working
abroad in Vienna.  When I became job-less in the wake of the 2008 financial
crisis, I am not sure whether I would have managed to stay in science
without her support. She gave me the motivation to carry on.

\hspace*{3mm} During my scientific career, I had the pleasure to
discuss physics and chemistry and collaborate with a number of highly
intelligent, knowledgable, and inspiring scientists, from whom I
learnt a lot. I want to mention a few names here: Erwin Sedlmayr,
Carsten Dominik, Peter Cottrell, Vincent Icke, Sime-Jan Parrdekoper,
Uffe Gr{\r a}e J{\o}rgensen, Wing-Fai Thi, Bill Dent, Inga Kamp,
Christian Rab, Christophe Pinte, Francois M{\'e}nard, Geoffroy Lesur,
Michiel Min, Manuel G{\"u}del, Ken Rice, Peter Hauschildt, Guillaume
Laibe, Will Rocha, Anders Johansen, Rens Waters, and Christiane
Helling.  I often found myself pondering about their views and
insights, and their experience with numerical methods, which motivated
me to try attacking the next level of physical consistency in my
models.

\hspace*{3mm} I would also like to thank my former and current Master
and PhD students, in particular Vasco Schirrmacher, Gilles Niccolini,
Ian Tilling, Oliver Herbort, Clayton Roberts, Aditya M.~Arabhavi,
Leoni Janssen, Marie-Luise Steinmeyer, Jayatee Kanwar, Thorsten
Balduin and Till K{\"a}ufer. They have found ways to develop and carry
out new ideas which I could not, and some of their work has
contributed to the development of \ProDiMo.

\hspace*{3mm} I want to thank the IWF for welcoming me as a new group
leader, in particular Werner Magnes, Aris Valavanoglou, Irmgard
Jernej, Wolfgang Voller, Franz Ginner, Luca Fossati, Manfred Stellar,
Ludmila Carone, Ruth Taubner, and Helmut Lammer. Their support made
our start in Graz much easier.

\hspace*{3mm} Finally, I would like to thank the Graz University of
Technology, in particular Prof.~Von der Linden, for supporting my
habilitation to obtain the venia docendi that will enable me to enrich
the teaching programme at the TU Graz in the future.

}}

\cleardoublepage

\pagestyle{empty}
\tableofcontents
\cleardoublepage

\pagestyle{headings}
\section{\bff Introduction to Planet-forming Discs}
\markboth{INTRODUCTION TO PLANET-FORMING DISCS}
         {1.\ \ INTRODUCTION TO PLANET-FORMING DISCS}
         
\subsection{Star, disc and planet formation}
\markright{1.1\ \ STAR, DISC AND PLANET FORMATION}

Star formation happens at the end of a line of complex developments
in the cold galactic medium towards ever smaller spatial structures,
driven by gravity, magnetic fields and turbulence, see reviews of the
Protostars and Planets VII conference \citep{Chevance2023, Zucker2023,
  Pineda2023}.  These structures span about five orders of magnitudes,
from spiral arms and giant molecular clouds ($\approx\!50$\,pc in
diameter) to bubbles, filaments, cores and discs ($\approx\!100$\,au).
When the cold cores of molecular clouds finally collapse under the
pull of their own gravity, new stars are born in their centres. As the
angular momentum of the in-falling matter is conserved and the
collapse reduces the spatial scales by several orders of magnitude,
the dust and gas begin to spin up, forming what is known as a
protoplanetary disc \citep[e.g.][]{Yorke1993,Dominik2015}.

\begin{figure}[!b]
  \vspace*{-1mm}
  \includegraphics[width=16cm]{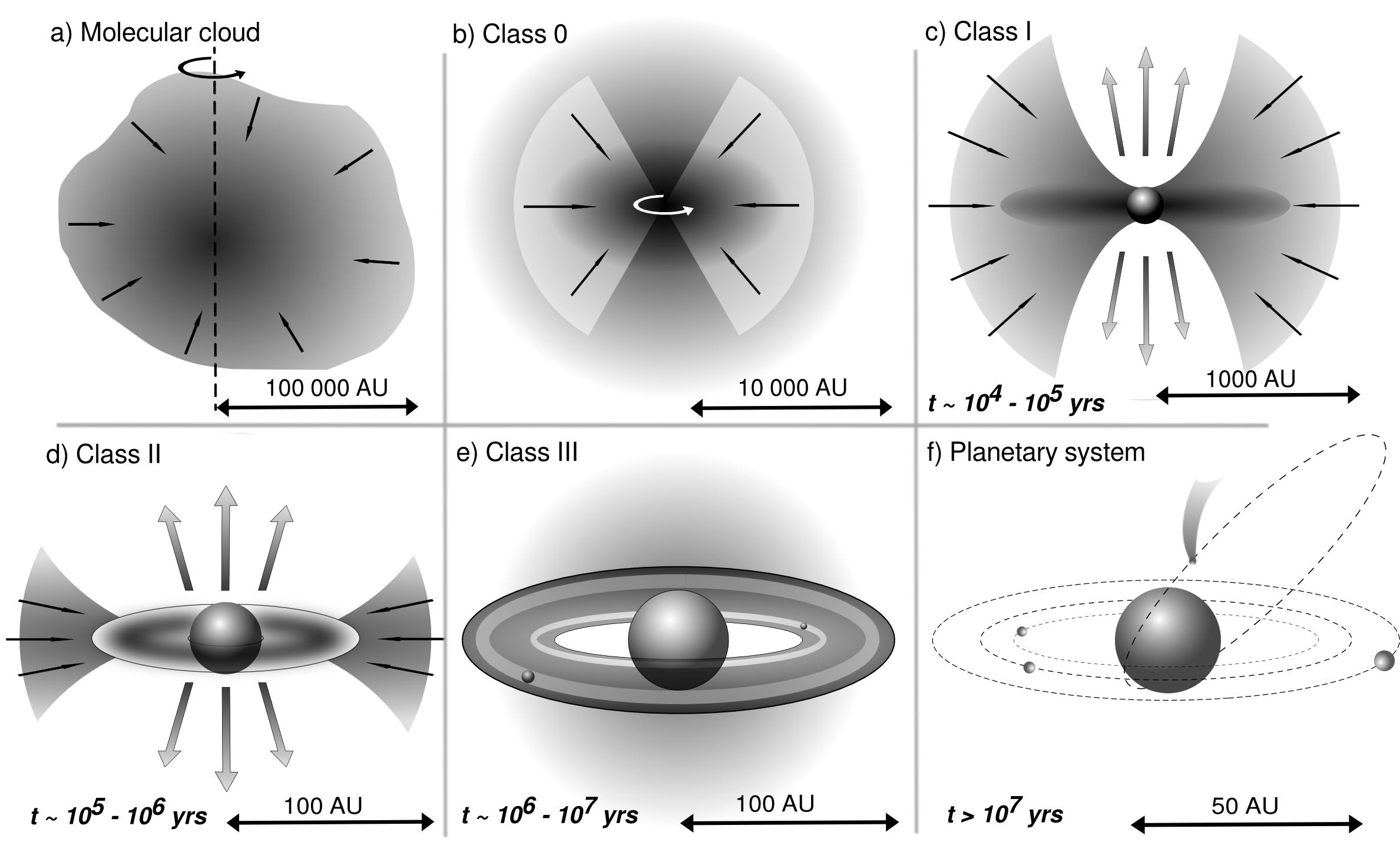}\\[-1mm]
  \resizebox{\textwidth}{!}{\parbox{18.2cm}{\caption{The phases of
        star and planet formation with timescales and spatial scales
        following the sketch developed by \citet{Drazkowska2022}.}
  \label{fig:discs}}}
  \vspace*{-1mm}
\end{figure}

These discs undergo several stages of evolution, as outlined in
Fig.~\ref{fig:discs}.  The following classification of these
evolutionary stages was developed originally by \cite{Lada1984},
\cite{Kenyon1995}, and \cite{Williams2011}: Class 0 objects are
short-lived and are characterised by the first appearance of
significant infall and angular motions, but yet without clear
observational evidence for the nascent star at the centre. In Class I
objects, the central star becomes visible and generates feedback by
driving fast jets along the rotation axes. Nevertheless, the
luminosity of the object is still generated mainly by frictional
forces in the in-falling disc, i.e., the discs are self-luminous
(active discs). In contrast, Class II objects are mainly powered by
the star (passive discs), there is only little ongoing accretion, and
the jets disappear. Class III is a designation for discs in the
clearing phase, where the matter in the disc is lost to space in the
form of disc winds or is consumed by planet formation.  Class III
objects include the so-called debris discs, in which collisions of
larger bodies re-create small dust particles and gas, causing
the disc to become observable again in the infrared.


However, the exact definition and duration of the different phases is
debated in the literature. The timescales stated in the sketch of
\citet[][Fig.~\ref{fig:discs}]{Drazkowska2022} are at the short end.
For example, many of the so-called Herbig Ae/Be objects (Class II
discs around massive stars of spectral type A or B) must be
significantly older than $10^6$\,years, because these stars start as
K-type stars and it takes at least 5\,Myrs to evolve into A or B stars
toward the main sequence \citep[e.g.][]{Siess2000}, yet their discs
belong to the most massive and most luminous Class II discs known.
Hydro-models, which simulate the disc evolution, generally predict
that the discs start small and then viscously spread in radius
\citep[e.g.][]{LyndenBell1974,Hartmann1998,Armitage2010,Vorobyov2020,
  Morbidelli2022}, which would mean that the Class I discs are
actually the smallest, however, they are still embedded into a much
larger envelope that feeds them, so from an observational point of
view, Class I discs are typically quite large.

\begin{figure}[!b]
  \vspace*{-4mm}
  \hspace*{-4mm}
  \resizebox{164mm}{!}{
  \begin{tabular}{cc}
    $\lambda\!=\!1.6\,\mu$m & $\lambda\!=\!1.3\,$mm\\
  \begin{minipage}{77mm}
  \vspace*{-14mm}    
  \includegraphics[width=8cm,trim=70 70 70 70,clip]{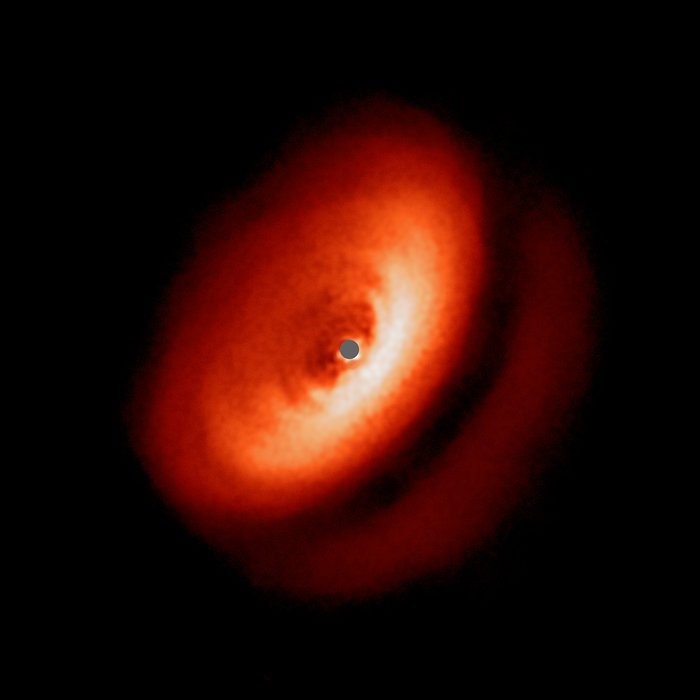} 
  \end{minipage} &
  \begin{minipage}{7cm}
    \centering
  \includegraphics[width=7cm,trim=0 0 0 0,clip]{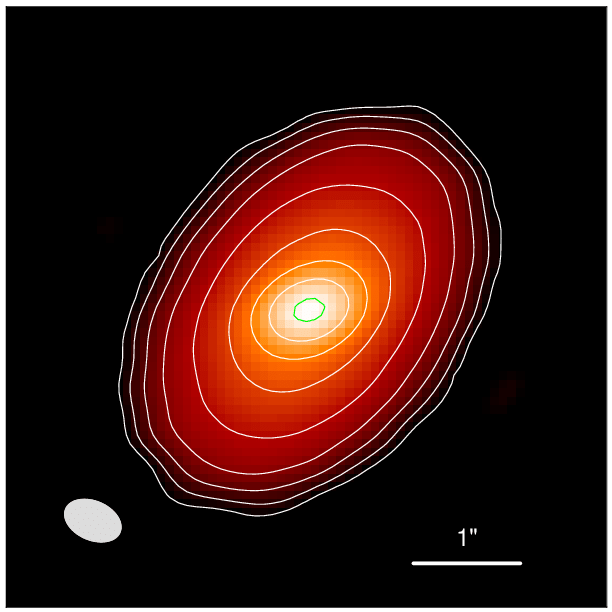}\\
  \includegraphics[width=2.5cm,trim=0 0 0 0,clip]{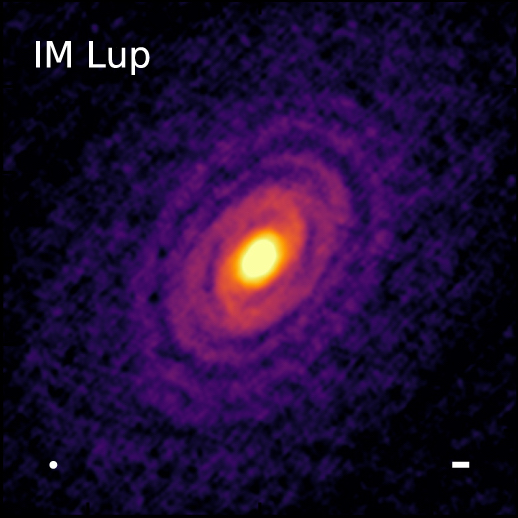}
  \end{minipage}  
  \end{tabular}}\\[0mm]
  \resizebox{\textwidth}{!}{\parbox{18.7cm}{\caption{Three continuum
        images from the dusty disc around the young T\,Tauri star
        IM\,Lupi. The left is a polarimetric H-band (1.6\,$\mu$m)
        image taken with the SPHERE instrument on ESO's Very Large
        Telescope \citep{Avenhaus2018}.  The right side shows two
        images at 1.3\,mm taken with ALMA, using two different spatial
        resolutions, with PSFs and spatial scales as shown.  The upper
        right shows the entire disc with a spatial resolution of about
        0.4'', with contours showing doubled intensities
        \citep{Pinte2018}. The lower right uses a different instrument
        setup (spatial resolution 0.045'', the short bar shows a
        stretch of 10\,au) to focus on the inner disc regions
        \citep{Andrews2018}. All three images are to scale, showing
        that the discs often appear significantly larger in gas than
        in dust. The left picture also shows that discs have a flared
        structure, whereas they appear to be flat at mm-wavelengths. 
        With a radial disc extension of about 400\,au in the left picture,
        IM\,Lupi is one of the largest T\,Tauri type discs known.}
      \label{fig:images}}}
  \vspace*{-2mm}
\end{figure}

Growing evidence points towards the fact that planets are born at
nearly the same time as their host stars in the same disc of material.
First observational evidence of a ringed inner disc structure at
millimetre wavelengths, which is interpreted as a signpost of ongoing
planet formation, has been found in the object HL\,Tau
\citep{ALMA2015} which has an age of only about $5\times 10^5$
\,years.  But the same effects are still clearly seen in TW\,Hya, the
age of which is estimated to be about 10 Myrs \citep{Vacca2011}.

\subsection{Observational appearance of discs}
\markright{1.2\ \ OBSERVATIONAL APPEARANCE OF DISCS}

\begin{figure}[!t]
  \hspace{3mm} $\lambda\!=\!1.3\,$mm
  \hspace{59mm} ${\rm
    CO}\ J\!=\!2\!\to\!1$\\
  \includegraphics[width=16cm]{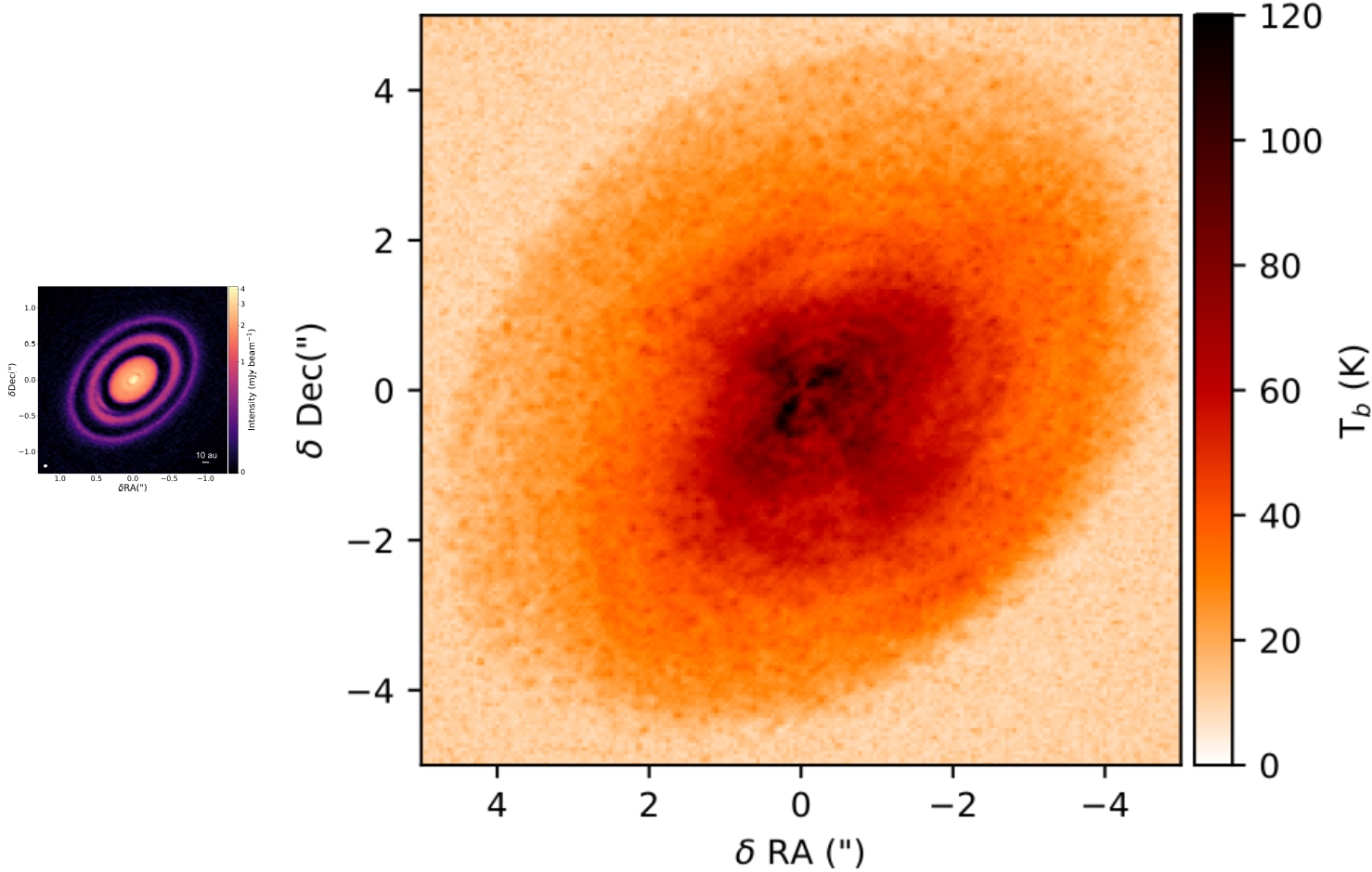}\\[-1mm]
  \resizebox{\textwidth}{!}{\parbox{18.2cm}{\caption{The Herbig Ae
        disc of HD\,163296 as seen by ALMA \citep{Isella2018}. The
        picture on the left shows the band\,6 continuum (about
        1.3\,mm) on a non-linear colour scale, with flat ring-like
        structures at radii of about 70\,au and 100\,au, similar to
        structures seen in other discs observed with high angular
        resolution by ALMA.  In contrast, the CO\,$J\!=\!2\!\to\!1$
        line map shows evidence for the CO gas being radially extended
        to at least 500\,au, without clear rings.  The images are to
        scale.}
  \label{fig:HD163296}}}
\end{figure}

Protoplanetary discs have been first imaged at UV and optical
wavelength, in particular the Hubble Space Telescope (HST) provided the
first disc images \citep{O'Dell1996}. Since 2014, the Atacama Large
Millimeter Array (ALMA) has started to provide disc images at
mm-wavelengths with exquisite spatial resolution, and more recently the
Spectro-Polarimetric High-contrast Exoplanet REsearch (SPHERE)
instrument at the Very Large Telescope (VLT) has imaged discs in the
H-band ($\sim\!1.6\,\mu$m), see a collection of continuum images of IM\,Lupi
in Figs.~\ref{fig:images} and \ref{fig:HD163296}.

\clearpage
\section{\bff The ProDiMo Model}
\markboth{THE PRODIMO MODEL}
         {2.\ \ THE PRODIMO MODEL}

The \underline{Pro}toplanetary \underline{Di}sc \underline{Mo}del
(\ProDiMo) project was started in 2007, and the major software
development was finished by myself in 2009 during my time as a
post-doc at the UK Astronomy Technology Center (ATC) in Edinburgh in
collaboration with Dr.\ Wing-Fai Thi (University of Edinbugh, now in
Garching) and Prof.\ Inga Kamp (Groningen).  The immediate scientific
goal at the time was to predict the forbidden far
infrared (far-IR) fine-structure emission lines of atoms and ions
observed by the Herschel Space telescope (Herschel) using the PACS and
SPIRE instruments, such as [OI]\,$2\!\to\!1$ (63.18\,$\mu$m),
[OI]\,$3\!\to\!2$ (145.53\,$\mu$m), and [CII]\,$2\!\to\!1$
(157.74\,$\mu$m), as well as a few far-IR high-$J$ rotational emission
lines of CO and \ce{H2O} between about 60\,$\mu$m and 180\,$\mu$m, see
\citet{Dent2013}.

The first \ProDiMo\ paper was published in Astronomy \&
Astrophysics in 2009 \citep[][see Publication\,1 on page
  \pageref{PUB1}]{Woitke2009a}, immediately followed by an application
to far-IR water emission lines in Herbig Ae discs \citep{Woitke2009b}.
The physical and chemical roots of \ProDiMo\ reach back to the
COSTAR-program formerly developed by Inga Kamp
\citep{Kamp2000,Kamp2001}, but \ProDiMo\ was completely re-written
in modern Fortran~90, with a modular parallel software architecture,
and includes many more physical and chemical processes, which are
consistently coupled to an in-built continuum and line radiative
transfer.  The basic setup and design of \ProDiMo\ is nowadays
known as a {\sl thermo-chemical disc model}.

The basic modelling idea of \ProDiMo\ is to assume a quasi-static 2D
axisymmetric gas density structure $\nH(r,z)$, where $\nH$ is the
hydrogen nuclei particle density, $r$ is the distance from the
rotation axis (often simply called ``radius''), and $z$ is the height
over the midplane of the disc.  The gas is assumed to be on stable
circular Keplerian (or sub-Keplerian) orbits, which are
pressure-supported to maintain their height $z$.  Additional
model parameters are the stellar irradiation properties, such
as the stellar effective temperature and luminosity, the stellar
non-photospheric UV and X-ray properties, and similar parameters
describing the irradiation of the disc from the outside, i.e.\ the
cosmic microwave background, interstellar UV irradiation, incident
X-ray and cosmic ray fluxes.  The interstellar irradiation properties
are assumed to be isotropic.  All parameters are assumed to be
constant in time (neglecting, for example, stellar variability), and
therefore, the resulting chemical and temperature structure in the
disc will eventually also become time-independent.  It is this disc
structure that we primarily want to calculate with \ProDiMo.

\begin{figure}[!t]
  \hspace*{-6mm}
  \includegraphics[width=17cm]{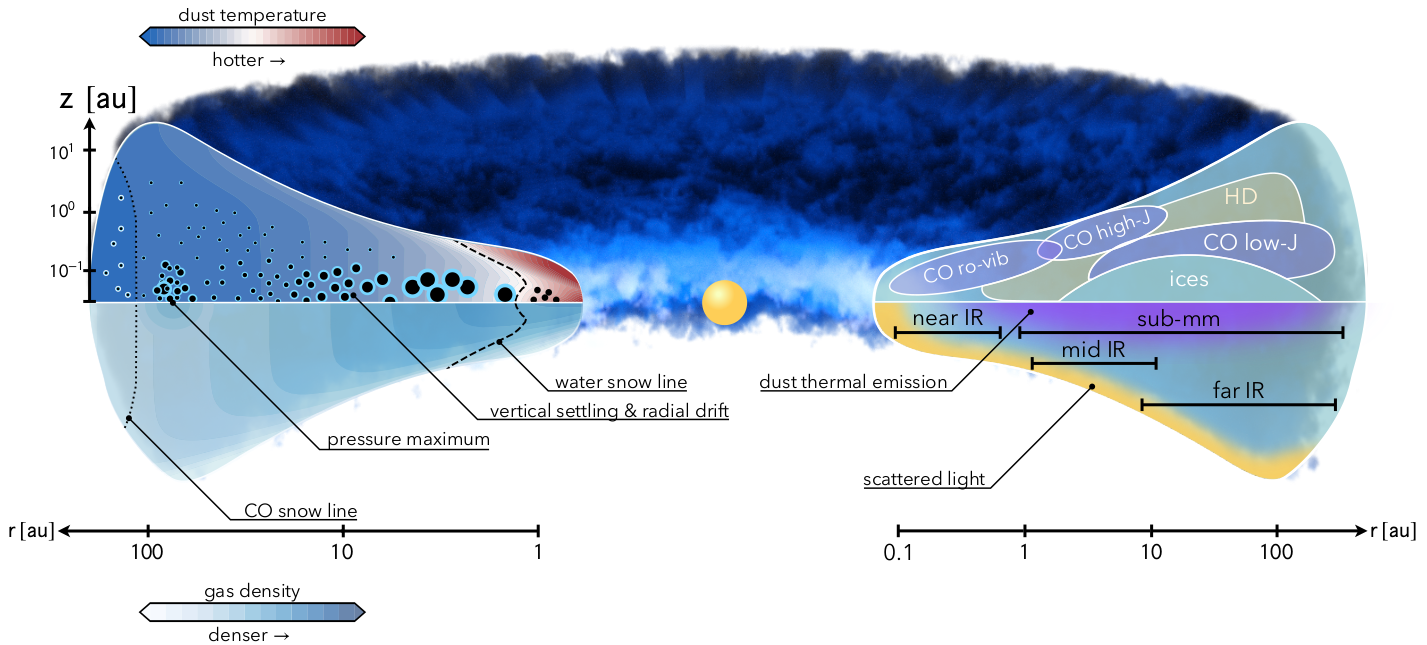}\\[-1mm]
  \resizebox{\textwidth}{!}{\parbox{18.2cm}{\caption{Sketch of the
        disc geometry and some of the important dust and gas phase
        physical and chemical processes. The right side shows which
        kind of continuum and line observations probe certain regions
        in the planet-forming discs \citep{Miotello2022}.}
  \label{fig:discsketch}}}
\end{figure}

The slowest relaxation process in disc is usually the chemistry, in
fact the transformation of some ice phases in the midplane into others
can take longer than typical disc lifetimes, so when we refer to a
``time-dependent \ProDiMo\ model'', we mean a disc model where
the densities and the radiation field in the disc is assumed to be
constant, but we advance the chemistry at all points in the disc to a
desired disc age \citep[e.g.][]{Helling2014}.  This concept can be
relaxed to have different epochs with different irradiation
parameters, for example to model the chemistry after an episodic
accretion event \citep{Rab2017b}.

\ProDiMo's basic modelling method can shortly be described as a 2D
XPDR-code, where PDR means photon dominated region
\citep[e.g.][]{Roellig2007,Visser2009,Wolfire2022} and the X means
additional X-ray physics and chemistry: In a given gas and dust
density structure, radiative transfer is carried out to get the local
radiation field and dust temperature. Based on these properties of the
medium, a kinetic chemistry for gas and ice is solved (or advanced in
time), and the gas temperature is calculated by solving the gas energy
balance including all relevant heating and cooling processes.

\ProDiMo\ has been further developed ever since, adding more
physics, chemistry and radiative transfer modules and enlarging its
capabilities to predict various kinds of continuum and line
observations based on the calculated chemical and
temperature-structure in the disc. This habilitation thesis summarises the
creation and subsequent development of \ProDiMo, it's
applications to various kinds of astronomical data, and the
conclusions drawn from these simulations about the chemical and
physical state of protoplanetary discs as the birth places of
exoplanets.

\subsection{Disc geometry and density setup}
\markright{2.1\ \ DISC GEOMETRY AND DENSITY SETUP}
\label{sec:discshape}

Figure~\ref{fig:discsketch}, taken from \citet{Miotello2022},
visualises the assumed disc geometry and some of the important dust
and gas phase physical and chemical processes that are included in the
\ProDiMo\ disc simulations.  The disc is assumed to have an
axisymmetric density structure, which can be setup in various ways,
either using the disc mass and various disc shape parameters, or by
using the output of other, e.g.\ hydrodynamical programs like FEOSAD
\citep{Vorobyov2020}, PLUTO \citep{Mignone2007} or MOCCASIN
\citep{Ercolano2005,Ercolano2008}. In the standard \ProDiMo\
models, we assume pressure-supported Keplerian orbits of gas and dust
with zero radial and vertical velocities, see Eqs.(1-6) in
\citep[][Publication\,1 on page \pageref{PUB1}]{Woitke2009a}.
However, \citet{Rab2017b} used a density structure derived from disc
evolution models for rotationally flattened in-falling envelopes, in
this case these velocity components are non-zero in \ProDiMo. The
radial column density structure $\Sigma(r)$ can be setup by a radial
powerlaw \citep{Woitke2009a} or a combination of powerlaw and
exponential tapering-off \citep[][see Publication\,4 on page
  \pageref{PUB4}]{Woitke2016}.  The vertical disc extension is either
assumed to follow a Gaussian distribution with given scale height that
only depends on radius (parametric radial powerlaw), or is calculated
consistently with the resulting gas temperature structure and mean
molecular weight \citep{Woitke2009b, Thi2011b}.

\subsection{Dust opacities and settling}
\markright{2.2\ \ DUST OPACITIES AND SETTLING}
\label{sec:dust}

\begin{figure}[!t]
  \centering
  \vspace*{-3mm}
  \includegraphics[width=15cm]{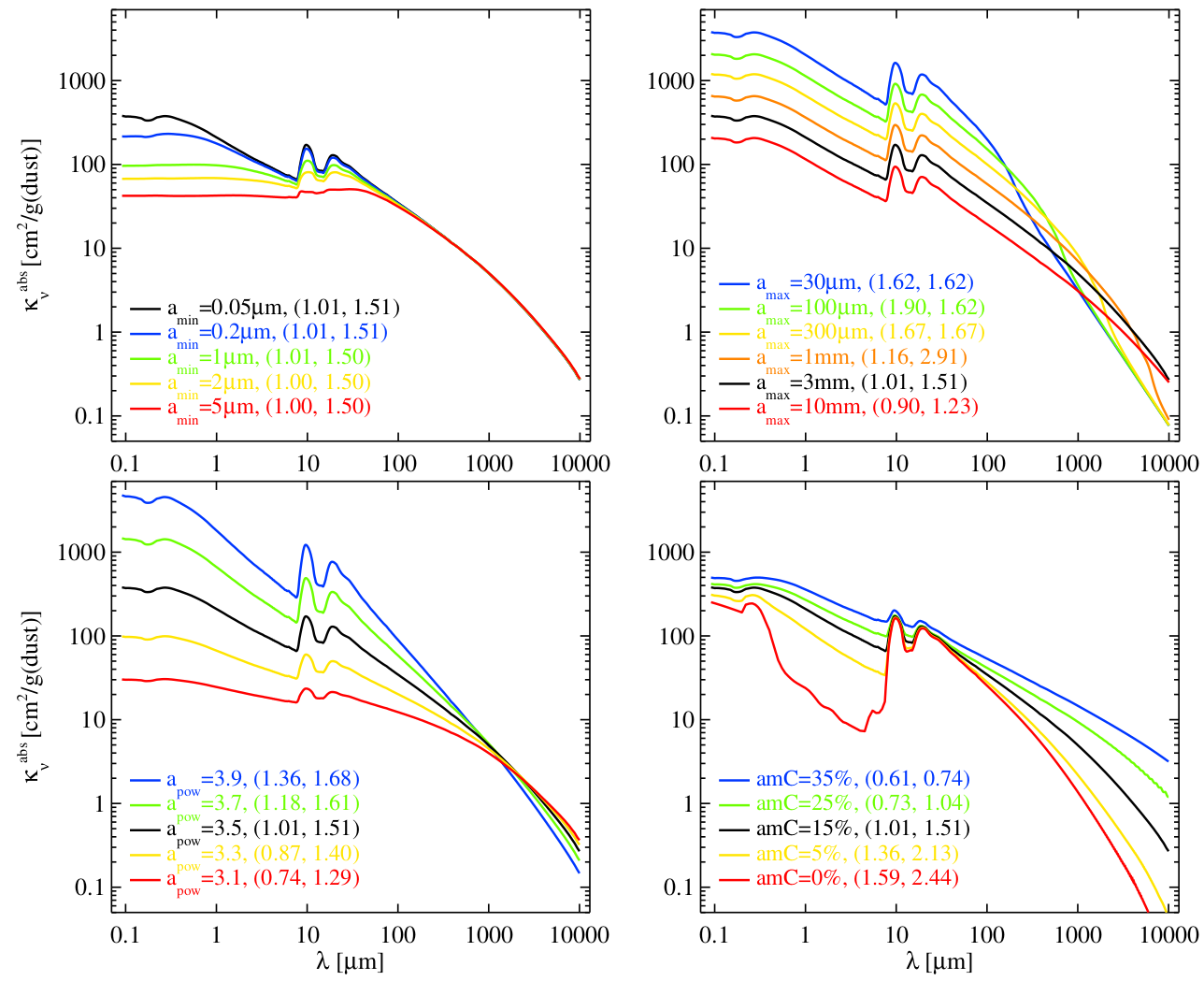}\\[-1mm]
  \resizebox{\textwidth}{!}{\parbox{18.2cm}{\caption{DIANA standard
        dust opacities, figure re-printed from Publication\,4
        \citep{Woitke2016}. All plots show the dust absorption
        coefficient per dust mass as function of dust size and
        material parameters. The upper two figures show the
        dependencies on minimum and maximum particle size, $a_{\rm
          min}$ and $a_{\rm max}$. The lower two plots show the
        dependencies on dust size powerlaw index $a_{\rm pow}$ and on
        the maximum volume fraction of amorphous carbon $\rm
        amC$. Fixed values for porosity (25\%) and maximum hollow
        volume ratio $V_{\rm hollow}^{\rm max}\!=\!0.8$ are assumed.
        The two numbers in brackets represent the log-log dust
        absorption opacity slopes between 0.85\,mm and 1.3\,mm, and
        between 5\,mm and 1\,cm. The black lines are all identical.}
  \label{fig:DIANAopacities}}}
\end{figure}

The assumptions about the dust opacities are described in \citep[][see
  Publication\,4 on page \pageref{PUB4}]{Woitke2016}.  We assume a
powerlaw dust size distribution function $f(a)$ between a minimum and
a maximum particle size, $\amin$ and $\amax$, respectively, typically
covering sub-micron to millimetre grains, about 4-5 orders of
magnitude in size space.  The grains are assumed to have the same
uniform material composition.  The wavelength-dependent scattering and
absorption properties of the grains with a single size $a$ are
calculated by applying Mie-theory based on the effective optical
constants given the assumed material mix, porosity, and shape.  In
\citet{Woitke2016}, we established the ``standard DIANA dust
opacities'' where we assume a mixture of two materials (silicate and
amorphous carbon), 25\% porosity, effective medium theory according to
\citep{Bruggeman1935}, and a shape distribution of hollow spheres
\citep{Min2011}.
These assumptions were guided by the previous experience to fit disc
continuum and line observations, and by a detailed study of optical
properties of dust aggregates \citep{Min2016}, where the Discrete
Dipole Approximation (DDA) is used to compute the interaction of light
with complexly shaped, inhomogeneous aggregate particles. These
computations are computationally too expensive to be included in
complex disc models, but the simple method described above avoids
several artefacts of Mie theory (spherical resonances), can account
for the most important shape effects, and captures the co-called
``antenna-effect'', where irregularly shaped inclusions of conducting
materials result in a considerable increase of mm-cm absorption
opacities, see Fig.~\ref{fig:DIANAopacities}.

These assumptions have partly been relaxed in \citet[][see
  Publication\,6 on page \pageref{PUB6}]{Woitke2019}, allowing for
several disc zones with different dust/gas ratios, and different
values for $\amin$ and $\amax$.  In \citet{Arabhavi2022}, the ``bare''
grains made of refractory materials are assumed to be covered by a
layer of ice of varying material and thickness.  The
position-dependent ice abundances are taken from the results of the
chemistry, where ice formation is part of a kinetic chemical network.
These models hence require iterations between the computation of the
dust opacities, the radiative transfer, and the chemistry. However, in
all other, more basic \ProDiMo\ disc models, we assume
temperature-independent dust opacities, which do not change with time,
so we need to calculate the dust opacities only once during
initialisation.

Another important model assumption is the treatment of vertical dust
settling, where we assume a balance of upward directed turbulent
mixing and downward directed gravitational settling, using the
theoretical concepts of either \citet{Dubrulle1995} or
\citet{Riols2018}. In both cases, all grains in a given disc column
are redistributed vertically such that the midplane becomes
dust-enriched, in particular considering the larger dust particles,
see \citet{Woitke2016} for details. In this way, the unsettled dust
size distribution $f(a)$ mentioned above becomes a position-dependent
quantity $f(a,r,z)$ in \ProDiMo. The advantage of the new
\citet{Riols2018}-setting is that the decrease of the midplane gas
density with height is taken into account, whereas the in old
\citet{Dubrulle1995}-settling treatment, that gas density is assumed
to be constant in one column.

The assumptions about dust size distribution, opacities and settling
determine how much UV radiation can penetrate into the disc, and how
much grain surface per H-nucleus is present.  Therefore, these
assumption are crucial not only for the predicted continuum
observations, but also for chemistry, heating, and line emission.  In
particular, other comparable thermo-chemical disc models \citep[see
  summary in Table~1 of][]{Woitke2016} often assume a single grain
size, or a distribution of small grains around\linebreak (sub-)micron
sizes \citep{Mathis1977}, which produces far-UV dust extinction
opacities that are larger by about a factor of 100 than our DIANA
standard dust opacities, see discussion in \citep{Woitke2016}.
Therefore, the various assumptions about the disc shape and the dust
opacity parameters can result in a very different spectral appearance
of the discs in line observations \citep[see Publication\,6 on page
  \pageref{PUB6}]{Woitke2019}.

  \def\kabs{\kappa_\nu^{\rm abs}}
  \def\ksca{\kappa_\nu^{\rm sca}}
  \def\kext{\kappa_\nu^{\rm ext}}
  \def\TPAH{T_{\rm PAH}}
  \def\nnn{\vec{n}}
  \def\rrr{\vec{r}}
  \def\rr0{\vec{r}_0}

\subsection{Continuum radiative transfer}
\markright{2.3\ \ CONTINUUM RADIATIVE TRANSFER}
\label{sec:RT}

The first modelling step in \ProDiMo\ is to solve the continuum
radiative transfer problem of an irradiated disc, which results in the
internal dust temperature structure $\Td(r,z)$ and the internal
radiation field $J_\nu(r,z)$. The basic equations to be solved are the
radiative transfer equation and the energy balance equation for the
dust grains. The radiative transfer equation is given by
\begin{equation}
  \frac{dI_\nu}{d\tau_\nu} = S_\nu - I_\nu
  \label{eq:RT}
\end{equation}
where $I_\nu\!=\!I_\nu(\vec{r},\vec{n})$ is the spectral intensity at
a certain 3D point $\rrr$ in the disc into a certain direction $\nnn$.
Assuming local thermodynamical equilibrium (LTE) and coherent
isotropic scattering, the source function $S_\nu$ is given by
\begin{equation}
  S_\nu = \frac{\kabs B_\nu(\Td)+ \ksca J_\nu}{\kext} \ ,
  \label{eq:source}
\end{equation}
and the optical depth
\begin{equation}
  \tau_\nu(s) = \int_0^s \kext(\rr0-s'\nnn)\,ds'
  \label{eq:tau}
\end{equation}
is measured backwards along a ray, where $\rr0$ is the point of
interest, $s'$ is the distance backward along the ray $\nnn$,
and $\rr0-s\nnn$ is the point where the ray enters the model
volume. When this point is reached during the numerical integration of
Eq.\,(\ref{eq:RT}), the incident radiation $I_\nu^{\rm
  inc}\exp(-\tau_\nu)$ is added, which can either be the stellar
irradiation, or the (isotropic) interstellar irradiation, depending on
whether the backward ray hits the star or not.
$J_\nu\!=\!\frac{1}{4\pi}\int I_\nu \,d\Omega$ is the mean intensity,
and $B_\nu$ the Planck function, and $\kabs$, $\ksca$ and
$\kext=\kabs+\ksca$ $[\rm cm^{-1}]$ are the dust absorption,
scattering and extinction coefficients, respectively.

The energy balance of the dust grains is written as 
\begin{equation} 
  \Gamma_{\rm dust} \,+ 4\pi\int \kabs\big(J_\nu-B_\nu(\Td)\big)\,d\nu
  ~=~ 0 \ ,
  \label{eq:RE}
\end{equation}
where $\Gamma_{\rm dust}\,\rm[erg/cm^3/s]$ is a non-radiative net
heating rate of the dust (for example via inelastic collisions with
gas particles having a different temperature).  In the standard
\ProDiMo\ models we assume $\Gamma_{\rm dust}\!=\!0$ and
Eq.\,(\ref{eq:RE}) simplifies to the condition of radiative
equilibrium for the dust, appropriate for passive (Class II and III)
discs.

\begin{figure}[!b]
  \begin{tabular}{cc}
  \hspace*{-7mm}  
  \begin{minipage}{83mm}
    \includegraphics[width=83mm]{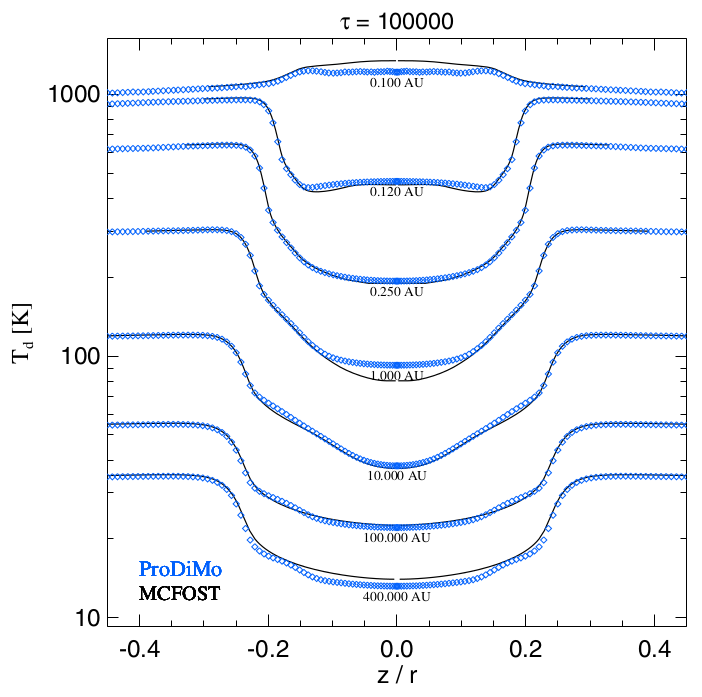}
  \end{minipage} &
  \hspace*{-5mm}  
  \begin{minipage}{83mm}
    \vspace{3mm}
    \includegraphics[width=83mm]{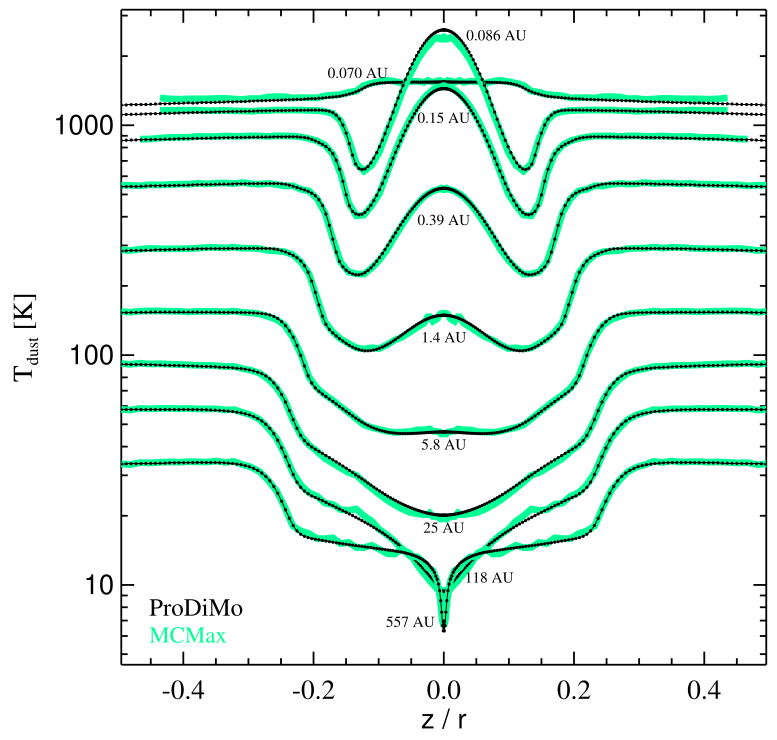}
  \end{minipage}
  \end{tabular}
  \resizebox{\textwidth}{!}{\parbox{18.2cm}{\caption{Resulting dust
        temperature structures $\Td(r,z)$ from \ProDiMo's continuum
        radiative transfer. The left plot shows a re-printed figure
        from Publication~1 \citep{Woitke2009a} for a passive disc
        without dust settling. The right plot shows a re-printed
        figure from \citep{Oberg2022} for an active disc with a mass
        accretion rate of $\dot{M}_{\rm
          acc}\!=\!10^{-8}\rm\,M_\odot/yr$ and Dubrulle settling
        ($\alpha_{\rm settle}\!=\!0.01$). Both results are benchmarked
        against the leading Monte-Carlo radiative transfer programs
        MCFOST \citep{Pinte2006} and MCMax \citep{Min2009}. The
        comparison also reveals the progress on improving the details
        in the continuum radiative transfer module of \ProDiMo\ over
        the years.}
  \label{fig:RTresults}}}
\end{figure}

Only very recently, we have included the process of viscous heating as
a source for dust heating $\Gamma_{\rm dust}$ by means of a diffusion
solver, see Appendix~A in \citep{Oberg2022}, to be able to apply 
\ProDiMo\ also to active protoplanetary and circumplanetary
discs. Figure~\ref{fig:RTresults} shows a comparison between the dust
temperatures calculated for a passive and non-settled T\,Tauri disc,
and a viscously heated active disc with dust settling.  Note the
re-increase of the temperature towards the disc midplane in the inner,
optically thick disc regions, due to viscous heating, and the 
sharp shadow that the inner disc casts onto the outer disc,
resulting in very low temperatures, and another shadow forming
in the outermost disc parts on the right, where the gas density is low
and hence the dust strongly settled.

The actual solution of the radiative transfer problem in discs is
numerically very challenging, because of the huge optical depths of
order $10^6$ involved, see e.g.\ the radiative transfer benchmark
papers by \citet{Pascucci2004} and \citet{Pinte2009}. \ProDiMo\
follows a simple concept here, based on formal solutions of the
radiative transfer equations on a given set of about $20\times 20$
rays that all emanate from a single point in the disc to cover the
$4\pi$ solid angle. Using a spatial grid of $100\times100$ radial and
vertical grid points in cylinder coordinates (each cell is a torus in
3D), this concept requires to trace about
$20\times20\times100\times100=4.000.000$ rays.  The opacities and
intensities are averaged over about 30 frequency bins that cover the
electromagnetic spectrum from 91.2\,nm (the far-UV threshold for neutral
hydrogen ionisation) to about 1\,cm.

Given an initial guess of $\Td(r,z)$ and $J_\nu(r,z)$, \ProDiMo\ (i)
calculates the source functions according to Eq.\,(\ref{eq:source}),
(ii) numerically solves Eq.\,(\ref{eq:RT}) backward along each ray
that originates in every grid point in all considered directions for
all wavelengths, while also integrating the optical depths
(Eq.\,\ref{eq:tau}), and (iii) re-calculates $J_\nu(r,z)$ and
$\Td(r,z)$ from the definition of the mean intensity and from
Eq.\,(\ref{eq:RE}). This procedure is repeated until the results
converge. While this simple concept (called $\Lambda$-iteration)
actually works well for optically thin discs, it fails miserably for
the usually optically thick discs, because the changes applied between
iterations become tiny as the optical depths become large. In order to
solve this problem, and a number of other numerical challenges,
\ProDiMo
\begin{itemize*}
\item uses a log-interpolation of $S_\nu(r,z)$ between spatial
  grid points as the rays are integrated along $s'$,
\item varies the spatial step size $\Delta s$ to make sure that
  the numerical deviations between one full step and two half steps
  are small,
\item carefully auto-adjusts the ray directions during the radiative
  transfer (RT) iterations to focus on the incoming ray directions
  that causes the local dust heating,
\item applies the geometrical extrapolation method from
  \citet{Auer1984} after each $4^{\rm th}$ RT iteration, to
  accelerate the convergence, so \ProDiMo's RT method is an {\sl
    accelerated $\Lambda$-iteration}.
\item (since 2022) uses a 2D diffusion solver for the optically thick
  core on the midplane of the disc.
\end{itemize*}
The numerical details are outlined in \citet[][Sect.~4]{Woitke2009a} and
\citet[][Appendix~A]{Oberg2022}. In summary, \ProDiMo\ uses a
ray-based method to solve the continuum radiative transfer problem in
discs with stellar and interstellar irradiation and internal viscous
heating. The results are well tested against the leading Monte Carlo
radiative transfer codes MCFOST and MCMax.  Compared to these MC codes,
ProDiMo's results are always noise-free and smooth, both spatially and
in wavelength space, which is a huge advantage when discussing
photo-chemistry, but it's performance is slower and the solutions
are limited by the assumption of isotropic scattering.

\subsection{PAHs and other gas opacities}
\markright{2.4\ \ PAHS AND OTHER GAS OPACITIES}
\label{sec:PAHs}

The inclusion of other opacities in \ProDiMo's radiative transfer is
explained in Sect.~8 and Appendix~B in \citet[][see Publication\,4 on page
  \pageref{PUB4}]{Woitke2016}.  Polycyclic Aromatic Hydrocarbon
molecules (PAHs) are observed via their strong mid-IR emission
bands in many Herbig Ae/Be stars \citep[e.g.][]{Maaskant2014}, whereas
detection rates in T\,Tauri stars are much lower \citep{Geers2006},
likely because T\,Tauri stars generate much less blue and soft UV
stellar radiation to heat the PAHs. PAHs in Herbig Ae/Be discs seem to
have sizes of at least 100 carbon atoms \citep{Visser2007}, and to be
somewhat less abundant than in the interstellar medium.

The PAH abundance in the disc is assumed to be given by the standard
abundance in the interstellar medium \citep[ISM,][]{Tielens2008},
modified by factor $f_{\rm PAH}$
\begin{equation}
 \frac{n_{\rm PAH}}{\nH} = 3\times 10^{-7}\,f_{\rm PAH}\,\frac{50}{N_{\rm C}} \ .
\end{equation}
Here, $n_{\rm PAH}\rm\,[cm^{-3}]$ is the PAH particle density,
$\nH\rm\,[cm^{-3}]$ is the hydrogen nuclei density and $N_{\rm C}$ is
the number of carbon atoms in the PAH. Values of $f_{\rm
  PAH}\!\approx\!0.1$ or lower seem typical in Herbig\,Ae discs.

The opacities of neutral and charged PAHs are calculated according to
\citep{Li2001} with updates in \citep{Draine2007}, including the
``graphitic'' contribution in the near-IR and the additional
``continuum'' opacities of charged PAHs. We normally assume $N_{\rm
  C}\!=\!54$ carbon atoms and $N_{\rm H}\!=\!18$ hydrogen atoms
(``circumcoronene''), resulting in a PAH mass of 666.7\,amu and a PAH
radius of 4.87\AA\ \citep{Weingartner2001}. However, $N_{\rm C}$ and
$f_{\rm PAH}$ are free model parameters, as well as a decision whether
to consider the neutral or charged PAH opacities, or a mixture of
both.

\begin{figure}
  \centering
  \includegraphics[width=115mm,height=90mm]{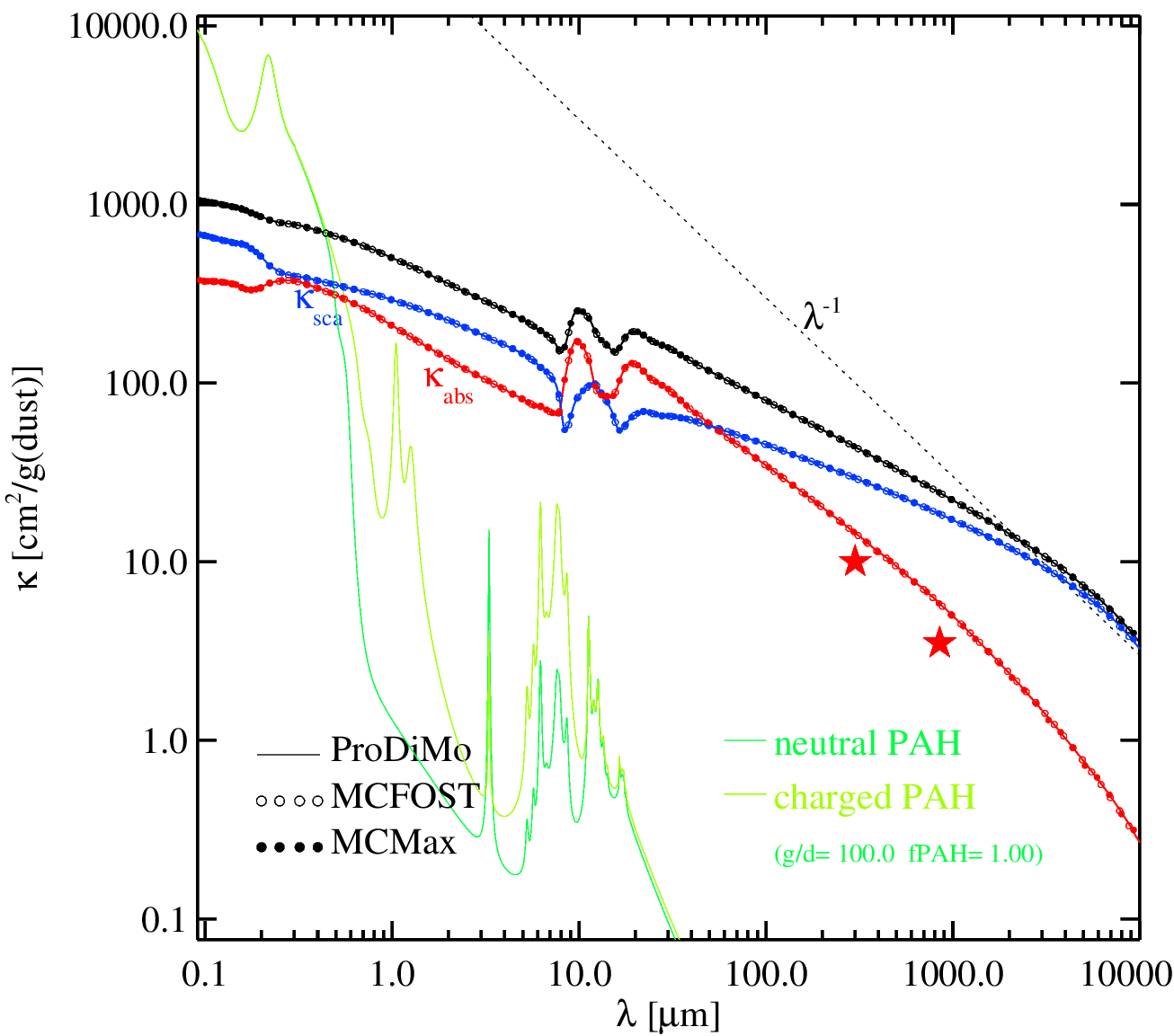}\\[-1mm]
  \resizebox{\textwidth}{!}{\parbox{18.2cm}{\caption{Comparison of
        unsettled DIANA standard dust opacities
        ($\amin\!=\!0.05\,\mu$m, $\amax\!=\!3\,\mu$m,
        $\apow\!=\!3.5$), as calculated by \ProDiMo, MCFOST and MCMax,
        to the opacities of neutral and charged PAH molecules,
        assuming $\rm gas/dust\!=\!100$ and $f_{\rm PAH}\!=\!1$.  The
        two red stars represent dust absorption opacities commonly
        used to derive disc masses from millimetre fluxes: $\rm
        10\,cm^2/g(dust)$ at 1000\,GHz \citep{Beckwith1990} scaled to
        $\rm 3.5\,cm^2/g(dust)$ at 850\,$\mu$m.}
  \label{PAHopac}}}
\end{figure}

To include PAHs in the radiative transfer, we neglect PAH scattering
and extend the source function by two terms describing PAH emission
and absorption
\begin{eqnarray}
 S_\nu &=& \frac{\kappa_\nu^{\rm dust,abs}B_\nu(\Td)
                +\kappa_\nu^{\rm PAH,abs}B_\nu(\TPAH)
                +\ksca J_\nu}
                {\kappa_\nu^{\rm dust,abs}+\kappa_\nu^{\rm PAH,abs}+\ksca} 
                \ ,
\end{eqnarray}
introducing the PAH temperature $\TPAH$. $\Td$ and $\TPAH$ are calculated as
\begin{eqnarray}
 0 &=& 4\pi \int \kappa_\nu^{\rm dust,abs}\big(B_\nu(\Td)-J_\nu\big)\,d\nu 
\ , \label{eq:dustRE}\\
 0 &=& 4\pi \int \kappa_\nu^{\rm PAH,abs}\big(B_\nu(\TPAH)-J_\nu\big)\,d\nu
 \ ,
 \label{eq:PAHRE}
\end{eqnarray}
assuming independent radiative equilibria for the dust grains and the
PAH molecules.  The validity of these assumptions is discussed in
Appendix~B of \citep{Woitke2016}.

For PAH abundances or order $f_{\rm PAH}\!=\!0.01$ the PAH opacities
can usually be neglected in the continuum radiative transfer, see
Fig.\,\ref{PAHopac}. However, when dust settling is taken into account,
the PAH UV opacities can dominate over the UV dust opacities in the
uppermost disc layers, depending on the assumptions about the dust
size distribution and settling. Because the PAHs are very small and
therefore very inefficient scatterers, PAHs can effectively shield the
disc from the stellar and interstellar UV by absorbing the UV and
re-emitting this energy in form of the infrared PAH features at
3.3\,$\mu$m, 6.2\,$\mu$m, 7.6\,$\mu$m, 8.6\,$\mu$m, 11.3\,$\mu$m and
13.5\,$\mu$m. This can have important consequences for the disc
internal gas temperature and chemical structure, see
Sect.~\ref{sec:heatcool}.

Apart from the direct impact of PAHs on the internal radiation field
via radiative transfer effects, PAHs can also be included in 
\ProDiMo' chemistry via five charge states
(Sect.~\ref{sec:chemistry}) and the photoelectric heating rates of
PAHs can be included in the gas energy balance
(Sect.~\ref{sec:heatcool}).  There are various options in 
\ProDiMo\ to include any combination of these effects.  In the most
consistent models, PAHs are included in chemistry and radiative
transfer, and their opacities and photoelectric heating rates are
calculated from the ambient radiation field and the calculated
distribution of their charging states, which requires iterative models.

Molecular UV gas opacities can also be included in 
\ProDiMo, but this is still an experimental, yet unpublished
feature.  The idea is to add the molecular opacities to
$\kappa_\nu^{\rm PAH,abs}$. In this case, $\TPAH$ describes the
thermal emission of a mixture of PAHs and molecules, but the molecules
do not play a significant role here, because we only include molecular
UV cross-sections, and the disc is certainly too cold for UV emission.

\subsection{Chemistry}
\markright{2.5\ \ CHEMISTRY}
\label{sec:chemistry}

The chemical rate network is the core of all thermo-chemical disc
models. In the most general case, we have a reaction-diffusion
problem to solve
\begin{equation}
  \frac{dn_i}{dt} = P_i ~-~ L_i ~-~ \nabla\!\cdot\Phi_i\ ,
  \label{eq:chemsys}
\end{equation}
where $n_i\ \rm[cm^{-3}]$ is the particle density of species $i$,
$t\rm\,[s]$ is the time, $P_i$ and $L_i\ \rm[cm^{-3}s^{-1}]$ are the chemical
production and destruction rates and $\Phi_i\ \rm[cm^{-2}s^{-1}]$ is
a diffusive particle flux. Since we treat the disc as quasi-static,
there is no advection term in Eq.~(\ref{eq:chemsys}).

The chemical production and destruction rates include $1^{\rm st}$
order processes, two-body and three-body reactions with rates
\begin{eqnarray}
  R_1 &=& k_1 \times n_{\rm A}  \label{eq:first-order}\\
  R_2 &=& k_2 \times n_{\rm A} \, n_{\rm B}  \\
  R_3 &=& k_3 \times n_{\rm A} \, n_{\rm B} \, n_{\rm C}
  \label{eq:three-body}\ ,
\end{eqnarray}
where A, B, C are the reacting chemicals, and $n_{\rm A}$, $n_{\rm
  B}$, $n_{\rm C}$ their particle densities.  The $1^{\rm st}$ order
processes include photoprocesses, cosmic ray and X-ray induced processes, and
spontaneous decay processes, each destroying A. For example, a
photoionisation or photodissociation process
\begin{equation}
  \ce{A} ~+~ h\nu ~\longrightarrow~ \ce{D} ~+~ \ce{E}
\end{equation}
has a rate coefficient $\rm[s^{-1}]$
\begin{equation}
  k_{\rm ph} = 4\pi \int \sigma_{\rm ph}(\nu)
  \frac{J_\nu}{h\nu}\;d\nu \ ,
  \label{eq:photorate}
\end{equation}
where $\nu$ is the photon frequency, $\sigma_{\rm ph}(\nu)$ is the
UV photo cross section and $J_\nu$ is the local
direction-averaged UV intensity taken from the radiative transfer
module (Sect.\,\ref{sec:RT}).  The corresponding rate $R=k_{\rm
  ph}\,n_{\rm A}\,\rm[cm^{-3}s^{-1}]$ needs to be added
three times, namely to the destruction rate of A, $L_{\rm A}$, and to
the two production rates of D and E, $P_{\rm D}$ and $P_{\rm E}$.  In
a similar way, the gas-kinetic two-body reaction rates are computed,
for example
\begin{equation}
  \rm A ~+~ B ~\longrightarrow~ D ~+~ E  \ ,
\end{equation}
using a modified Arrhenius law
\begin{equation}
  k_2 = \alpha \left(\frac{T}{300\,{\rm K}}\right)^{\!\beta}
  \exp\left(-\frac{\gamma}{T}\right) \ .
\end{equation}
In this case, the rate coefficient $k_2$ has units $\rm[cm^3s^{-1}]$,
and the resulting rate $R=k_2\times n_{\rm A}\,n_{\rm B}$ has again
units $\rm[cm^{-3}s^{-1}]$.  This rate is added to $L_{\rm
  A}$, $L_{\rm B}$, $P_{\rm D}$ and $P_{\rm E}$.  Concerning the
three-body reactions $~\rm A ~+~ B ~+~ M ~\longrightarrow~ D ~+~ E ~+~ M~$
there is, in most cases, an unspecified collision partner M
involved, which stabilises an activated complex, following this scheme
\begin{equation}
  \rm A ~+~ B ~\stackrel{\displaystyle\longleftarrow}{\longrightarrow}
              ~ (AB)^\star
              ~\stackrel{M}{\longrightarrow}~ D ~+~ E \ .
\end{equation}
Such reactions have low and high pressure limits, depending on the
particle density of the collision partner $n_{\rm M}$.  However, the
critical density of M of these reactions, where the low pressure limit
changes gradually into the high pressure limit, is typically
$10^{18}-10^{21}\rm\,cm^{-3}$.  Such high densities are typically not
present in discs, so we can use the low pressure limit throughout, in
which case the effective rate is given by $R=k_3\times n_{\rm
  A}\,n_{\rm B}\,n_{\rm M}\,\rm[cm^{-3}s^{-1}]$, and
$k_3\,\rm[cm^6s^{-1}]$ can again be expressed with a modified
Arrhenius law.

The \ProDiMo\ project started in 2009 with a small reaction
network of 71 species and 911 reactions from the UMIST 2006 database
\citep{UMIST2007}, which was originally designed by
\citet{Kamp2000} to compute the concentrations of the small species
relevant for the heating and cooling in discs, such as \ce{H2},
OH, CO, and water.  Since then, many publications using \ProDiMo\
have expanded and refined the chemical network as we use it
today. Here we list some of the most significant chemical processes added: 
\begin{itemize*}
  \item A simple freeze-out chemistry with ice adsorption and
    thermal/photo-/Xray desorption was introduced by
    \citet{Woitke2009a}, using the concept of an ``active'' surface
    layer \citep{Aikawa1996}, where only the top $\sim\!2$ atomic
    layers of an ice mantel can be desorbed.
  \item An X-ray ionisation chemistry was added by \citet{Aresu2011},
    with primary X-ray ionisations and secondary ionisations via fast
    electrons, based on \citet{Meijerink2005} with doubly ionised
    atoms.
  \item The small and large DIANA standard chemical networks were
    introduced by \citet{Kamp2017}, which have 100 and 235 chemical
    species, respectively, with reaction rates mostly taken from the
    UMIST 2012 database \citep{McElroy2013}. The paper also summarises
    the rates used for electronically excited molecular hydrogen
    \ce{H2}$^\star$, which is a product of ``failed'' photoionisations
    of \ce{H2}, and establishes our standard element abundances
    \citep[see Table~5 in][]{Kamp2017}, which are close to solar
    except for the depleted metals Na, Mg, Si, S and Fe.
  \item The key process of \ce{H2}-formation on grain surfaces is
    calculated according to \citet{Cazaux2002,Cazaux2004} with updates
    described in \citet{Cazaux2009,Cazaux2010}.
  \item \citet{Thi2019} introduced a network for 5 charging states of
    PAH molecules of adjustable size, with the default choice being
    {\sl circumcoronene} \ce{C54H18}.  The total abundance of the PAHs is
    treated as a free parameter.
  \item Instead of using the UMIST database for the base reactions, it
    is now also possible to use the KIDA standard network 2014
    \citep[Kinetic Database for
      Astrochemistry,][]{Wakelam2012,Wakelam2013}, or the OSU~2009
    database (Ohio State University chemical network) from Eric
    Herbst.
  \item Surface chemistry was introduced by \citet{Thi2020a,Thi2020b}
    and benchmarked against a standard setup for cold molecular clouds
    \citep{Semenov2010}. The application of surface chemistry in discs
    is subject of active research, many of the energy barriers are only
    poorly known, which becomes increasingly more relevant for higher
    temperatures.  \citet{Thi2020b} have applied and tested
    the various surface chemical concepts to the formation of \ce{H2}
    and HD on warm grain surfaces, and \citet{Thi2020a} have developed
    a model for the formation of phyllosilicates on warm grain
    surfaces.
  \item \citet{Thi2019} and \citet{Balduin2023} have developed an
    advanced chemical network for thousands of individual charging
    states of dust grains of different sizes.  
  \item Kanwar\etal (2023, submitted to A\&A) have developed an
    extended hydro-carbon network to simulate the formation of larger
    hydrocarbon molecules like benzene (\ce{C6H6}) in the warm inner
    disc. 
\end{itemize*}
Some of the processes listed above do not follow the simple rate
formulation (Eqs.~\ref{eq:first-order}-\ref{eq:three-body}), for
example the active layer concept for ice desorption involves the
relative composition of the surface layer \citep{Woitke2009a}, where
all ice species appear in the denominator, and if one ice species is
dominant, for example water ice, its desorption rate becomes
independent of the ice concentration itself ($0^{\rm th}$-order rate)
as long as the active surface layer is fully occupied. These special
cases need to be treated carefully when it comes to the computation
of the Jacobian matrix.

\begin{figure*}
  \centering
  \vspace*{-3mm}
  \begin{tabular}{cc}
    \hspace*{-5mm}
    \includegraphics[width=79mm,height=54mm,trim=0 0 0 0,clip]
                    {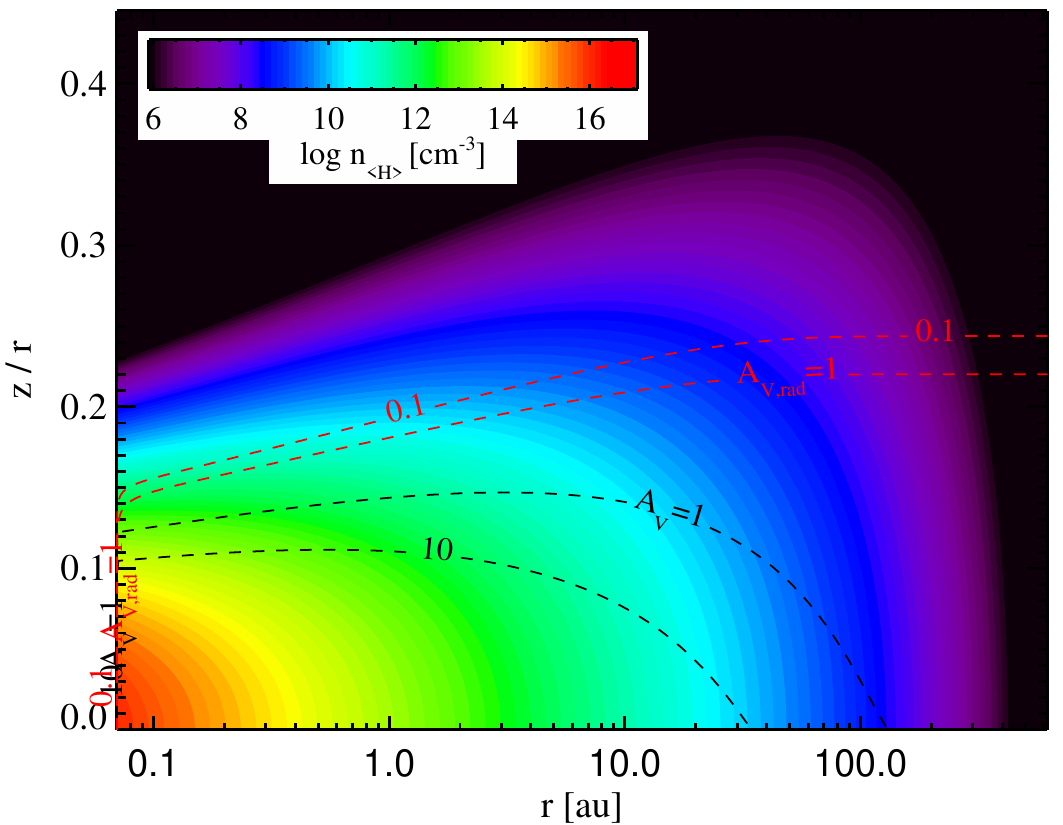} &
    \hspace*{-5mm}
    \includegraphics[width=79mm,height=54mm,trim=0 0 0 0,clip]
                    {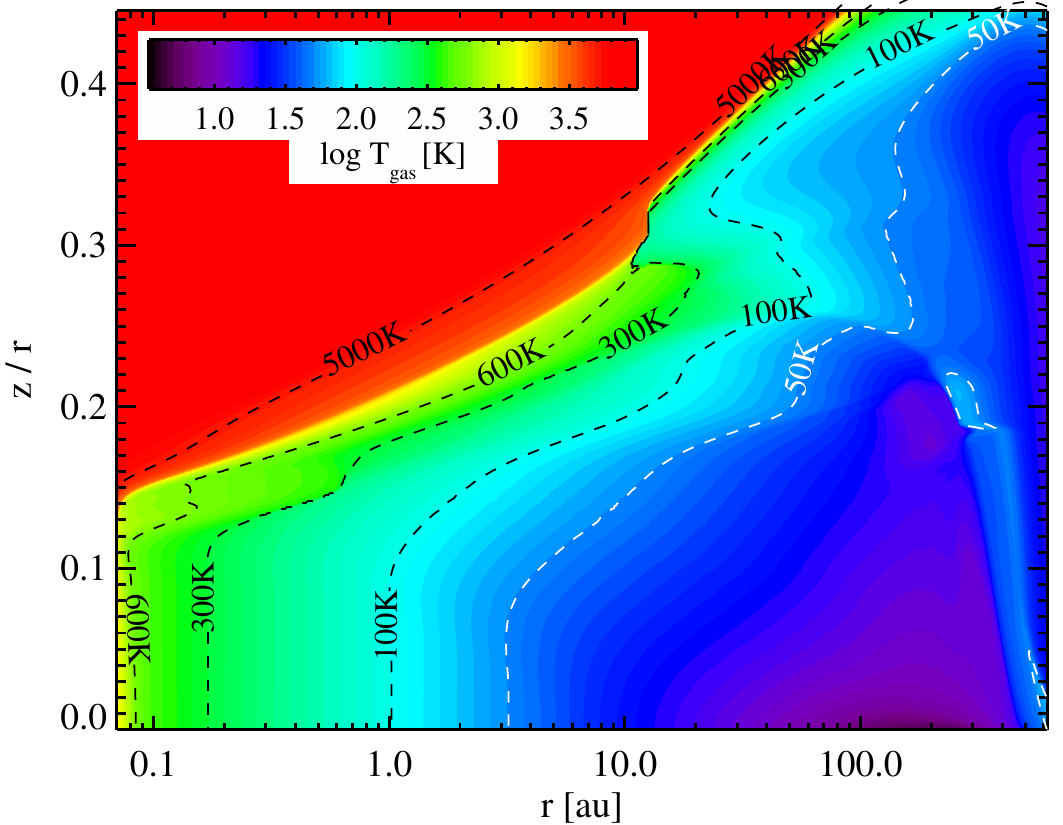}\\[-4mm]
    \hspace*{-5mm}
    \includegraphics[width=79mm,height=54mm,trim=0 0 0 0,clip]
                    {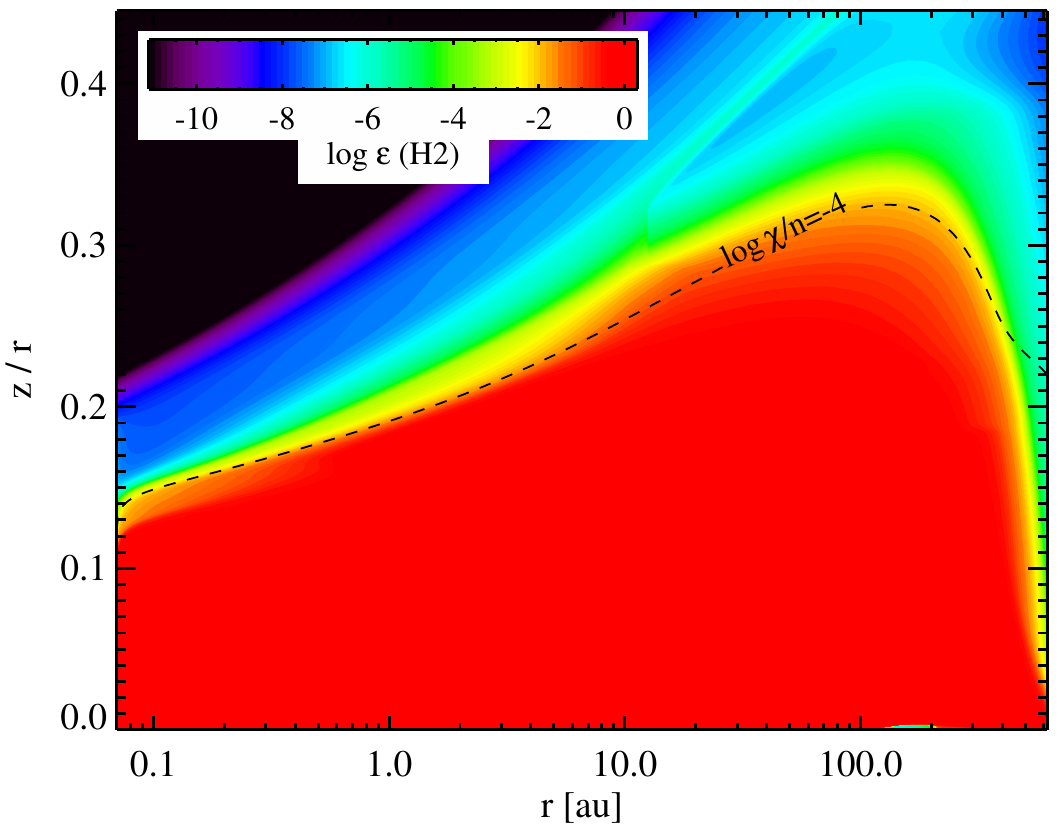} &
    \hspace*{-5mm}
    \includegraphics[width=79mm,height=54mm,trim=0 0 0 0,clip]
                    {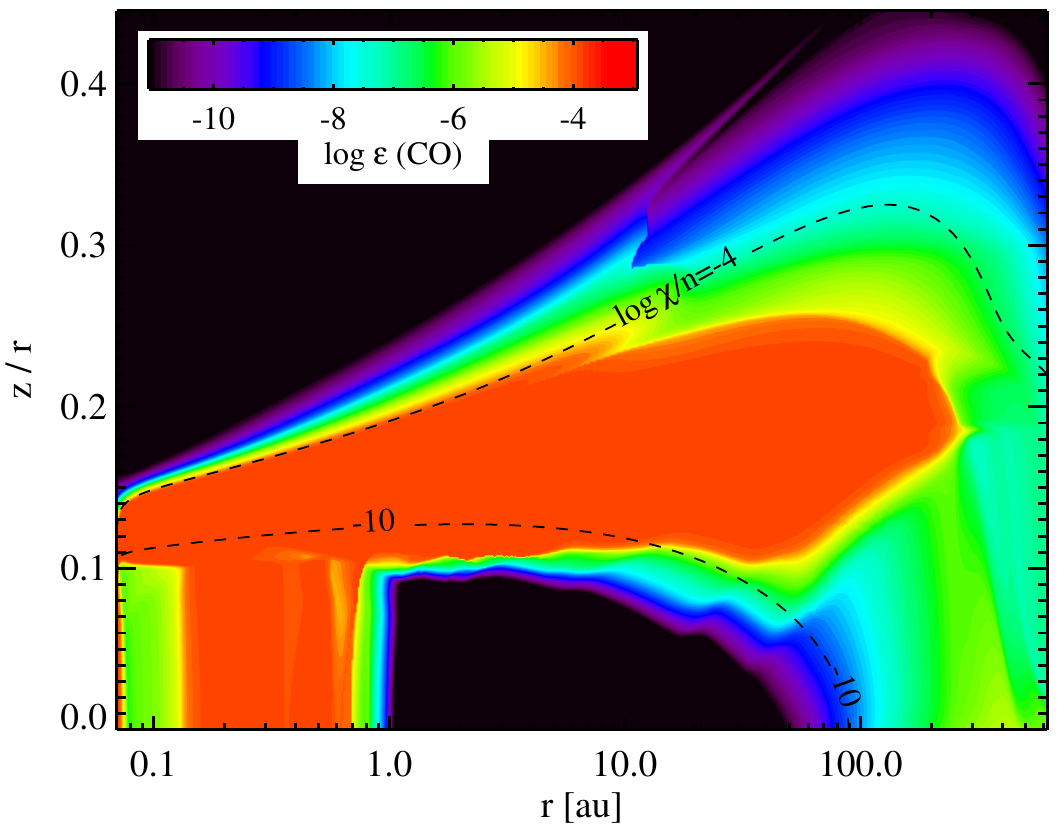}\\[-4mm]
    \hspace*{-5mm}
    \includegraphics[width=79mm,height=54mm,trim=0 0 0 0,clip]
                    {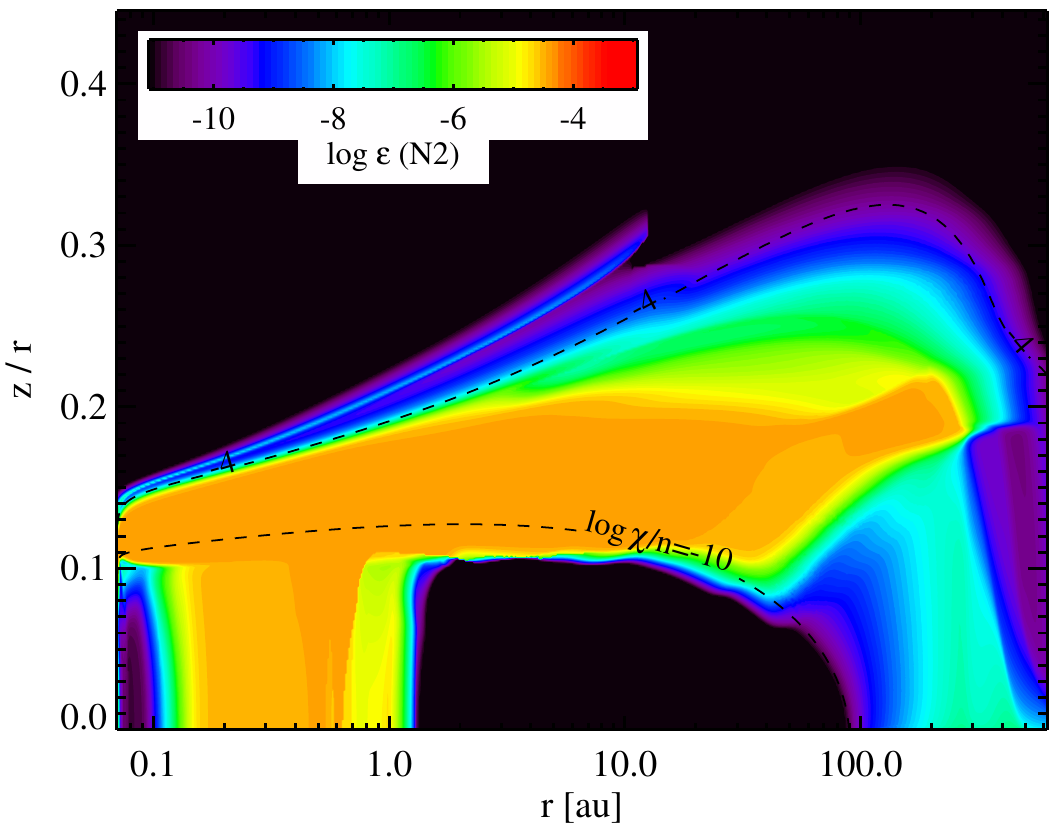} &
    \hspace*{-5mm}
    \includegraphics[width=79mm,height=54mm,trim=0 0 0 0,clip]
                    {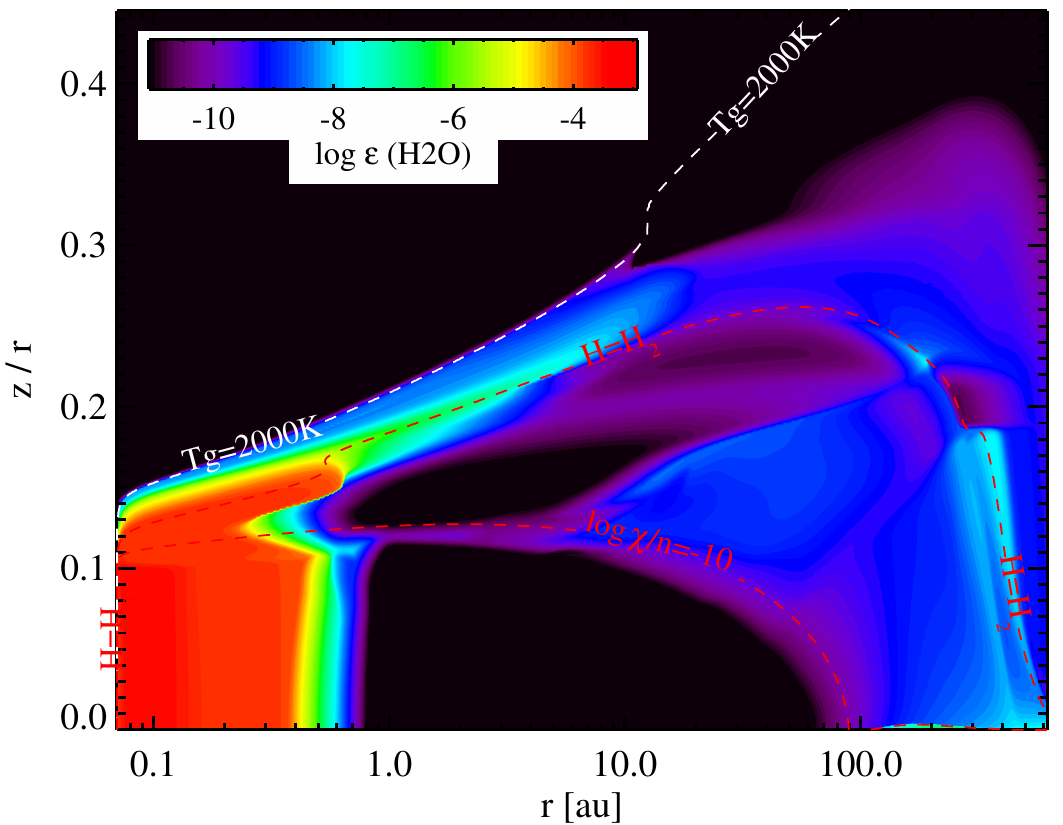}\\[-4mm]
    \hspace*{-5mm}
    \includegraphics[width=79mm,height=54mm,trim=0 0 0 0,clip]
                    {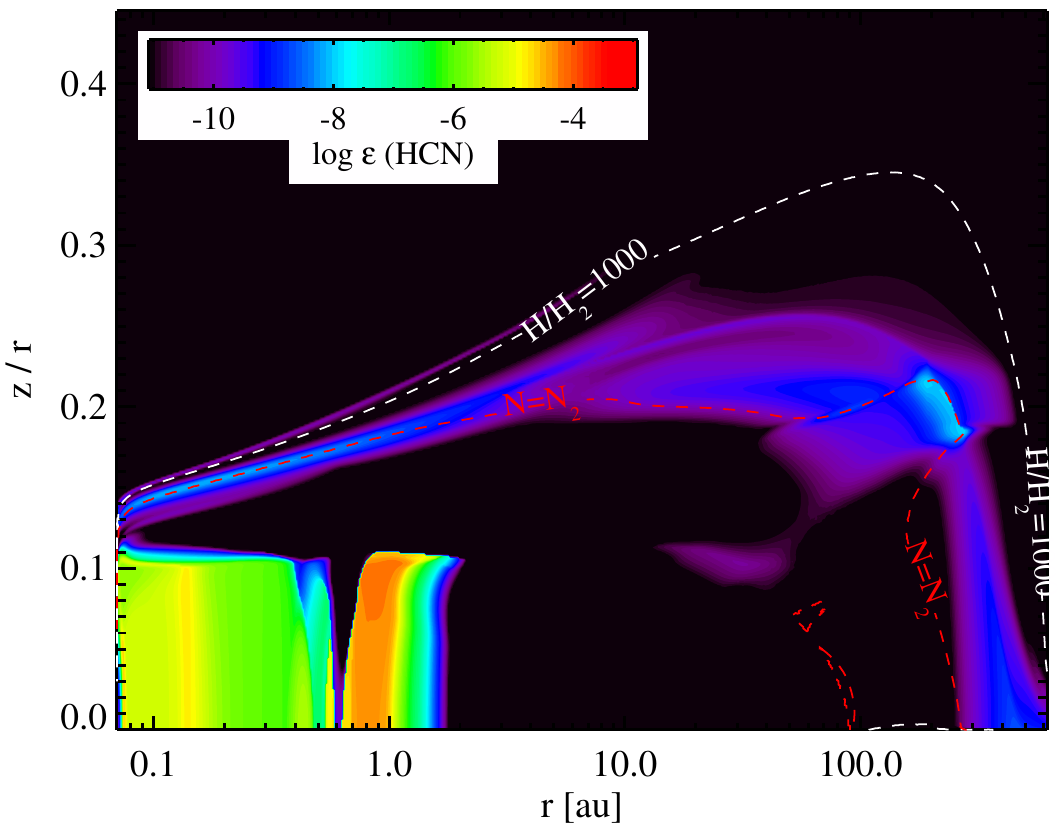} &
    \hspace*{-5mm}
    \includegraphics[width=79mm,height=54mm,trim=0 0 0 0,clip]
                    {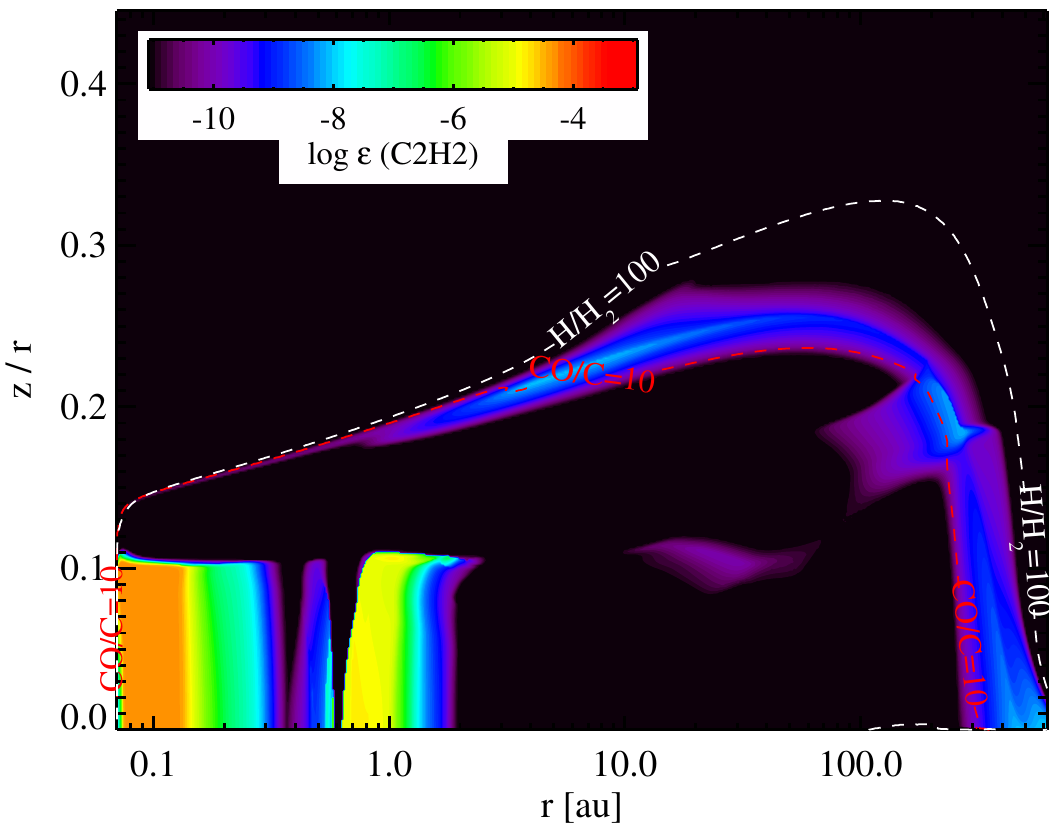}\\[-1mm]
  \end{tabular}
  \resizebox{\textwidth}{!}{\parbox{18.2cm}{\caption{Selected results
        from the chemistry of a standard T\,Tauri model.  The plots in
        the top row show the assumed gas density structure (left) and
        the calculated gas temperature structure (right) as function
        of radius $r$ and relative height $z/r$ over the midplane. The
        other plots show the calculated concentrations of the four
        major molecules \ce{H2}, \ce{CO}, \ce{N2}, \ce{H2O}, and the
        two spectroscopically interesting molecules HCN and \ce{C2H2},
        which only occur with lower concentrations in certain disc
        regions and layers. Additional contour lines include the
        vertical optical extinction $A_V$, the radial optical
        extinction $A_{V,\rm rad}$, and the ionisation parameter
        $\chi/n$ (i.e.\ the UV field strength divided by gas particle
        density).}
  \label{fig:standard}}}
\end{figure*}

\noindent Ordered by increasing complexity, the following chemical
models can be computed:
\begin{equation*}\begin{array}{clrcl}  
  (1) & \mbox{\sf thermo-chemical equilibrium:}
  & \multicolumn{3}{l}{\mbox{~~minimisation of Gibb's free energy}} \\ 
  (2) & \mbox{\sf kinetic chemical equilibrium:}
  & 0 &\!\!=\!\!& P_i - L_i\\ 
  (3) & \mbox{\sf time-dependent chemistry:}
  & ~~\displaystyle\frac{dn_i}{dt} &\!\!=\!\!& P_i - L_i\\[-1mm]
  (4) & \mbox{\sf 1D time-independent reaction-diffusion:}
  & 0 &\!\!=\!\!&\displaystyle P_i - L_i - \pabl{\Phi_i}{z}
\end{array}\end{equation*}
The first variant is based on the principle of minimisation of the
total local Gibb's free energy, given the local pressure and
temperature, either just for the gas, or with condensation, using the
chemical and phase equilibrium code {\sf GGchem} by
\citet{Woitke2018}. This mode requires an equality between gas and
dust temperature. It has rarely been used so far
\citep[e.g.][]{Thi2020a}, but is likely to be very useful to discuss
the stability of refractory condensates in the inner disc.

The second option is the default, the work-horse of \ProDiMo.  Most
publications using \ProDiMo\ have assumed kinetic chemical
equilibrium.  Indeed, the chemical relaxation timescale \citep[see
  Sect.~8.3 in][]{Woitke2009a} is much shorter than typical disc
evolutionary timescales in the observable, line-emitting, warm disc
surface, which justifies this approximation when it comes to comparing
disc models with observations. However, this simplification is not
valid in the optically thick midplane, where the chemical
transformation of one ice species into others can by far exceed the
disc lifetime.  Figure~\ref{fig:standard} shows a few selected results
of our standard T\,Tauri disc model in kinetic chemical equilibrium,
with physical disc parameters listed in Table~3 of \citet{Woitke2016}.
This model uses the large DIANA standard chemistry with 235 species
and 3067 reactions.

For the third option, we fix the physical conditions at each point in
the disc model in terms of temperatures, densities, and radiation
field conditions, and solve the chemistry time-dependently, usually
setting the initial chemical conditions by the results of a molecular
cloud model.  In this case, the disc chemistry evolves with disc age,
as for example in \cite{Helling2014}. \cite{Rab2017b} considered the
time-dependent chemistry after an outburst of a young Class-I object
going through episodic accretion events.

The forth option is the latest feature added to \ProDiMo.
\citet{Woitke2022} explored the effects of vertical turbulent mixing
on the chemical structure of protoplanetary discs. As the more
abundant species are mixed upwards and downwards, at the
locations where these chemicals are finally destroyed, for example by
photo-processes, the release of reaction products has important
consequences for all other molecules. This generally creates a more
active chemistry, with a richer mixture of ionised, atomic, molecular,
and ice species, and new chemical pathways that are not relevant in
the unmixed case.

\subsection{Gas heating/cooling balance}
\markright{2.6\ \ GAS HEATING/COOLING BALANCE}
\label{sec:heatcool}

\begin{figure}
  \vspace*{-3mm}
  \centering
  \begin{tabular}{ll}
  \hspace*{12mm}{\large\bf LTE} & \hspace*{12mm}{\large\bf non-LTE}\\[1mm]
  \hspace*{-10mm}  
  \begin{minipage}{86mm}
    \includegraphics[width=84mm]{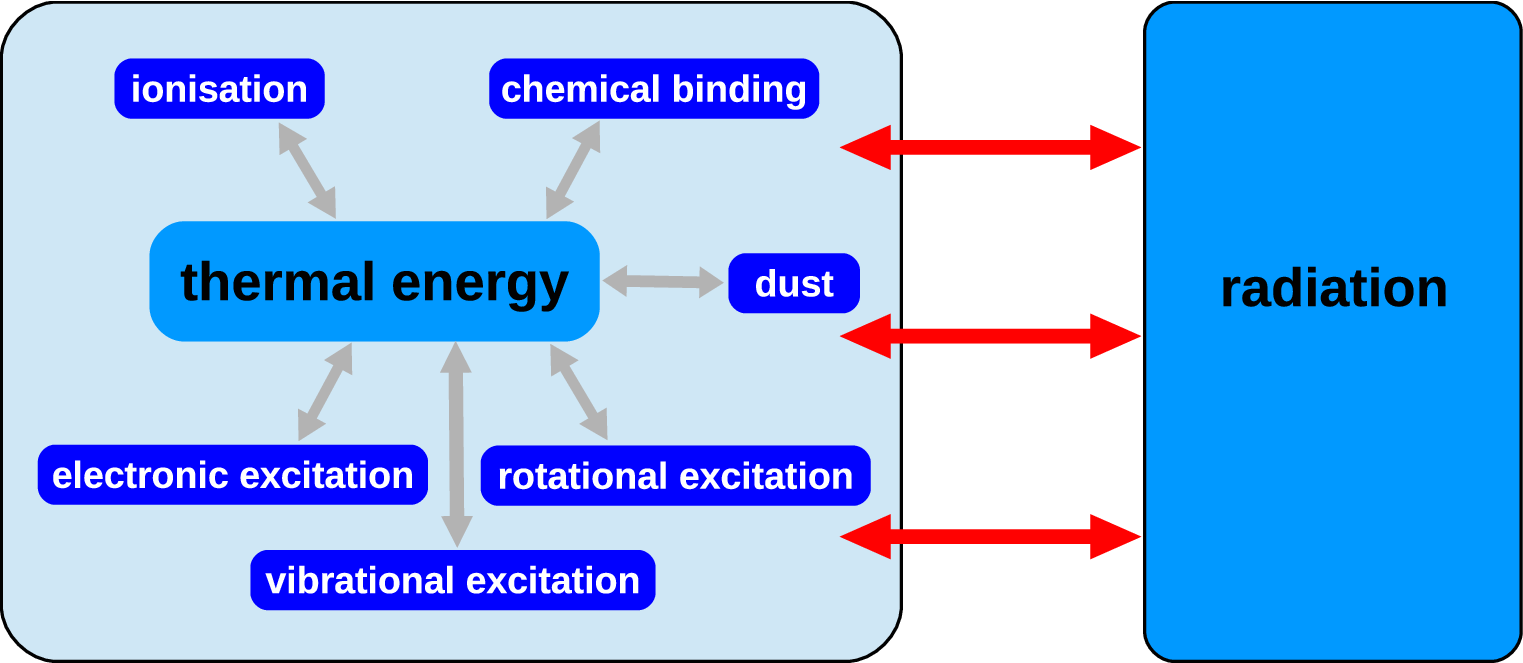}
  \end{minipage} 
  &
  \begin{minipage}{86mm}
    {\ }\\[0mm]
    \includegraphics[width=84mm]{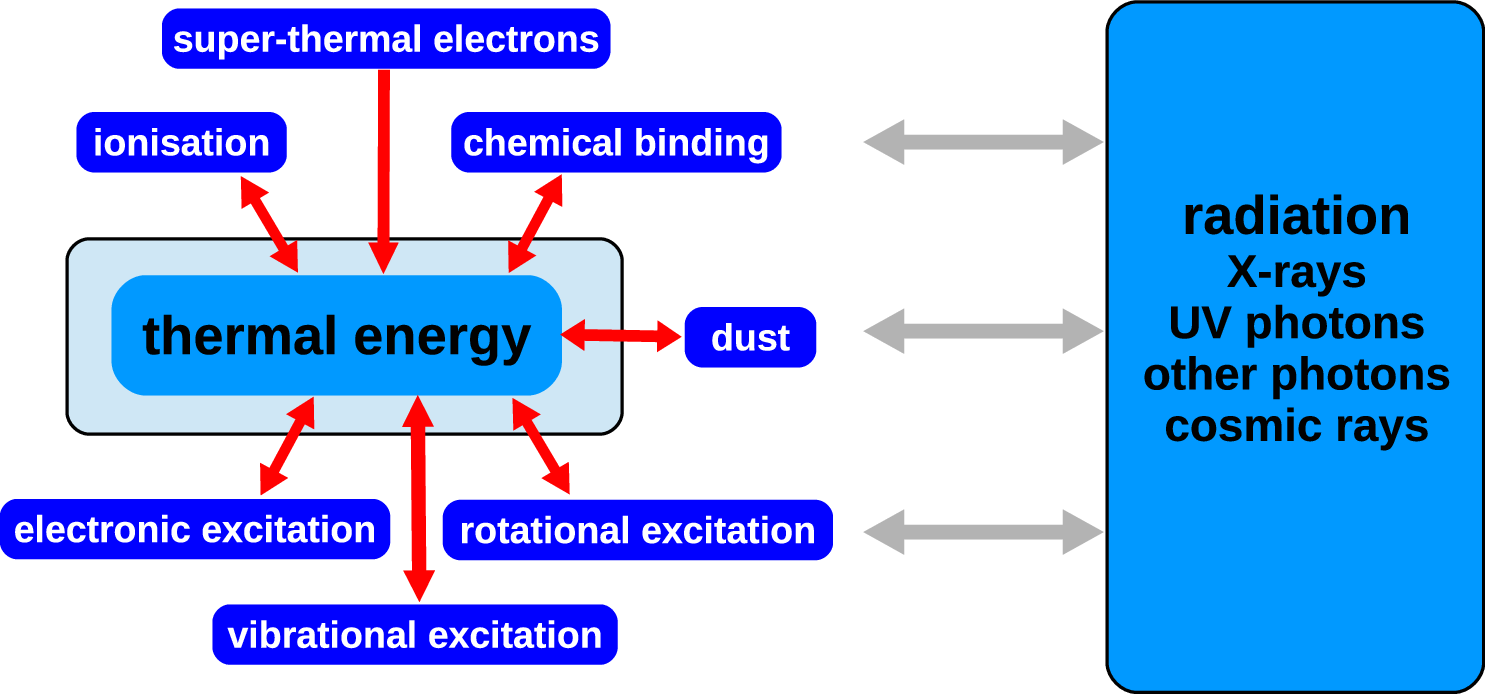} 
  \end{minipage}\\
  & \\[-6mm]
  \hspace*{-10mm} $ e = \frac{3}{2} n\,k\Tg + E_{\rm rot} + E_{\rm vib}
                   + E_{\rm ion} - E_{\rm diss} + ... $ &
  \hspace*{14mm}$ e = \frac{3}{2} n\,k\Tg $
  \end{tabular}
  \hspace*{-7mm} \resizebox{173mm}{!}{\parbox{185mm}{\caption{\small
        Two different approaches to apply the first law of
        thermodynamics. Following the LTE approach (left), the
        internal energy $e$ contains all possible ways to store energy
        in a gas.  According to the non-LTE approach (right) $e$ is
        given by the thermal kinetic energy of the gas particles
        only. The red arrows visualise the energy fluxes considered to
        calculate the net heating/cooling rate $Q$.  In contrast, the
        grey arrows visualise processes that are not directly included
        in the computation of $Q$. Figure reprinted from
        \cite{Woitke2015}.}
  \label{fig:pools}}}
\end{figure}

The determination of the temperature of the gas in protoplanetary
discs is an integral part of any disc modelling, and constitutes one
of the four pillars of \ProDiMo\ (radiative transfer, chemistry,
heating \& cooling, and hydrostatic structure), see
Fig.\,\ref{fig:pillars}.  The resulting gas temperature is important
for (i) the chemical processes, (ii) the production of emission lines
connected to the observability of the gas phase, and (iii) the
prediction of the hydrostatic vertical stratification and shape of the
disc. The first law of thermodynamics is applied to the gas as shown
on the right side of Fig.\,\ref{fig:pools} as
\begin{equation}
  \frac{d}{dt} \Big(\frac{e}{\rho}\Big) ~=~ -\,p\frac{dV}{dt} 
       ~+~ \frac{1}{\rho}\Big(\sum_i \Gamma_i -\sum_k \Lambda_k\Big) \ ,
  \label{eq:firstlaw}
\end{equation}
where $e\,\rm[erg/cm^3]$ is the internal energy per volume,
$p\,\rm[erg/cm^3]$ the gas pressure, $V\!=\!1/\rho\,\rm[cm^3/g]$ the
specific volume, $\rho\,\rm[g/cm^3]$ the mass density, and $\Gamma_i$
(``gain'') and $\Lambda_k$ (``loss'') are the various heating and
cooling rates per volume $\rm [erg/cm^3/s]$.

The $-pdV$ work is zero as long as the gas moves on stable,
pressure-supported Keplerian orbits.  In this case, $\rho$ and $V$ are
constants, and Eq.\,(\ref{eq:firstlaw}) simplifies to
\begin{equation}
  \frac{de}{dt} = Q(\Tg) = \sum_i \Gamma_i - \sum_k \Lambda_k = 0\ ,
  \label{eq:energybalance}
\end{equation}
which states the condition of thermal energy equilibrium.  Since the
net heating rate $Q$ depends on the gas temperature,
Eq.~(\ref{eq:energybalance}) states an implicit equation for the
determination of the gas temperature in thermal equilibrium $T_{\rm
  g}^{\,0}$.

In order to assess the importance of the $-pdV$ work, we can compute
the cooling relaxation timescale
\begin{equation}
  \tau_{\rm cool} = -\frac{\partial e}{\partial\Tg}\Bigg\vert_{T_{\rm g}^0} 
                   \;\bigg{/}\;\;
                   \frac{\partial Q}{\partial\Tg}\Bigg\vert_{T_{\rm g}^0} \ ,
  \label{eq:taucool}
\end{equation}
and compare it with the dynamical timescale $\tau_{\rm dyn}$ given by
the evolutionary timescale of the disc of the order of millions of
years.  However, in the case of winds, or hydrodynamical instabilities
such as spiral waves or vortices, $\tau_{\rm dyn}$ will be much
shorter, but not much shorter than the orbital timescale.  The
\ProDiMo\ models show that, even in such strongly time-dependent cases,
$\tau_{\rm cool} \ll \tau_{\rm dyn}$ often remains valid, such that
Eq.\,(\ref{eq:energybalance}) is justified.  However, this may not be
valid in highly optically thick regions in the midplane of massive
discs where the excess energy can only escape diffusively, which is
subject of ongoing research.

\ProDiMo\ is known for its particularly detailed treatment of heating
and cooling, taking into account various physical and chemical
processes, and evaluating about 100 heating rates $\Gamma_i$ and about
100 cooling $\Lambda_k$, which is summarised below.

\subsubsection{Line heating/cooling}
\label{sec:lineEscPro}

\begin{figure}
  \centering
  \vspace*{-1mm}
  \includegraphics[width=110mm]{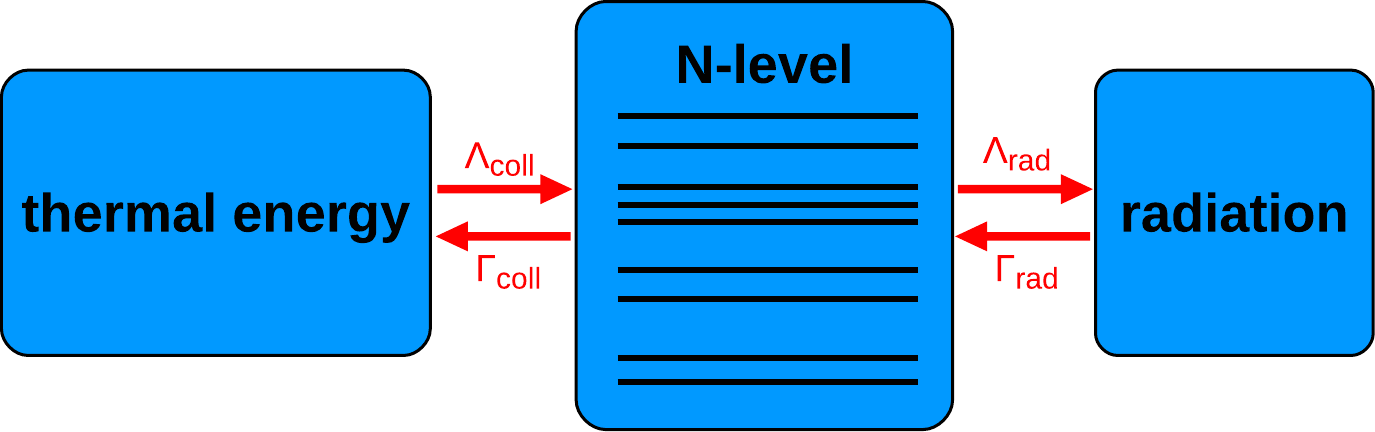}\\[-1mm]
  \resizebox{\textwidth}{!}{\parbox{18.2cm}{\caption{$N$-level heating
        and cooling rate.  Energy fluxes from left to right are
        cooling, and from right to left heating.  Subscript ``coll''
        means collisional, and subscript ``rad'' means radiative.}
  \label{fig:NlevelCooling}}}
\end{figure}

For the basic $N$-level bound-bound (line) heating and cooling rates,
we use a non-LTE formulation with escape probability theory, see
\cite{Woitke2009a} and \cite{Woitke2015}. It is important to realise
that spectral lines do not only cause cooling by collisional
excitation followed by line emission, but can also do the reverse,
namely heating by line absorption followed by collisional
de-excitation, see Fig.~\ref{fig:NlevelCooling}.  In statistical
equilibrium, the net energy exchange rate between the gas thermal
energy and the radiation field $\rm [erg/cm^3/s]$ can be measured on
either side of the energy buffer labelled with $N$-level
\begin{eqnarray}
  Q_{N-{\rm level}} &=& \Gamma_{\rm rad}-\Lambda_{\rm rad} 
                 \;=\; \Gamma_{\rm coll}-\Lambda_{\rm coll} \ ,
  \label{eq:NlevelCool} \\
  \Gamma_{\rm col} &=& \sum_{l=1}^{N-1}\sum\limits_{u=l+1}^N 
    n_u\,C_{ul}\,\Delta E_{ul} \label{eq:Qline1}\\
  \Lambda_{\rm col} &=& \sum_{l=1}^{N-1}\sum\limits_{u=l+1}^N 
    n_l\,C_{lu}\,\Delta E_{ul} \\
  \Gamma_{\rm rad} &=& \sum_{\rm lines} n_l \Delta E_{ul} 
                    \Pp_{ul}B_{lu}\Jcont_{\nu_{ul}} \\
  \Lambda_{\rm rad}&=& \sum_{\rm lines} n_u \Delta E_{ul} 
            \,\big(\Pe_{ul}A_{ul} +
            \Pp_{ul}B_{ul}\Jcont_{\nu_{ul}}\big) \label{eq:Qline4}\ .
\end{eqnarray}
$\Delta E_{ul}=E_u-E_l$ is the energy difference between an upper
state $u$ and a lower state $l$, $C_{\rm ul}=\sum_p
n_p\,\gamma_{ul}^p(\Tg)\rm\,[s^{-1}]$ are the collisional
de-excitation rates, $C_{\rm lu}$ the collisional excitation rates,
$n_p$ is the density of a collision partner $p$ and
$\gamma_{ul}^p(\Tg)\!=\!\langle v_P\sigma_{ul}\rangle$ its collisional
rate coefficient $\rm[cm^3\,s^{-1}]$ for de-excitation $u\!\to\!l$.
$A_{ul}$, $B_{lu}$ and $B_{ul}$ are the Einstein coefficients for
spontaneous emission, absorption and stimulated emission.

$\Pe_{ul}$ and $\Pp_{ul}$ are the escape and pumping probabilities,
expressing the probability that a locally emitted line photon is not
re-absorbed again in the neighbourhood in the same line, and the
probability that a continuum photon makes it to the considered location
given the existence of line opacity in the disc, respectively.  Both
probabilities become very small when the line is optically thick.
Various authors have presented different ideas and formula for
$\Pe_{ul}$ and $\Pp_{ul}$, for example \cite{Gorti2009},
\cite{Woitke2009a}, \cite{Woods2009}, \cite{Du2014} and
\cite{Bruderer2014} use different assumptions about disc geometry,
and approximations about velocity gradients, and line broadening.
Recently, Woitke\etal (2023, A\&A submitted) have developed a new
escape probability theory for 2D disc geometries, where local
continuum radiative transfer effects in the line resonance regions are
taken into account, leading to expressions for $\Pe_{ul}$ and
$\Pp_{ul}$ that depend on radial and vertical line and continuum
optical depths.

In order to evaluate the non-LTE equations (\ref{eq:Qline1}) to
(\ref{eq:Qline4}), the respective atomic and molecular data are
required.  The data includes the level energies $E_i$ and statistical
weights $g_i$, the radiative data (transition wavelengths
$\lambda_{ul}$ and Einstein coefficients $A_{ul}$), and, in
particular, the collisional data for all important collision partners,
including \ce{e-}, H, He, and \ce{H2}, as function of temperature,
which are often incomplete or entirely missing.  Over the years, we
have collected such data from various sources and databases, combining
experimental data, or quantum-mechanical predictions, or a combination
of both.  A selection of atomic data sources is listed in Table~A.1 of
\cite{Woitke2009a}, updated with larger model atoms in Table~4 of
\cite{Woitke2011}, using the
\href{https://www.chiantidatabase.org/}{CHIANTI database (Atomic
  Database for Spectroscopic Diagnostics of Astrophysical Plasmas)}
and the \href{https://www.nist.gov/pml/atomic-spectra-database}{NIST
  database (National Institute of Standards and Technology)}.

Concerning the molecules, we mostly use the
\href{https://home.strw.leidenuniv.nl/~moldata/}{LAMDA database
  (Leiden Atomic and Molecular Database)}, but there have been
additional efforts to collect more extensive data in form of
ro-vibronic data (i.e.\ electronically, vibrationally and rotationally
excited states) of the two key molecules \ce{H2} and CO, see
\cite{Thi2013b}.  The data situation is particularly challenging for
the ro-vibrational transitions of larger molecules, and for some
molecules like \ce{CO2}, \ce{C2H2} and \ce{HCN}, we are simply forced
to assume LTE, in which case a line list is sufficient.  For this
purpose, we mostly use the \href{https://hitran.org/}{HITRAN 2020
  database (High-resolution Transmission Molecular Absorption
  Database)}.

\subsubsection{Bound-free and free-free heating/cooling}

In addition to the interaction of the gas with line photons via
excited states, there are a couple of bound-free and free-free
processes to be taken into account, for example photo-ionisation
of neutral C, Fe, Si, and \ce{H-}
\begin{equation}
  {\rm A} \,+\, h\nu \;\;\begin{array}{c} {k_f} \\[-0.6ex]
                                           \rightleftharpoons \\[-0.6ex]
                                           {k_r}  
                          \end{array}\; {\rm A^+ \,+\, e^-} \ ,
\end{equation}
where $k_f$ is the photo-ionisation rate (see Eq.\,\ref{eq:photorate})
and $k_r$ the direct recombination rate.  Here, it is the excess
kinetic energy of the created free electron that is thermalising with
the gas.  From the threshold frequency $\nu_{\rm thr}$ and the energy
conservation $h\nu = h\nu_{\rm thr} + E_{\rm th}$ we find the
photo-ionisation heating rate to be
\begin{equation}
  \Gamma = n_{\rm A}\;4\pi\!\int\limits_{\nu_{\rm thr}}^\infty \,J_{\nu}\, 
           \frac{\nu-\nu_{\rm thr}}{\nu}\,\sigma^{\rm bf}(\nu)\,d\nu \ ,
  \label{heat}
\end{equation}
where $\sigma^{\rm bf}(\nu)$ is the bound-free cross section. The
reverse process, the recombination cooling rate, is slightly more
difficult to understand.  A direct recombination destroys one thermal
electron, so the recombination cooling rate should be of order
\begin{equation}
  \Lambda = n_{\rm A^+} n_{\rm e^-} k_r(\Tg) \langle E_{\rm th}
  \rangle \ ,
\end{equation}
where $\langle E_{\rm th} \rangle$ is the mean thermal energy of the
captured electron.
By considering an equilibrium between photo-ionisation and direct
recombination in thermal equilibrium, where $J_\nu\!=\!B_\nu(\Tg)$,
called Milne relations, it is possible to derive the recombination
cooling rate just from the bound-free cross sections as
\begin{equation}
  \Lambda = \frac{n_{\rm A^+}\,n_{\rm e^-}\,k_r(\Tg)}
                 {k_f\big\vert_{J_\nu=B_\nu(\Tg)}}\;
            4\pi\!\int\limits_{\nu_{\rm thr}}^\infty
            B_{\nu}(\Tg)\,\frac{\nu-\nu_{\rm thr}}{\nu}
            \,\sigma^{\rm f}(\nu)\,d\nu \ .
  \label{eq:recomb_cool}          
\end{equation}
Depending on the spectral shape of $\sigma^{\rm bf}(\nu)$,
Eq.\,(\ref{eq:recomb_cool}) typically results in $\langle E_{\rm th}
\rangle\!\approx\!(1.5-2.5)\,k\Tg$.  In applications to discs, we find
that the photo-ionisation of neutral carbon is an important heating
process in the outer disc, and the photo-attachment of \ce{H-}, i.e.\ $\rm H
+ e^- \to H^- + h\nu$, can be an important cooling process in the
inner disc.

\begin{table}
  \caption{High-energy heating and miscellaneous heating/cooling
    processes in \ProDiMo}
  \label{tab:heatcool}
  \vspace*{1mm}
  \resizebox{\textwidth}{!}{
  \begin{tabular}{p{4cm}|p{7.5cm}|l}
    \hline
    process & physical description & formula (reference) \\
    \hline
    dust thermal accommodation &
      Inelastic collisions of gas particles with dust grains cause
      an equilibration of $\Tg$ and $\Td$ -- can be heating or
      cooling.  
      & \cite{Burke1983}\hspace*{-4mm}\\            
    heating by formation of \ce{H2} on dust &
      The \ce{H2} molecules forming on grain surfaces are ejected with
      super-thermal velocities in vibrationally highly excited
      states. The kinetic part is assumed to get thermalised by
      collisions. 
      & \cite{Black1976} \\
    background/formation heating by \ce{H2} &
      In addition to the radiative and collisional excitation of its
      ro-vibrational states, the vibrational states of \ce{H2} are
      pumped by the formation of \ce{H2} on grain surfaces. The excess
      energy on ejection is partly thermalised by de-exciting
      collisions.  This effect is part of the non-LTE modelling of
      \ce{H2} in \ProDiMo.
      & \cite{Duley1986} \\
    heating by collisional de-excitation of $\rm H_2^\star$ &
      The fluorescent excitation of \ce{H2} by UV photons pumps
      the vibrational levels of \ce{H2}, which can be converted into
      thermal energy by collisions. 
      & \cite{Tielens1985}\hspace*{-4mm} \\
    dust{\color{white}-}photo-electric\linebreak heating &
      UV photons impinging on dust grains can eject electrons with
      super-thermal velocities (photoeffect) which then thermalise
      through collisions with the gas. The efficiency of this process
      decreases strongly with grain charge.
      & \cite{Kamp2000}\\
    PAH heating &
      Photoeffect on PAH molecules, similar to dust photo-electric
      heating, depends on PAH charging.
      & \cite{Thi2019}\\
    X-ray Coulomb heating &
      The absorption of X-rays in the $K$-shells of the various
      elements dissociates and ionises the gas, which produces fast
      electrons. If the degree of ionisation in the gas is high,
      these super-thermal electrons predominantly undergo elastic
      Coulomb collisions with ambient electrons, which can very
      efficiently heat the gas. 
      & \cite{Dalgarno1999}\\                        
    X-ray \ce{H2} dissociation heating &
      If the gas is mostly molecular, the fast electrons produced
      X-ray interactions mostly interact with \ce{H2}, creating
      \ce{H2+}, which will recombine or produce \ce{H3+}.
      These reactions are exotherm, heating the gas. 
      & \cite{Meijerink2005}\hspace*{-4mm} \\
    \hline
  \end{tabular}}
\end{table}

\addtocounter{table}{-1}
\begin{table}
  \caption{continued}
  \vspace*{1mm}  
  \resizebox{\textwidth}{!}{
  \begin{tabular}{p{4cm}|p{7.5cm}|l}
    \hline
    process & physical description & formula (reference) \\
    \hline
    cosmic ray heating & 
      Energetic particles (cosmic rays or SEPs) loose their energy
      in small steps, characterised by the energy to create an ion
      pair. We use a simple formula for the energy thermalised by the
      subsequent physical and chemical processes per ionised H-atom
      and \ce{H2}-molecule.
      & \cite{Jonkheid2004} \\
    viscous heating &
      The transport of angular momentum in the disc is physically
      connected to the creation of heat by frictional forces.  It also
      leads to mass accretion.  We use a formula that expresses the
      heat production due to viscous heating as function of the mass
      accretion rate.   
      & \cite{Dalessio1998} \\                             
    photo-dissociation\linebreak heating &
      Similar to bound-free ionisation heating, but more complicated,
      as only a small fraction of the excess energy over binding
      energy is liberated in form of kinetic energy. However, the
      reaction products can be in excited ro-vibronic states, the
      energy of which might get thermalised by collisions.
      & \cite{Glassgold2015} \\                   
    chemical heating &
      Chemical reactions generally liberate or consume chemical
      binding energy.  We multiply the reaction rates of all
      exotherm and endotherm gas phase reactions with their
      reaction enthalpies and sum them up, excluding the UV, cosmic
      ray and X-ray reactions.  The stable
      molecules destroyed by these reactions typically undergo a
      series of exotherm reactions before they eventually
      re-form. 
      & \cite{Woitke2011}\\
    Lyman-$\alpha$ cooling &
      Once an H-atom is collisionally excited to its $n\!=\!2$ level,
      it will emit a Lyman-$\alpha$ photon.  These photons are
      absorbed and re-emitted many times, until they eventually leave
      the disc (resonance scattering).   
      & \cite{Tielens1985}\hspace*{-4mm}\\
    \hline
  \end{tabular}}
\end{table}

\begin{figure*}[!t]
  \centering
  \hspace*{-6mm}
  \begin{tabular}{cc}
    \hspace*{-5mm}
    \includegraphics[width=88mm,height=120mm,trim=30 0 30 0,clip]
                    {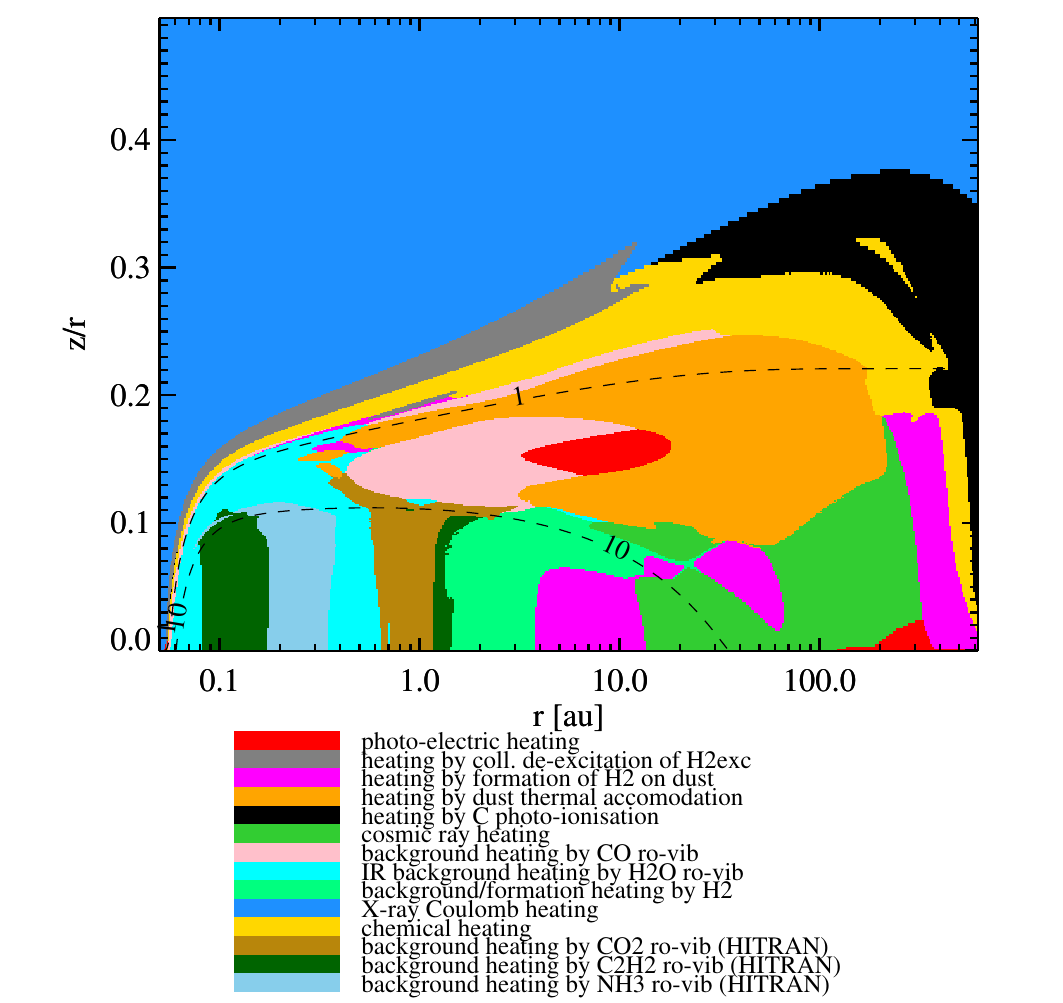} &
    \hspace*{-10.5mm}
    \includegraphics[width=88mm,height=120mm,trim=30 0 30 0,clip]
                    {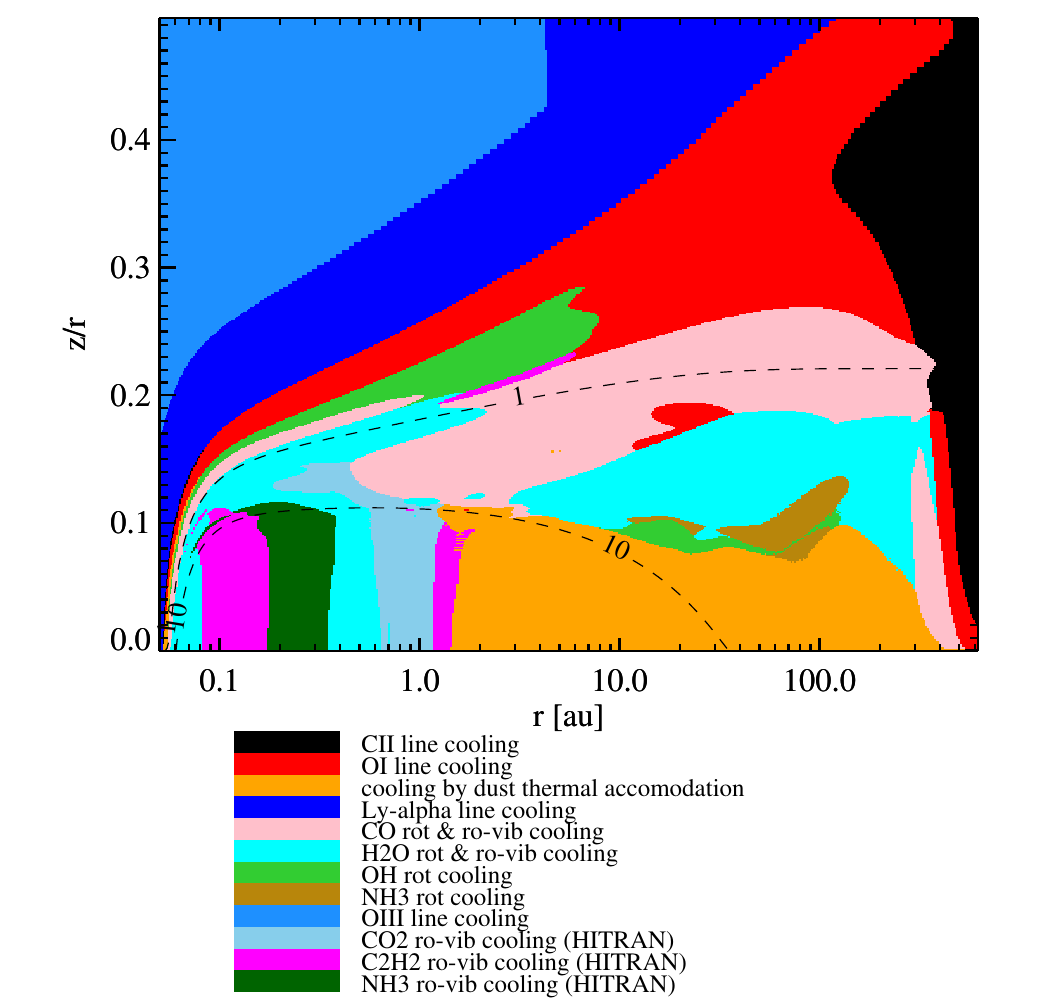}\\[-1mm]
  \end{tabular}
  \resizebox{\textwidth}{!}{\parbox{18.2cm}{\caption{The leading
        heating process (left) and the leading cooling process (right)
        in a \ProDiMo\ standard T\,Tauri disc model. Processes are
        only plotted when they fill in at least 1\% of the depicted
        area in these plots. The dashed lines mark the visual disc
        surface ($A_{V,\rm rad}\!=\!1$) and a vertical optical
        extinction of $A_V\!=\!10$. The model includes 103 heating and
        95 cooling processes, with altogether 64910 spectral lines.}
  \label{fig:HeatCool}}}
\end{figure*}

\subsubsection{Heating and cooling processes involving dust grains}

The presence of dust particles in the disc, which are ``outside'' of
the therodynamical system considered for the gas internal energy and
the first law of thermodynamics (Fig.~\ref{fig:HeatCool}), leads to a
number of additional heating and cooling processes.  Such processes
include inelastic collisions of gas particles with dust grains, and
the injection of super-thermal electrons or gas particles just formed
by certain physical and chemical processes taking place on the surface
of the dust graiuns, such as UV photon absorption or surface
chemical processes. Table~\ref{tab:heatcool} summarises the
heating/cooling processes included in ProDiMo.

\subsubsection{High-energy heating processes}

The young star in the centre of the disc is usually active and
accreting, therefore it is a strong source of UV radiation, X-rays, and
stellar energetic particles (SEPs). When this high-energy radiation
interacts with the upper disc surface, a number of additional
high-energy processes are triggered, which usually dominate the
heating of these regions, see Table~\ref{tab:heatcool}. In addition,
the disc is also irradiated from the outside in form of interstellar
UV photons and cosmic rays. The hard X-rays and in particular the cosmic rays
can penetrate quite deep into the disc, where they provoke some
chemical activity even in the UV-shielded regions that can still be
relevant for the heating of the gas.

Figure~\ref{fig:HeatCool} shows which of these heating and cooling
processes are found to be the relevant ones in a standard
\ProDiMo\ model for a T\,Tauri disc without viscous heating and
PAHs. The hot upper disc regions are controlled by X-ray Coulomb
heating versus Lyman-$\alpha$ and forbidden [O\,III] and [O\,I] line
cooling. The warm surface layers above the $A_{V,\rm rad}\!=\!1$-line
are heated by UV-induced processes, in particular the heating by
collisional de-excitation of $\rm H_2^\star$, chemical heating by
exothermal reactions, photo-dissociation, and photo-ionisation of
neutral carbon. This heating is balance by line cooling in particular
of OH, water and CO.  Below the $A_{V,\rm rad}\!=\!1$ line, the
efficient line cooling creates a situation where the gas temperatures
are slightly lower than the local dust temperatures, and the gas is
heated by thermal accommodation in addition to the heating via
absorption of continuum photons by the CO, water and \ce{CO2} lines.
In the optically thick part of the disc ($A_V\!\ga\!10$), where
$J_\nu\!\approx\!B_\nu(\Td)$ and $\Tg\!\approx\!\Td$, dust and gas
exchange energy mostly via photons, i.e.\ the gas absorbs continuum
photons in the lines, and the dust absorbs the line photons emitted by
the gas.

\subsection{Numerical solution methods}
\markright{2.7\ \ NUMERICAL SOLUTION METHODS}

The previous sections have laid out how the radiative, physical and
chemical processes are interlinked in protoplanetary discs. The
chemistry can only be calculated when the radiation field and the
local dust and gas temperatures are known, and the calculation of the
heating/cooling rates requires to know the radiation field, the gas
and dust temperatures, and the local concentrations of all relevant
atoms, ions, electrons and molecules.  Finally, the hydrostatic
structure of the disc can only be calculated when the gas temperatures
and the mean molecular weights are known, and that structure changes
the disc shape and hence the results of the continuum radiative
transfer.

\begin{figure}
  \vspace*{-5mm}
  \begin{tabular}{cc}
  \begin{minipage}{77mm}
    \includegraphics[width=77mm]{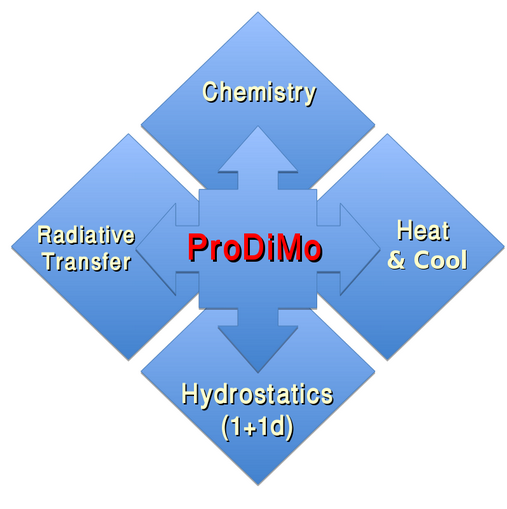}
  \end{minipage} &
  \begin{minipage}{77mm}
    \includegraphics[width=77mm]{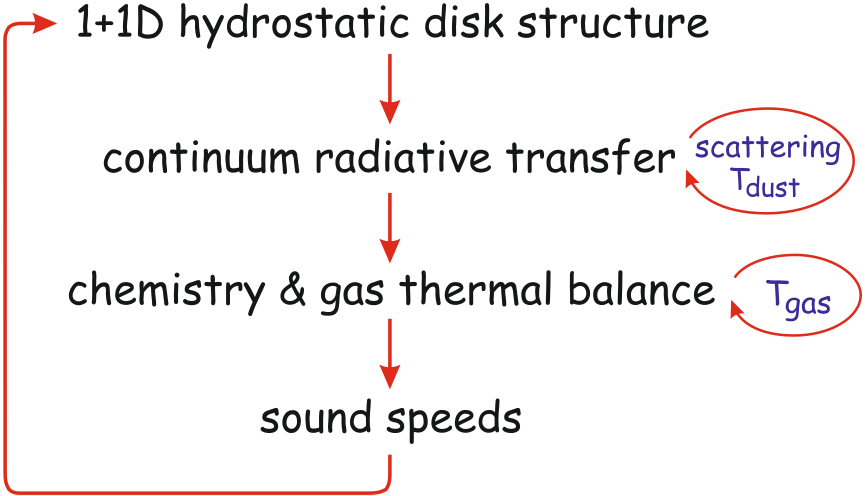}
  \end{minipage}\end{tabular}\\[-5mm]
  \resizebox{\textwidth}{!}{\parbox{18.2cm}{\caption{Schematic
        description of the thermo-chemical disc modelling code ProDiMo
        and the course of actions and global iterations. The right
        figure is adopted from \citet{Woitke2009a}.}
  \label{fig:pillars}}}
\end{figure}

In order to solve these dependencies, \ProDiMo\ uses a sequence of
internal and global iterations, see Fig.~\ref{fig:pillars}. We start
by setting the stellar and interstellar irradiation properties and (a
guess of) the disc density structure $\rho(r,z)$.  After determining
the dust settling (Sect.~\ref{sec:dust}), we calculate the dust
absorption and scattering opacities (Sect.~\ref{sec:dust}) and solve
the continuum radiative transfer problem (Sect.~\ref{sec:RT}), which
involves an iteration to find the correct dust temperature structure
and source functions including scattering.

Next, we solve the chemistry on a given point for a given gas
temperature, and determine the net gas heating function $Q(\Tg)$.  We
then vary the gas temperature, re-calculate the chemistry and net gas
heating, and repeat this process until the gas temperature $\Tg^{\,0}$
in thermal balance is found, where $Q(\Tg^{\,0})\!=\!0$, which on
average requires about 5-10 calls of the chemistry module per point.
Due to dependencies of the photo-dissociation rates on radial and
vertical molecular column densities (shielding factors) in the
chemistry, and the dependencies of the line escape probabilities on
radial and vertical line optical depths for the gas heating/cooling
balance, the chemistry on the various spatial grid points must be
solved in a particular order. A new point can only be calculated once
all points above that point, and all points inside of that point are
already completed. But different processors can still work on several
disc columns in parallel, which leads to a filling of the computed
points in a ``diagonal'' way, where the computations start with the
closest highest grid point, and end with the most distant midplane
point.

Once the chemistry and the gas energy balance have been determined on
all grid points, we can compute the local sound speeds and then solve
the following equation for the vertical hydrostatic equilibrium by
numerical integration to obtain the vertical disc structure
\begin{eqnarray}
  p &=& \sum n_i\,k\Tg ~=~ c_S^2(r,z)\;\rho \\
  \frac{1}{\rho}\frac{dp}{dz} &=& c_S^2(r,z)\,\frac{d \ln p}{dz}
  ~=~ -\,\frac{z\;GM_\star}{\big(r^2+z^2\big)^{3/2}}
  \label{eq:hydrostat} \ .
\end{eqnarray}
Such models require an additional outer iteration, in which the
resulting disc density structure $\rho(r,z)$ is passed back to the
continuum radiative transfer, see Fig.~\ref{fig:pillars}. Such
hydrostatic disc models are very slow and tend to have convergence
problems, so they are rarely used in practise, see however
Sect.~\ref{sec:discshape} for published examples.

But even if the disc density structure $\rho(r,z)$ is fixed in a
parametric way, some disc models still require global iterations, in
cases where the opacities used in the continuum radiative transfer
depend on the results of the chemistry, for example
\begin{itemize*}
  \item inclusion of X-ray gas opacities in the radiative
    transfer \citep{Rab2018}.
  \item PAH opacities in the radiative transfer with consistent
    charging \citep{Thi2019},
  \item inclusion of position-dependent ice opacities
    \citep{Arabhavi2022}.
\end{itemize*}

\subsection{Line transfer and predictions of observations}
\markright{2.8\ \ LINE TRANSFER AND PREDICTIONS OF OBSERVATIONS}

Once a disc model has been completed, we solve the line and continuum
radiative transfer equation in 3D, along a bundle of parallel rays, to
produce observable quantities.  The basics of this ray-tracing
technique are explained in Appendix A.7 of Publication~3
\citep{Woitke2011}, and the setup of the parallel rays forming the
image plane, to observe the disc under an inclination angle $i$, is
described in Section 2.3 of \cite{Thi2011b}.  This ray-tracing
technique has been used to connect the disc models to various kinds of
observations, and is the key to our endeavour to compare and verify our
modelling results with observations.

\begin{figure}[!b]
  \vspace*{-2mm} \centering
  \includegraphics[width=85mm]{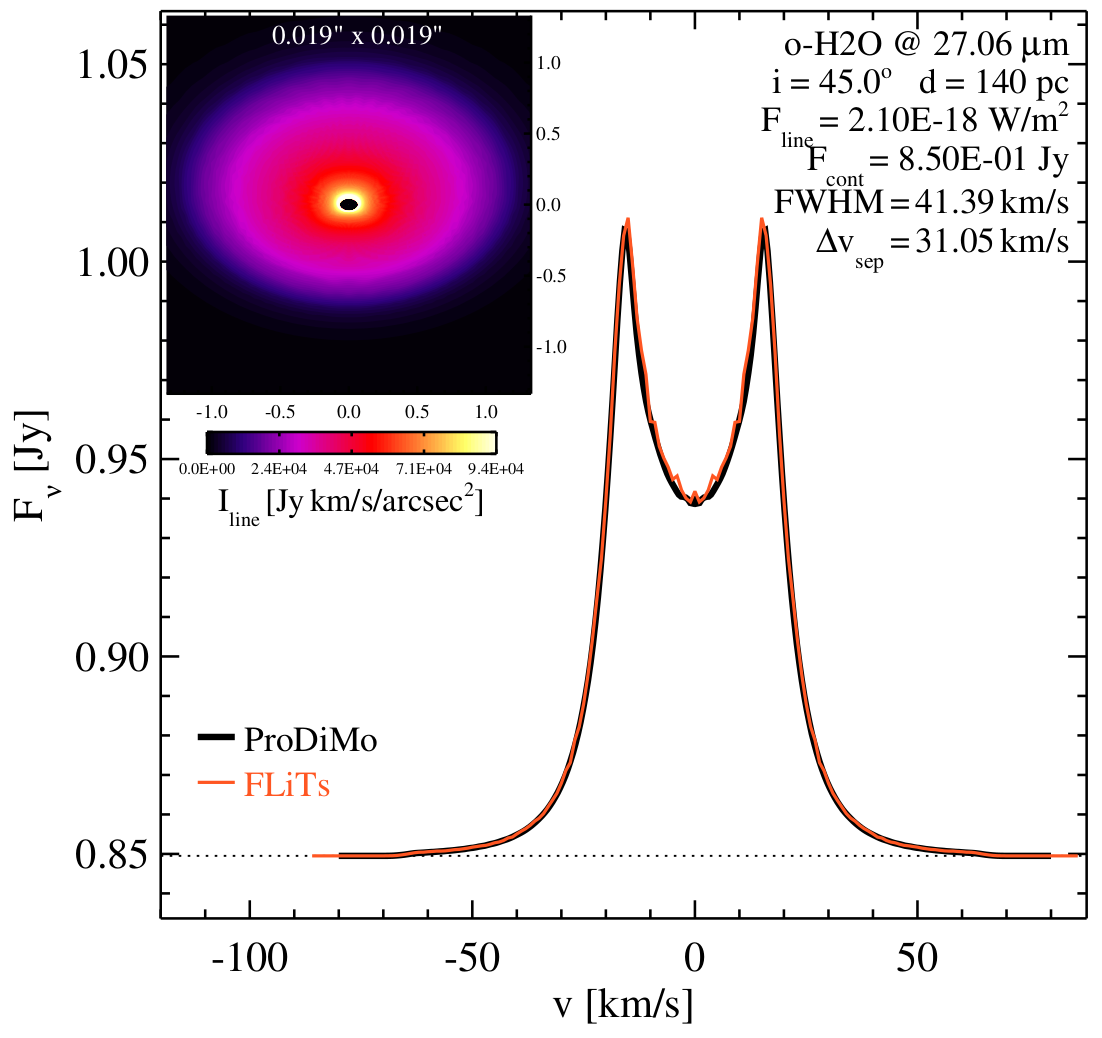}\\[-2mm]
  \resizebox{\textwidth}{!}{\parbox{18.4cm}{\caption{The o-\ce{H2O}
        rotational line $8_{5,4}\!\to\!8_{2,7}$ in the mid-IR at
        $27.06\,\mu$m emitted from a T\,Tauri disc as simulated with
        \ProDiMo\ and FLiTs. The line is emitted from the disc surface
        within about 1\,au. The Keplerian velocity field creates a
        symmetric double-peaked profile. The line sits on a continuum
        of about 0.85\,Jy, figure re-printed from Publication~5
        \citep{Woitke2018}.}
  \label{fig:line1}}}
  \vspace*{-1mm}
\end{figure}

The observations predicted this way are the spectral energy
distribution \citep{Thi2011b}, continuum images and radial continuum
intensity profiles \citep{Woitke2019}, infrared molecular line spectra
\citep{Woitke2018}, line and continuum visibilities
\citep{Woitke2016,Woitke2019}, line velocity profiles
\citep[e.g.][]{Woitke2011,Tilling2012,Thi2013a} and channel maps
\citep[e.g.][]{Rab2020}. Applying additional instrument-specific input
data, such as filter transmission curves and baseline configurations,
we can simulate the observations available on different platforms,
with different instruments and observational techniques this way, for
example photometric fluxes from UV to millimetre wavelengths for the
SED, the Submillimeter Array (SMA) for continuum images at
sub-millimetre images, VLT/CRIRES for velocity-resolved CO fundamental
line spectra, VLT/MIDI and VLT/PIONIER for visibilities, the Spitzer
Space Telescope and JWST for IR spectra, and the Atacama Large
Millimeter Array (ALMA) for channel maps.

\begin{figure}
  \vspace*{-1mm}\hspace*{-6mm}
  \includegraphics[width=17.2cm]{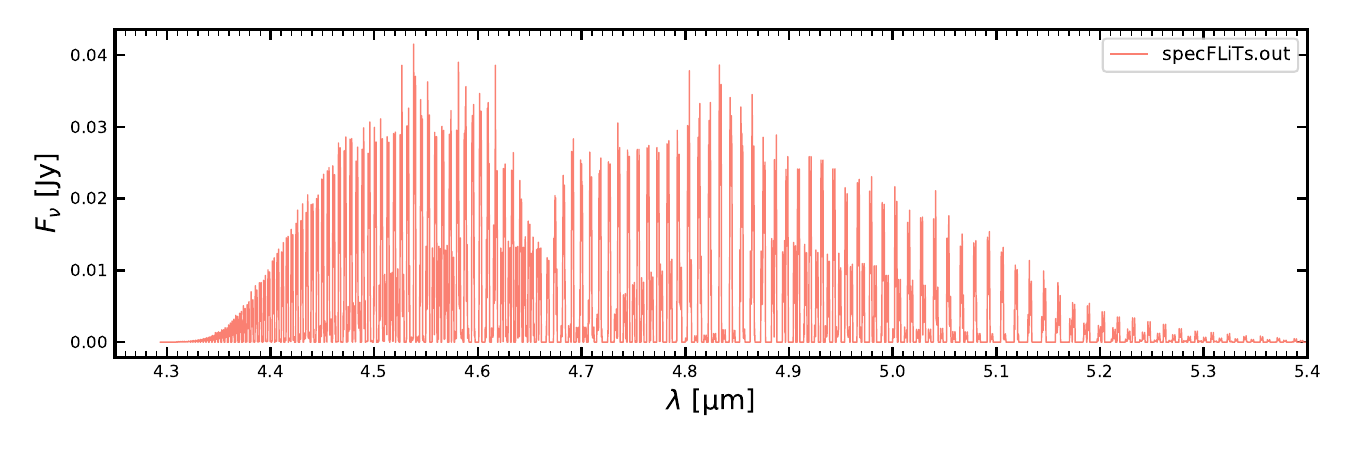}\\[-6mm]
  \resizebox{\textwidth}{!}{\parbox{18.4cm}{\caption{The CO
        fundamental ro-vibrational spectrum around 4.7\,$\mu$m emitted
        from a T\,Tauri disc simulated with \ProDiMo\ and FLiTs. We
        see the $P$ and $R$ branches of the vibrational $\rm
        v\!=\!1\!\to\!0$, $\rm v\!=\!2\!\to\!1$ and $\rm
        v\!=\!3\!\to\!2$ bands in emission after continuum
        subtraction. One ro-vibrational line creates a narrow,
        symmetric double-peaked profile as in Fig.\,\ref{fig:line1},
        but the superposition of all lines from the three different
        vibrational bands can make the individual lines appear
        asymmetric.}
  \label{fig:line2}}}
\end{figure}

\begin{figure}
  \vspace*{2mm}\hspace*{-5mm}
  \includegraphics[width=16.9cm]{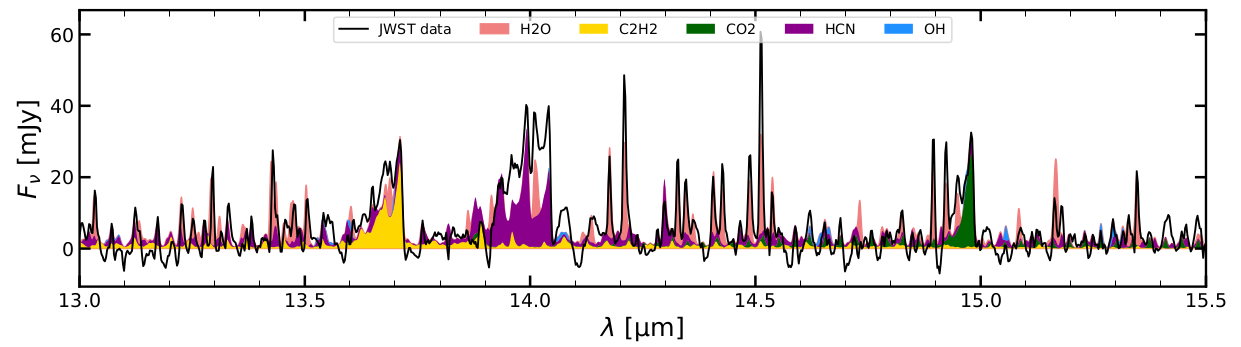}\\[-1mm]
  \resizebox{\textwidth}{!}{\parbox{18.4cm}{\caption{The
        continuum-subtracted spectrum of the T\,Tauri
        star EX\,Lupi observed with JWST \citep[][black
          line]{Kospal2023} compared to fitted \ProDiMo\,--\,FLiTs 2D
        disc model, convolved to a spectral resolution of
        $R\!=\!2500$, showing the contributions of the various
        molecules \citep{Woitke2023}.}
  \label{fig:line3}}}
  \vspace*{-1mm}
\end{figure}

The physical assumptions going into the continuum and line transfer at
this stage are (i) pressure-supported Keplerian velocity field, (ii)
Gaussian line profile function with thermal $+$ turbulent broadening,
(iii) non-LTE population of the molecular states computed with escape
probability, and (iv) isotropic and coherent continuum scattering.
The assumption of Keplerian orbits can be relaxed (yet without
consistent feedback on the escape probabilities), for example to
discuss the asymmetric line profiles caused by disc winds
\citep{Rab2022}.  The non-LTE populations are computed as laid out in
Sect.~\ref{sec:lineEscPro}, when the gas line heating/cooling rates
are determined, and the continuum source functions are calculated as
laid out in Sect.~\ref{sec:RT} for the continuum radiative
transfer.

A major improvement of this techniques was the development of the Fast
Line Tracer (FLiTs) by Michiel Min \citep[see][]{Woitke2018}, which
allows us to compute thousands of spectral lines at a time, to
efficiently produce the complicated infrared molecular line spectra
that we can now observe with the James Webb Telescope (JWST). Figures
\ref{fig:line1}, \ref{fig:line2} and \ref{fig:line3} show three
examples. FLiTs uses the level populations, dust opacities and
continuum source functions from ProDiMo, on the same $(r,z)$-grid, and
applies a fast numerical technique to trace all lines simultaneously,
including line overlaps and a clever randomisation of the ray
positions to avoid Nyquist-like artefacts.

\section{\bff Results and Implications}
\markboth{RESULTS AND IMPLICATIONS}
         {3.\ \ RESULTS AND IMPLICATIONS}

The following section provides a brief historical overview of the
development of the \ProDiMo\ radiation thermo-chemical disc model, its
application to various physical processes in discs, and the results
obtained with \ProDiMo\ to analyse and interpret various observational data
obtained with different instruments in different wavelengths domains.

\subsection{Physical Processes and Code Development}
\markright{3.1\ \ PHYSICAL PROCESSES AND CODE DEVELOPMENT}

\myparagraph{Code Basics:} The first published results of \ProDiMo\ appeared in
\citet{Pinte2009}, who performed a dust radiative transfer benchmark
for optically thick discs.  It was an important milestone for the
development of \ProDiMo\ to pass these tests, showing that the
ray-based continuum radiative transfer provides reliable results.  The
full modelling concept of \ProDiMo\ was then presented by \citet[][see
  Publication\,1 on page \pageref{PUB1}]{Woitke2009a}.  The paper
introduces the considered axi-symmetric disc geometry, the foundations
of the dust radiative transfer, the formulation of the chemical rate
network including UV photo-rates and a simple freeze-out ice
chemistry, and the treatment of the gas energy balance via the
calculation of various heating and cooling rates. This paper has
meanwhile reached over 300 citations.

\myparagraph{First line predictions:} In \citet{Woitke2009b}, far-IR
water emission lines from Herbig~Ae discs were calculated in
consideration up the upcoming ESA Herschel Space Observatory mission
(Herschel). To predict these lines, we utilised the non-LTE Monte
Carlo radiative transfer program RATRAN by \citet{Hogerheijde2000},
and discussed the differences between LTE population, non-LTE
population calculated by \ProDiMo's escape probability theory, and the
full non-LTE computations performed internally by RATRAN. These
investigations showed that the escape probability model (in the 2009
formulation) slightly underpredicts the water emission lines by about
5-20\%, whereas assuming LTE leads to critical line overpredictions by
a factor of about 2-3.  These same technique was used by
\citet{Kamp2010} to make detailed predictions for the
[OI]\,$2\!\to\!1$ (63.18\,$\mu$m), [OI]\,$3\!\to\!2$
(145.53\,$\mu$m), and [CII]\,$2\!\to\!1$ (157.74\,$\mu$m)
forbidden atomic lines targeted by the Herschel open time key
program GASPS ``{\em GAS evolution in Protoplanetary Systems}''.

\myparagraph{Deuteration of water:} \citet{Thi2010a} expanded
\ProDiMo's chemical network by a small set of deuterated species
$\rm\{D^+,D,HD,HD^+,OD,OD^+,HDO,HDO^+,H_2DO^+\}$ to discuss the
chemical pathways to form deuterated water and the resulting
HDO/H$_{2}$O fractionation ratio in discs.  The results show that it
is possible to obtain a surprisingly high water fractionation ratio of
order 10\% when the gas is warm ($\Tg\!\ga\!200\,$K), with interesting
consequences for the delivery of water to Earth.

\myparagraph{SEDs and inner rims:} In \citet{Thi2011b}, we presented
the first post-processing tool included in \ProDiMo\ to predict
observational quantities from the results of the disc models.  The
continuum radiative transfer is numerically solved along a bundle of
parallel rays, allowing us to calculate the integrated spectral flux
under inclination angle $i$ and predicting monochromatic images.  In
this particular paper, we consider hydrostatic disc models which have
tall inner rims in the case of Herbig~Ae discs. These ``puffed-up''
inner rims cast a shadow onto the outer disc, which is important for
near-IR excess and the shape of the spectral energy distribution (SED)
at longer wavelengths.

\myparagraph{Detailed line-transfer as post-processing:} In Appendix
A.7 of \citet{Woitke2011}, we introduced a method to calculate 
the line and continuum radiative transfer equation along a bundle of
parallel rays intersecting the disc midplane under inclination angle
$i$. This tool is similar to the SED tool presented by
\citet{Thi2011b}, but on a fine, velocity-resolving frequency grid,
where the Doppler-shift of the line-emitting gas according to a
Keplerian velocity field is taken into account. The line-transfer
post-processing tool is very robust and fast, but only applicable to
spectroscopically isolated lines.  The results typically show a
double-peaked line profile as e.g.\ observed by SMA and ALMA in the
(sub-)mm spectral region. The tool replaced the previously used calls
of RATRAN or MCFOST in all publications after 2011.  Besides properly
predicting the line fluxes, it allows us to calculate spatially
resolved line intensity maps and channel maps.  Both the SED and the
line tools are actually equipped with Fourier transform algorithms to
calculated visibilities for user-selected baselines, which so far was
only used once in \citet{Woitke2016}.

\myparagraph{X-ray chemistry:} \citet{Aresu2011} introduced an X-ray
chemistry to \ProDiMo\ with primary ionisations and secondary
ionisations by fast electrons, based on \citet{Meijerink2005}, with
doubly ionised atoms. These high-energy processes are particularly
relevant for T\,Tauri stars that have high X-ray to UV luminosity
ratios. The penetration of the stellar X-rays is treated by a radial
extinction law using X-ray gas opacities from the literature.  The
X-rays generally reach only slightly deeper layers in the disc when
compared to far-UV radiation, but cannot change the midplane
ionisation. However, X-rays can specifically unblock the CO, which can
lead to the formation of molecules like HCN and \ce{C2H2} in an
oxygen-rich environment, which has now become an active research topic
in view of the first results from the James Webb Space Telescope
(JWST), see \citet{Woitke2023}.  \citet{Meijerink2012} and
\citet{Aresu2012} presented a grid of X-ray disc models for T\,Tauri
stars and studied the impact of the X-rays on on the forbidden far-IR
lines of [OI] and [CII], as well as on the [NeII] line at
12.8\,$\mu$m, which has been frequently observed with the Spitzer
Space Telescope (Spitzer).  Neon cannot be ionised by far-UV radiation,
so this line is a prime indicator for X-ray driven chemistry in
discs. The same code was later used by \citet{Meijerink2013} to
discuss the time-chemistry chemistry in Active Galactic Nuclei (AGN)
discs, making line ratio predictions, such as HCN/HCN$^+$ for the
Atacama Large Millimeter Array (ALMA).

\myparagraph{CO ro-vibronic model molecule:} \citet{Thi2013b} collected
molecular data for the CO energy levels, radiative transitions, and
rates for collisional excitation by \ce{H2}, H, He, and \ce{e-}, for
the ground $X^1\Sigma^+$ and electronic $A^1\Pi$ state, up to 9
vibrational states each, and up to 50 rotational sub-states, creating
a large ro-vibronic CO model molecule.  This implementation allows us
to discuss the fluorescent pumping by UV light of the $X^1\Sigma^+$
vibrational states, which is an important observational diagnostic of
the physical state of the gas around the inner rim and in the upper
surface layers of Herbig~Ae discs close to the star, which often show
rotationally ``cool'' but vibrationally ``hot'' line spectra.

\myparagraph{Ice evolution:} In \citet{Helling2014}, we studied the
evolution of ices in the midplane of protoplanetary discs.  By setting
up time-dependent disc models with chemical abundances from molecular
clouds as initial condition, we simulated how ices form in the disc,
such as \ce{H2O\#}, \ce{CO\#}, \ce{CO2\#}, \ce{CH3OH\#}, \ce{CH4\#},
\ce{N2\#} and \ce{NH3\#}. Since the midplane is rather cold, the
molecules with stable ice phases can freeze out very quickly (within a
few 100\,yrs), but then it takes millions of years to convert the
remaining gas phase molecules without stable ice phases into those
which can freeze out, and this conversion is triggered by cosmic rays
to break the bonds of already existing molecules such as \ce{O2}. The
gas composition reflects these fast and slow changes, finally leading
to a gas that is devoid of oxygen and nitrogen (i.e.\ carbon-rich)
between the water iceline and the CO iceline, and devoid of all
elements with stable ice phases (i.e.\ pure hydrogen and noble gases)
beyond the CO iceline. These are the chemical pre-conditions for
planet formation.

\myparagraph{DIANA basics:} Between 2012 and 2016, I have been leading
the FP7-SPACE collaboration project ``DiscAnalysis'' (DIANA), bringing
together disc experts from five different European countries.  In
\citet[][see Publication\,4 on page \pageref{PUB4}]{Woitke2016} we
summarised our agreements about the modelling foundations of
protoplanetary discs, such as a powerlaw surface density profile with
exponential tapering-off, dust settling, standard dust opacities and
treatment of PAHs.  With over 230 citations to date, this is my second
best cited publication.  \citet{Kamp2017} is a follow-up publication
where we published standard element abundances and the setups of our
enlarged chemical networks for the disc modelling, with excited
molecular hydrogen, with PAHs in five different charging states and
X-ray processes.  All neutral gas species (except for H, \ce{H2} and
noble gases) can freeze out, and all molecules can be protonated.
These rate networks became later known as the ``small and large DIANA
standard'' chemical networks with 100 and 235 chemical species,
respectively.

\myparagraph{High-Energy Physics:} The FP7-DIANA project triggered
three important inclusions of high-energy physics into
\ProDiMo. \citet{Rab2017a} included Stellar Energetic Particles (SEPs)
and compared their impact on T\,Tauri disc models to the impact of
Cosmic Rays (CRs) and radioactive decays.  \citet{Rab2017b} studied the
time-dependent chemistry after episodic accretion events, where the
stellar luminosity suddenly increases by a factor of 100, showing that
the ices in the inner disc regions all sublimate quickly, but it takes
tens of thousands of years for the chemistry in the outer disc regions
to return to quiescent state, with implications on ALMA line
observations. The paper also introduces a way to include an in-falling
extended envelope to the disc modelling.  Furthermore, \citet{Rab2018}
included X-ray radiative transfer including Compton scattering to
\ProDiMo, showing that the scattering enhances the X-ray impact on
disc ionisation in deeper disc layers.  \citet{Brunn2023} have used
\ProDiMo\ to predict the midplane ionisation of inner T\,Tauri
discs affected by magnetic re-connection events, where the magnetic
field of the star interacts with the magnetic field of the disc,
which produces SEPs that ionise the disc.

\myparagraph{IR Molecular Emission Lines:} In \citet[][see
  Publication\,5 on page \pageref{PUB5}]{Woitke2018} we introduced the
Fast Line Tracer (FLiTs) developed by Michiel Min, which allows us to
calculate thousands of velocity-resolved IR molecular emission lines
at a time as post-processing. This tool, which uses the disc
structure, dust opacities and molecular abundances calculated by
\ProDiMo, is nowadays key to interpret JWST spectra. In this paper, we
showed that the IR molecular emission lines are expected to be emitted
from different layers close to the star, from OH (highest) -- CO --
\ce{H2O}-- \ce{CO2} -- \ce{HCN} -- \ce{C2H2} (deepest), and discuss
the differences between molecular emissions coming from the disc
surface and form distant walls behind disc gaps.
\citet{Greenwood2019a} studied the IR line-emitting regions of
\ce{H2O}, OH, \ce{CO2}, HCN and \ce{C2H2} in T\,Tauri discs by means
of \ProDiMo\ models, and reported on the changes of the predicted line
spectra with the stellar UV and X-ray luminosities, the disc flaring
angle, and dust settling.

\myparagraph{Dust Evolution:} \citet{Greenwood2019b} coupled
\ProDiMo\ to the dust evolution model called ``two-pop-py'' by
\citet{Birnstiel2012,Birnstiel2015}. This 1D-model includes dust
growth, fragmentation and radial drift by means of a simplified
version of \citet{Birnstiel2010}, only using two dust size bins.  The
small grains remain coupled to the gas, whereas the large grains are
affected by radial drift and can eventually disappear. As the dust
population evolves and diminishes, the continuum optical depths in the
disc surface are reduced, which can increase the IR line fluxes of
\ce{H2O}, OH, \ce{CO2}, \ce{HCN} and \ce{C2H2} by up to a factor of
100.

\myparagraph{Brown Dwarf Discs:} \citet{Greenwood2017} extended
\ProDiMo's range of applicability to brown dwarf discs, using the
Drift-Phoenix brown dwarf synthetic spectra by
\citet{Witte2009,Witte2011}, and discussed observable ALMA line ratios
such as $^{13}$CO 3-2/\ce{HCO+} 4-3, and their difference to T\,Tauri stars.

\myparagraph{Dust charging, MRI, and Lightning in Discs:} \citet{Thi2019}
introduced a moment method to include dust grains with arbitrarily
large positive or negative charges into \ProDiMo's chemical rate
network. In the midplane, the dust grains collect most of the free
electrons created by cosmic rays, leading to highly negatively charged
grains, which has a significant impact on the midplane ionisation.
The paper discusses further how the gas ionisation is connected to the
dynamical coupling between gas and magnetic field, and how it
therefore changes the effective viscosity parameter $\alpha$ of the
magneto-rotational instability (MRI), which drives mass accretion and
disc evolution.  \citet{Balduin2023} have recently expanded this model
to include size-dependent charging of dust particles, showing that the
grains eventually reach a charge proportional to their size, where
$\langle Q\rangle/a$ ranges from a few hundreds of negative charges
per micron in the midplane to several thousands of positive charges
per micron in the upper irradiated disc layers. The paper also
includes a simple model for the electrification of the disc due to
turbulent motions of the charged dust grains, concluding however, that
this mechanism is too inefficient to create electric fields
of the order of the critical electric field for a run-away collisional
ionisation (lightning). 

\myparagraph{Surface Chemistry and Phyllosilicates:} \citet{Thi2020b}
implemented a state of the art surface chemical rate network in
\ProDiMo, benchmarked against \citet{Semenov2010}. This paper, however,
concentrates first and foremost on the \ce{H2} and HD formation on
warm grain surfaces.  \citet{Thi2020a} then expanded the network to
discuss the formation and stability of chemisorbed water, and simulated
how chemisorbed water might diffuse into the silicate lattice on
timescales of millions of years to form phyllosilicates (hydrated
silicates), which are stable up to about 700\,K. In a related work by
\citet{DAngelo2019}, \ProDiMo\ models have been used to estimate the
content of chemisorbed water present in form of hydrated
silicates. They conclude that, only taking into account the first
mono-layer on 0.1\,$\mu$m grains, 0.5-10 Earth oceans of chemisorbed
water can be present in protoplanetary discs, showing how water might
have been delivered to planet Earth.

\myparagraph{Miscellaneous Chemical and Physical Processes}
\citet{Meisner2019} quantified Arrhenius parameters for gas phase
reactions that are enhanced by atomic tunnelling at low temperatures.
They used \ProDiMo\ models to test the impact of their improved rated
on disc chemistry and mid-IR water lines.  \citet{Oberg2020} have used
\ProDiMo\ to model the dust radiative transfer in circumplanetary
discs, for the conditions expected for the young Jupiter, using a
hierarchical approach, where a model for the circumstellar disc
provides the boundary conditions for the circumplanetary disc
model. Using the resulting FUV radiation field, photoevaporative mass
loss rates from the Jovian circumplanetary disc are computed, causing
a truncation outside of the orbit of Callisto.  In \citet{Oberg2022}
the ice formation and viscous evolution of circumplanetary discs is
simulated, and further conclusions are drawn about the formation of
Jupiter's icy moons.  The paper also introduced a diffusion solver for
the continuum radiative transfer that has become standard for all
\ProDiMo\ disc models after 2022.  \citet{Rocha2023} simulated the
UV-irradiation of \ce{CH3OH} ice (UV photolysis) under laboratory
conditions and compared the theoretical \ProDiMo\ results with lab
data, considering the time-dependent destruction of the methanol ice,
the growth curves of the photolysis products, including branching
ratios, and the re-formation of methanol ice, which turns out to be
critical for the interpretation of the lab data.
\citet{Guadarrama2022} studied the effect of metallicity on molecules
in the disc observable with ALMA, in particular CO, HCN, CN, \ce{HCO+}
and \ce{N2H+} with \ProDiMo.

\citet{Arabhavi2022} included position-dependent ice opacities in
\ProDiMo. Depending on the results of \ProDiMo's ice chemistry, the
thickness and composition of the ice layers of the grains are
calculated, the opacities are interpolated in pre-computed
multi-dimensional opacity tables, and these opacities are fed back to
the radiative transfer (RT) calculations.  This procedure requires
iterations between RT and chemistry, but is found to converge after
just a few global iterations.  The paper discusses the observability
of near and mid-IR ice features with JWST. Without mixing (see below),
the \ProDiMo\ models only show abundant ices deep in the disc
(vertical optical extinction $A_{\rm V,ver}\!\ga\!10$, because of
UV-desorption. According to these models, the ices hardly produce any
observable spectral features.  \citet{Woitke2022}: explored the
effects of vertical turbulent mixing on the chemical structure of
protoplanetary discs. The mixing leads to an additional vertical
transport of molecules and icy grains into lower and higher disc layers
where these species do not form in-situ.  At the locations where they
are finally destroyed, for example by photo-processes, the
release of reaction products creates a more active chemistry, with a
richer mixture of ionised, atomic, molecular, and ice species, and new
chemical pathways that are not relevant in the unmixed case. In
particular, the paper shows that icy grains can be transported upwards
faster than photo-desorption can sublimate the icy mantles, and in this
case, already for slow mixing $\alpha\!=\!10^{-4}$, the ices start to
produce spectral emission features observable with JWST. Another
principle effect is that molecules and atoms, which form in higher,
warmer and more illuminated layers, can be transported downward into
the midplane, which can be relevant for ALMA observations.

\subsection{Analysis of Herschel Observations}
\markright{3.2\ \ ANALYSIS OF HERSCHEL OBSERVATIONS}

Since \ProDiMo\ was originally developed to interpret the far-IR gas
emission line observations of the Herschel open-time key program GASPS
(see review by \citet{Dent2013}), the first applications of
\ProDiMo\ to actual observational data in the early years (2010-2014)
were almost exclusively centred around these data.

\myparagraph{Preparations for the Herschel mission:} Even before the
first Herschel observations were carried out, \citep[][see
  Publication\,2 on page \pageref{PUB2}]{Woitke2010} prepared a large
grid (called the DENT grid) of 300000 disc models where 11 stellar,
disc and dust parameters were varied systematically, including the
total disc mass, several disc shape parameters and the dust-to-gas
ratio. In this grid, the Monte Carlo radiative transfer program MCFOST
\citep{Pinte2009} was used to perform the dust continuum RT, the
resulting radiation field and dust temperatures were then passed on to
ProDiMo to calculate the gas chemistry and determine the gas
temperature structure, and those results were passed back to MCFOST to
do the line transfer calculations. For each model, dust continuum and
line radiative transfer calculations were carried out to predict 29
far-IR and (sub-)mm lines of [OI], [CII], CO and \ce{H2O} under five
disc inclination angles.

\myparagraph{Early Herschel/GASPS results:} The first Herschel/GASPS
results about the [OI]\,63.18\,$\mu$m, [OI]\,145.53\,$\mu$m and [CII]
157.74\,$\mu$m emission lines reported by \citet{Mathews2010} were
within the expectations from the DENT grid for the T\,Tauri stars and
HD\,169142, see \citep{Pinte2010}, but the non-detections of these
lines for the bright Herbig~Ae star HD\,181327 were puzzling.  Later,
\citet{Lebreton2012} concluded that HD\,181327 is likely an object
which has passed the stage of gaseous planet formation already and the
observations led to a classification of the disc around HD\,181327 to
be an icy Kuiper belt. \citet{Meeus2010} presented the first serious
attempt to derive the gas and dust masses of the disc around the young
Herbig Ae star HD\,169142 from the Herschel/GASPS data together with a
collection of auxiliary photometric and CO millimetre line data. A
\ProDiMo\ model was convincingly fitted to these observations which
could explain the SED, fitted all Herschel/GASPS line flux detections
and non-detections, and the CO 2-1 and $^{13}$CO 2-1 lines.  The model
featured an inner disc, a gap, and a continuous disc between 20 and
200\,au with a gas-to-dust ratio of about 20--50.  In a similar way,
\citet{Thi2010b} presented a \ProDiMo\ disc model for the
$\sim$\,10\,Myrs old T\,Tauri star TW\,Hya using the Herschel/GASPS
data together with auxiliary photometric and CO millimetre line data.
The paper concluded that the gas-to-dust ratio has decreased to only
about 3--25 which is significantly lower that standard interstellar
value of 100, suggesting that a significant fraction of the primordial
gas has already disappeared.  \citet{Thi2011a} reported on the
detection of a series of rotational lines of CH$^{+}$ at from
$J\!=\!5\to 4$ at 72.16\,$\mu$m down to $J\!=\!2\to 1$ at
179.59\,$\mu$m by Herschel in HD\,100546. They provided a
\ProDiMo\ disc model for HD\,100546, to fit these and other Herschel
and sub-mm CO line fluxes, deriving a gas mass of about
$10^{-3}\,M_\odot$, which corresponds to a gas-to-dust ratio of about
8.

\myparagraph{Subsequent analysis of observations and models:} Based on
the DENT grid created for the Herschel/GASPS open-time key program,
\citet{Kamp2011} presented a thorough analysis of the correlations
found between the predicted line fluxes in the models, and the trends
and dependencies of these line fluxes on continuum fluxes and disc
parameters. The paper aimed at providing simple diagnostic tools to
derive integrated disc properties based on the measured Herschel and
sub-mm line fluxes, but a simple formula to derive e.g.\ the disc mass
from these line observations was shown to be difficult.
\citet{RiviereMarichalar2012} presented the results of the
Herschel/GASPS program concerning the T\,Tauri stars in the
Taurus-Auriga star formation region. From the 68 objects, 33 were
detected in the forbidden [OI] 63.18\,$\mu$m fine structure line, and
8 T\,Tauri stars were detected in the \ce{o-H2O} emission line close
by at 63.32\,$\mu$m as a by-product. These detection rates are clearly
limited by the S/N ratio of the Herschel/PACS observations.
\ProDiMo\ models were used to interpret these data in terms of the
line excitation temperatures and line emission regions.  Additional
Herschel observations and fitted disc models have been published for
the debris disc 49\,Cet \citep{Roberge2013}, the active T\,Tauri star
DG\,Tau \citep{Podio2013}, the young luminous B9.5 star 51\,Ophiuchi
\citep{Thi2013a}, the low-mass disc of HD\,141569A \citet{Thi2014},
and the exceptionally bright source T\,Tau\,N with extended envelope
\citep{Podio2014}. Despite the very different nature of these objects,
\ProDiMo\ models have been shown to be a powerful tool to analyse and
interpret the Herschel line observations.

\myparagraph{Analysis of larger data samples:} In later stages of the
Herschel data analysis, more in-depth studies were presented, and
larger data samples were considered. For example, \citet{Kamp2013}
challenged the \ProDiMo\ models to see how the chemical network, the
assumed metallicity and the carbon-to-oxygen ratio might need to be
modified to improve our understanding of the water line formation in
TW\,Hya, combining the data from several Herschel observational
programs. \citet{VanDerWiel2014} examined the Herschel SPIRE
spectroscopic observations between 240 and 660\,$\mu$m towards 18
protoplanetary discs, publishing series of high-$J$ emission line
fluxes of CO and $^{13}$CO from 11--10 down to 4--3, beside additional
line fluxes of water, HCN, \ce{CH+}, and [CI].  \ProDiMo\ models were
used to fit the CO spectral line energy distributions (SLEDs) and to
understand which disc regions are possibly populated with warm gas
that is required to excite these lines.

\citet{Aresu2014} analysed [OI]\,63.2\,$\mu$m line data obtained with
Herschel GASPS and studied the impact of the UV and X-ray luminosities
of the central stars on these line fluxes, both observationally and
using \ProDiMo\ models.  According to the models, the gas temperatures
above the H$\to$\ce{H2} transition, which is crucial for the
[OI]\,63.2\,$\mu$m line emission, should critically depend on the UV
and X-ray irradiation by the star, however, the sample of observations
does not seem to support this hypothesis. \citet{Keane2014} collected
Herschel line and continuum data for 21 transitional discs in Taurus,
Chameleon, Lupus and Ophiuchus. They used statistical tests to search
for correlations with known stellar properties, such as effective
temperature, stellar mass, accretion rate and UV luminosity, and disc
properties like the radius of the inner cavity. The DENT grid of
\ProDiMo\ models was used to extract the expected trends from the disc
models. The study remained largely inconclusive.

\citet{Antonellini2015} used \ProDiMo\ models to compare the predicted
behaviour of the far-IR water lines of T\,Tauri discs observable with
Herschel to those in the mid-IR observable with Spitzer. The paper
discussed which stellar, disc and dust parameters are important for
the line formation for the two families of water lines.
\citet{Antonellini2016} studied a puzzling result obtained with
Herschel and Spitzer, namely that water line detection rates for
T\,Tauri stars are remarkably lower than those for the more luminous
Herbig Ae/Be stars.  The explanation given in the paper is that the
brighter stars also produce a brighter IR continuum flux, and the line
data reduction techniques actually depend critically on the relative
continuum flux noise, so lines with the same flux on a brighter
continuum could not be detected.  However, this explanation might not
entirely explain this clear observational trend.

\subsection{Analysis of ALMA Observations}
\markright{3.3\ \ ANALYSIS OF ALMA OBSERVATIONS}

\begin{figure}
  \hspace*{-3mm}
  \begin{tabular}{cc}
  \includegraphics[width=69mm]{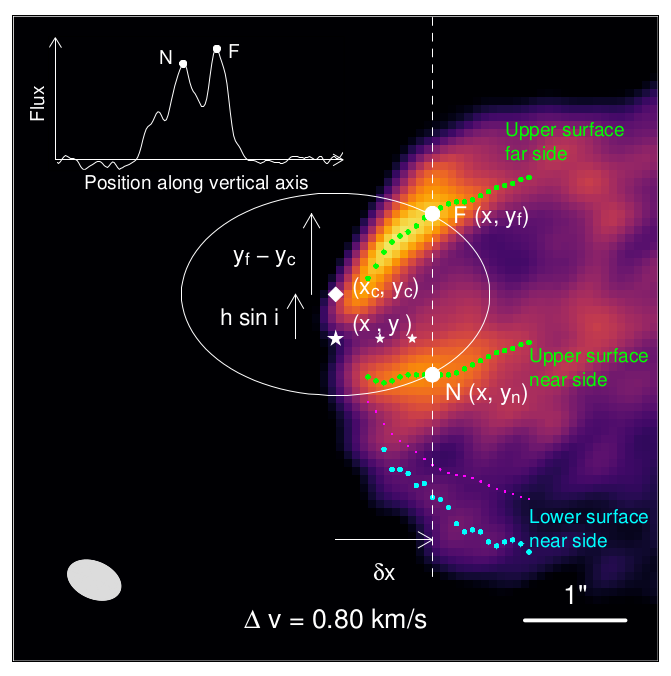} &
  \includegraphics[width=93mm]{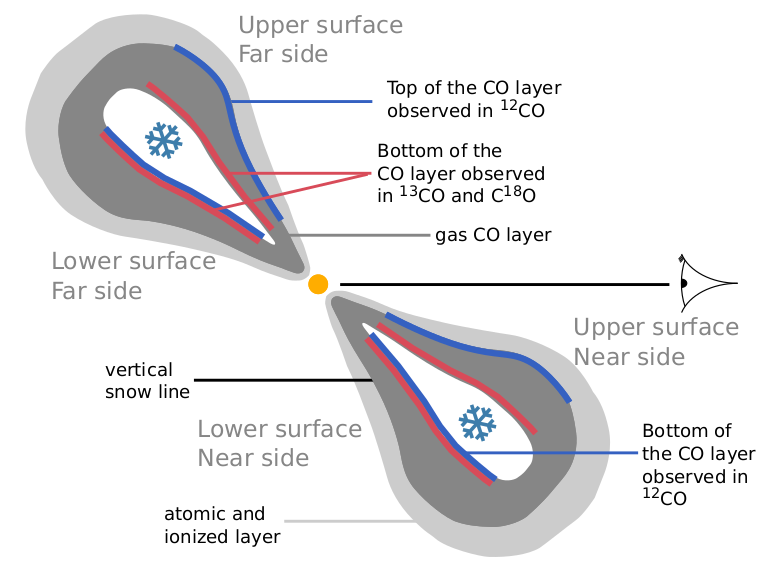}
  \end{tabular}\\[-1mm]
  \resizebox{\textwidth}{!}{\parbox{18.4cm}{\caption{Left: observed
        channel map of $^{12}$CO 2--1 of the T\,Tauri star IM
        Lupi. Right: schematic view of the CO emitting surfaces of IM
        Lupi, re-printed from \citet{Pinte2018}.}
  \label{fig:IMLupi}}}
\end{figure}

\citet{Pinte2018} published a spectacular paper in which the accurate
ALMA channel maps of three CO 2-1 isotopologue lines of the T\,Tauri
star IM Lupi allowed for a direct determination of the geometrical
height over the midplane from where the CO is emitting, as well as
its emission temperature as function of radius.  There are two
emitting CO surfaces (see Fig.~\ref{fig:IMLupi}), corresponding to the
close disc side and the far disc side that we see through the
midplane. These two layers are separated by the icy midplane in which
the CO is frozen out, and the upper boundary of the CO emitting layer
corresponds to CO photo-dissociation.  Since the CO-rich layer is
optically thick, we only see the surface, which is warm where the CO
photo-dissociates, and cold where it freezes out, and the ALMA
observations allow us to measure these two temperatures directly from
the observations, as function of radius.  Many of these observed CO
properties are very close to the expectations from \ProDiMo models, for
example the authors derive $21\pm 2$\,K for the CO freeze-out
temperature. In the \ProDiMo\ models we see values between about 20
and 25\,K.

\citet{Rab2019} modelled circumstellar discs around the T\,Tauri stars
known to host planet-mass companions (PMCs), which are likely still
embedded in the disc and have circumplanetary discs (CPDs) around
them.  By using hypothetical circumplanetary disc properties, for
example a planet mass of 20 Jupiter masses and a PMC luminosity of
0.01 to 0.001\,$L_\odot$, they discussed whether these CPDs might be
detectable with VLT/SPHERE and with ALMA in continuum and in CO lines.
\citet{Rab2020} presented a detailed study of the rings observed with
ALMA in the disc around the Herbig~Ae star HD\,163296. By analysing
the CO line photons coming from the far side of the disc through the
dust gaps (which are dark and optically thin in the continuum), the
authors were able to derive the gas column densities in the gaps,
which are much less deficient than the dust column densities.

\subsection{Analysis of Other Observations}
\markright{3.4\ \ ANALYSIS OF OTHER OBSERVATIONS}

\myparagraph{Optical and Near-IR spectroscopy:} \citet{Rab2022} have
used the density structure of 2D XUV photoevaporative disc wind models
calculated with the PLUTO hydrodynamics code \citep{Mignone2007} in
\ProDiMo.  In addition, the complex 3D wind velocity structure was
used during the line post-processing to predict o-H$_2$\,2.12\,$\mu$m
and [OI]\,6300\AA\ line fluxes and asymmetric line profiles. The
results are compared to high-resolution observations of the Telescopio
Nazionale Galileo ({\bff TNG}) in the GIARPS observing mode for the
near-IR, and the high-resolution spectrograph {\bff HARPS-N} for the
optical \nocite{Gangi2020}. The results are found to be generally
consistent with the observed wind signatures for both the
[OI]\,6300\AA\ and the o-H$_2$\,2.12\,$\mu$m spectral lines.

\myparagraph{L-band and M-band spectroscopy:} \citet{Fedele2011}
obtained high-resolution ($R\!\approx\!10^5$) L-band spectroscopic
data with {\bff VLT/CRIRES} of 11 Herbig AeBe stars, detecting a
number of hot water and OH lines. The authors discussed the principal
chemical formation and destruction paths of OH and \ce{H2O}, and
studied the line forming regions on the basis of the predictions from
by means \ProDiMo\ models.  Their conclusion was that there must be
some kind of particular depletion of water in Herbig AeBe discs.
\citet{HeinBertelsen2014} obtained VLT/CRIRES observations of CO
fundamental ro-vibrational lines from HD\,100546. They used
\ProDiMo\ models with the large ro-vibronic CO model molecule of
\citet{Thi2013b} to determine the CO line excitation mechanism, and
compare these models to the observations. The authors concluded about
the spatial extent of the CO emissions, with the final goal of using
the CO ro-vibrational lines as a diagnostic tool to determine the
inner disc structure in the context of planet formation.
\citet{HeinBertelsen2016b} observed CO fundamental ro-vibrational
emission lines from six Herbig~Ae stars with VLT/CRIRES and found
different types of line profiles including narrow single-peaked, broad
single-peaked and double-peaked. \ProDiMo\ models were used to relate
these different line profiles to flared and self-shadowed disc
geometries, and the presence of large gas holes or gaps.
\citet{HeinBertelsen2016a} observed CO fundamental ro-vibrational
emission lines from HD\,163296 for different epochs 10 years apart.
They found the central line profiles to change from double-peaked to
single peaked, whereas the line wings remained
similar. \citet{Antonellini2020} used archival VLT/CRIRES observations
of fundamental ro-vibrational CO emission lines from 37 T\,Tauri
stars, and compared the behaviour of these lines, in form of line flux
ratios and line widths, to the predictions from \ProDiMo\ models.
They concluded that the complex shapes of the inner disc, in
particular disc gaps and inner cavities, need to be considered to
understand these observations. 
\citet{Oberg2023} have studied the possibilities to observe
CO fundamental ro-vibrational emission lines with the mid-infrared ELT
imager and spectrograph ({\bff METIS}), which provides an exquisite
combination of high spectral with high angular resolution.
Circumplanetary discs have been identified as very promising targets,
as their gas rotation in the circumplanetary disc leaves detectable
signals in the background of the gas performing circumstellar motions.
These simulations are performed with two \ProDiMo\ disc models, one
for the circumstellar disc and one for the circumplanetary disc.

\myparagraph{Mid-IR spectroscopy:} \citet{Antonellini2017} used
\ProDiMo\ models to analyse the behaviour of three water emission line
blends observed by {\bff Spitzer/IRS} around 15.2$\,\mu$m,
17.2$\,\mu$m and 29.9$\,\mu$m. The authors discussed the predicted
line emission regions and expected trends with dust and stellar
parameters such as spectral type and stellar luminosity. They found a
correlation with the amplitude of the 10\,$\mu$m silicate dust
emission feature. \citet{DeLaRoche2021} used the {\bff VLT/VISIR}
instrument, and its upgrade VLT/VISIER-NEAR, to observe
high-resolution mid-IR images between 8.5 and 11.3\,$\mu$m.  They
measured the apparent sizes of the discs at these wavelengths covering
several PAH features.  The authors used the DIANA standard model of
HD~100546 \citep{Woitke2019} and compared their observations with
these model predictions.  After slight modifications of the inner
radius of the outer disc and the PAH abundance, they obtained a good
fit. \citet{Ercolano2022} have studied the potential of the exoplanet
missions {\bff TWINKLE} and {\bff ARIEL} to detect PAHs in the
atmospheres of discs and exoplanets. Using the PAH opacities from
\ProDiMo, the 3.3\,$\mu$m feature was identified as the most promising
PAH indicator in transmission spectroscopy, possibly related to haze
cloud layers.  Similar detections might be possible with {\bff
  JWST/NEARspec} or, at longer wavelengths, with {\bff
  JWST/MIRI}. \citet{Woitke2023} have published the first fit of a
JWST/MIRI spectrum by a full 2D thermo-chemical disc model. The JWST
observations of the T\,Tauri star EX~Lupi showed emission lines by CO,
\ce{H2O}, OH, \ce{CO2}, HCN, \ce{C2H2} and \ce{H2}. The fitted
\ProDiMo\ model can roughly reproduce all these lines, and provides
the disc structure, all molecular concentrations as function or
$(r,z)$, the gas and dust temperature structures and, in case of CO,
\ce{H2O} and OH, the non-LTE level populations in a consistent
way. According to the fitted model, the inner disc has a slowly
ascending surface density profile combined with strong dust
settling. The HCN and \ce{C2H2} molecules form in the close midplane
via X-ray chemistry in the oxygen-rich gas. All emission lines
observable with JWST originate from a layered molecular structure in
the inner disc $<\!0.7$\,au.

\myparagraph{Sub-mm line Observations:} \citet{DrabekMaunder2016}
observed the \ce{HCO+} 4--3 line at 840.38\,$\mu$m of Lk\,Ca\,15 with
the James Clerk Maxwell Telescope ({\bff JCMT}) and showed that the
extended wings of this emission line probe the gas inside of the
well-known dust rim at the inner cavity at about 50\,au.  They used
\ProDiMo models to provide a detailed 2-zone fit of the SED and the
\ce{HCO+} line flux and profile, concluding that the gas-to-dust mass
ratio in the inner cavity must be as large as $10^6$ to $10^7$.
\citet{White2019} obtained {\bff APEX} observations of the CO gas in
the envelope of the young FUor-type star V883\,Ori. They used \ProDiMo
models to test several envelope configurations.  They found that an
envelope with a mass of 0.2--0.4\,$M_\odot$ and a mass-infall rate of
$\dot{M}\!=\!(1-2)\times 10^{-6}\rm\,M_\odot/yr$ fits the observations
best, however the outflow structure around V883\,Ori is likely caused
by multiple outburst events, consistent with episodic accretion.

\subsection{Application to Multi-$\lambda$ observational data sets}
\markright{3.5\ \ APPLICATION TO MULTI-$\lambda$ OBSERVATIONAL DATA SETS}

\citet[][see Publication\,3 on page \pageref{PUB3}]{Woitke2011}
published a detailed \ProDiMo\ disc model for the low-mass T\,Tauri
star ET~Cha, fitting a large set of continuum and line
observations. The observational data included new optical/near-IR
ANDICAM photometric measurements, Herschel/PACS line flux results for
10 far-IR lines (only [OI] 63.2\,$\mu$m was detected), APEX/LABOCA
observations for the 870\,$\mu$m continuum and the CO 3--2 line,
archival data for the o-H$_2$\,2.12\,$\mu$m line, and a
high-resolution optical spectrum taken with the Anglo-Australian
Telescope AAT and University College London coud{\'e} Echelle
Spectrograph (UCLES) showing blue-shifted emission lines
[OI]\,6300\,\AA, [SII]\,6731\,\AA, [SII]\,6716\,\AA\ and
[NII]\,6583\,\AA.  Three archival HST/COS and HST/STIS spectra were
used to determine the incident far-UV spectrum of the star. The
non-detection of CO 3-2, which is very surprising for this type of
objects, was explained in the paper by assuming an extremely small
disc with an outer radius of only (6-9)\,au.  This conclusion was
later beautifully confirmed by ALMA CO 3--2 observations
\citep{Woitke2013} showing a weak but very broad line from which we
were able to confirm a disc outer radius of about 5\,au.

\citet{Tilling2012} developed a detailed \ProDiMo\ disc model for the
bright and very well-known Herbig~Ae star HD\,163296. They fitted a
compilation of photometric fluxes for the SED, a number of UV spectra
to determine the UV input spectrum, a near-IR SpeX/IRTF spectrum, a
mid-IR Spitzer/IRS spectrum, the measured Herschel fluxes of 13 [OI],
[CII], o/p-\ce{H2O} and CO lines between 63 and 181\,$\mu$m (all
non-detections but [OI]\,63.2\,$\mu$m), and three (sub-)mm CO and
$^{13}$CO lines. The fit showed the necessity to assume a large disc
with exponentially tapering-off surface density profile. Ian Tilling
was a former PhD student in Edinburgh (UK) under my supervision.

\begin{figure}[!b]
  \centering
  \begin{tabular}{cc}
  \includegraphics[width=71mm]{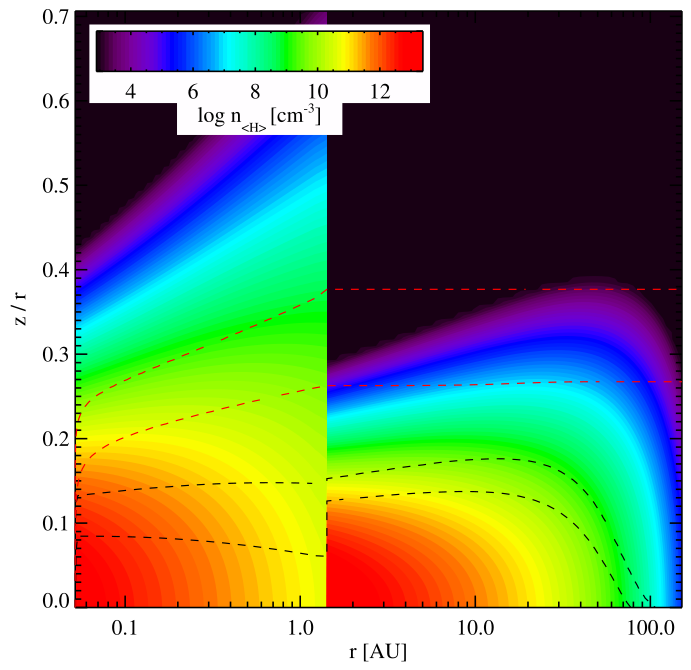} &
  \includegraphics[width=84mm]{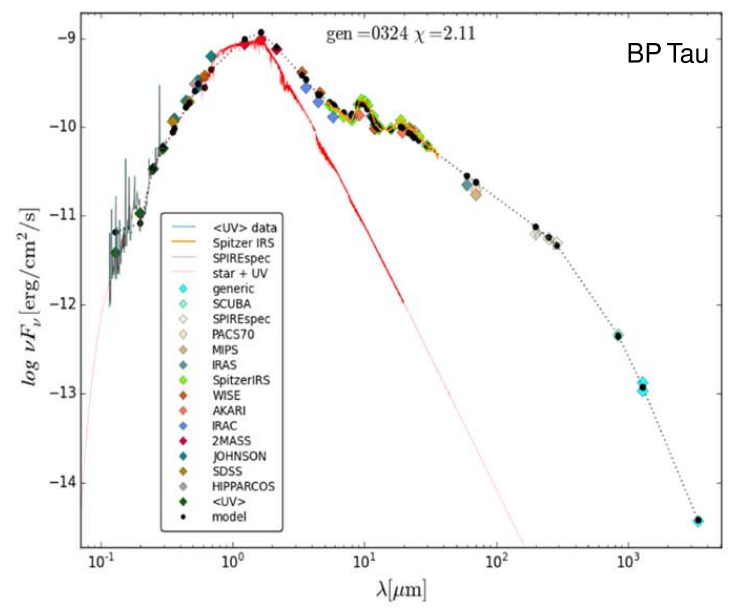}
  \end{tabular}\\[-1mm]
  \includegraphics[width=135mm]{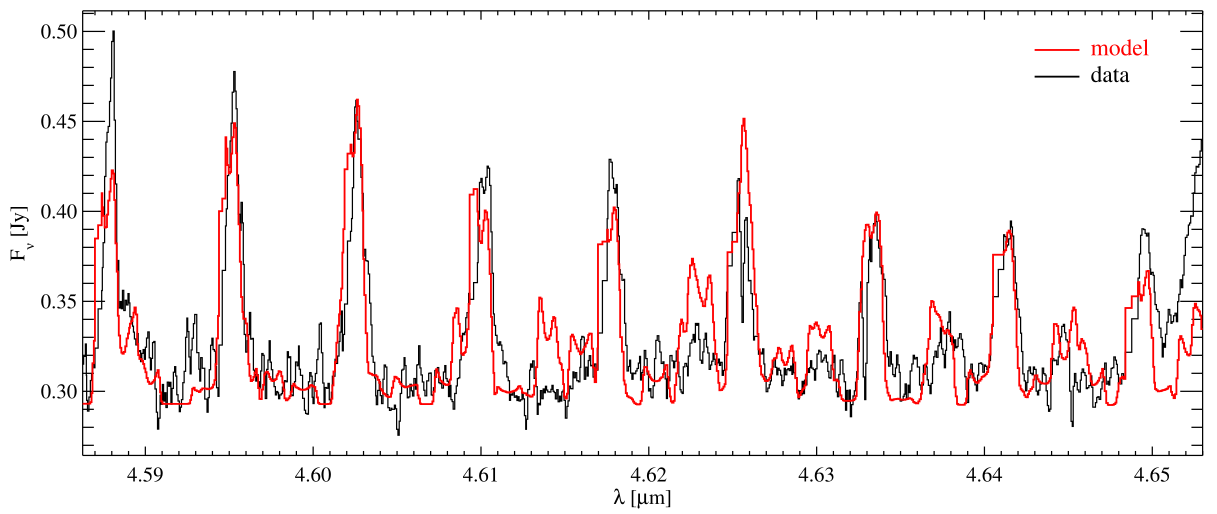}\\[1mm]
  \hspace*{-2mm}
  \includegraphics[width=163mm]{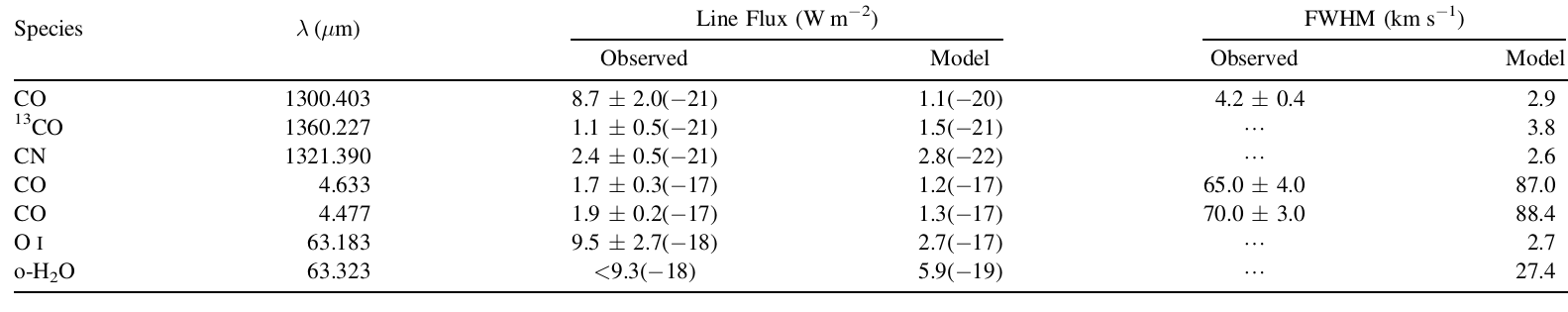}\\[-1mm]
  \resizebox{\textwidth}{!}{\parbox{18.4cm}{\caption{Multi-wavelength
        fit of various observations of the disc of T\,Tauri star
        BP\,Tau. The upper row of figures shows the assumed disc
        density structure $\nH(r,z)$ and the computed spectral energy
        distribution in comparison to multi-$\lambda$
        observations. The central figure shows a small part of the
        M-band IR spectrum of fundamental ro-vibrational lines of CO
        with FLiTs-fit, and the lower table shows the fit of other
        lines and line widths.}
  \label{fig:BPTau}}}
\end{figure}

\citet{Carmona2014} collected continuum and line observations for the
transitional Herbig Ae disc of HD\,135344B from 14 different
instruments, among them HST/COS, VLT/UVES, VLT/CRIRES, VLTI/PIONIER,
Subaru/HICIAO, Herschel/PACS, SMA and ALMA.  The observational data
included the SED, near-IR polarised dust scattered light images, IR
visibilities and closure phases, [OI]\,6300\,\AA, CO fundamental
ro-vibrational observations including spectro-astrometry signals,
11.2\,$\mu$m PAH emission, [OI]\,63.2\,$\mu$m, [OI]\,145.5\,$\mu$m,
[CII]\,157.7\,$\mu$m, CO 6--5, CO 3--2 and CO 2--1.  Using both MCFOST
and \ProDiMo\ disc models, the authors fitted the disc geometry
including a disc gap, the dust-to-gas ratio and a carbon-enriched
inner disc. One of their conclusions was that the CO fundamental
ro-vibrational lines must be partly emitted from the gas in the gap
between an inner and an outer dusty disc.

\citet{Garufi2014} used \ProDiMo\ models to fit a large
multi-$\lambda$ observational data set for the young low-mass T\,Tauri
star FT\,Tau that included an optical spectrum from TNG (La Palma
Observatory), a J-band spectrum taken with the long-slit Intermediate
Resolution Infrared Spectrograph at the WHT (La Palma Observatory), a
K-band spectrum taken with NOTCAM (Nordic Optical Telescope CAMera), a
high-resolution M-band Keck/NIRSPEC observations with CO fundamental
ro-vibrational emission lines, and diverse photometric, Herschel/PACS
and Spitzer/IRS observations. The modelling of FT\,Tau with
\ProDiMo\ revealed a massive disc ($0.02\rm\,M_\odot$) with a close
inner rim (0.05\,au) that is obscured by interstellar dust clouds
(optical exinction $A_V\!=\!1.8$).

\citet[][see Publication\,6 on page \pageref{PUB6}]{Woitke2019}
summarised the main modelling results of the FP7 DIANA project.
Similar to the papers described above, we collected large
multi-$\lambda$ observational data sets for 27 well-known
protoplanetary discs. The data collection included X-ray stellar
properties and modelled X-ray input spectra as seen by the disc, low
resolution UV-spectra, spectra from various instruments to cover the
optical and near-IR wavelengths, high-resolution M-band spectra for
the ro-vibrational CO lines, continuum images, interferometric data,
emission line fluxes, line velocity profiles and line maps. These
observations probe the dust, polycyclic aromatic hydrocarbons (PAHs)
and the gas in these objects. A standardised modelling procedure was
presented to fit these by our disc modelling codes (\ProDiMo, MCFOST,
MCMax), solving the continuum and line radiative transfer, disc
chemistry, and the heating and cooling balance for both the gas and
the dust.  \citet[][see Publication\,7 on page
  \pageref{PUB7}]{Dionatos2019} published all collected observations
in an open-access database.  The aim was to fit all available
observations of one object by one single disc model each.

A daunting question initially hanging over the DIANA project was {\em ``does
  this work?''} The answer was a surprisingly clear yes. In reality,
discs are very complicated objects, individual, time-dependent and not
strictly axi-symmetric. Yet, by allowing just for two disc zones, an
inner and an outer disc, we could find parameter combinations for our
models, which predict observations that resemble most of the continuum
and line data we could find, simultaneously.  One of the resulting
fits is shown in Fig.~\ref{fig:BPTau}.  As a result of these efforts,
we were able to discuss the average disc geometry of the fitting disc
structures, the scale heights, the disc masses, the dust opacities,
the dust-to-gas ratios and the PAH properties.  For example, the mean
viscosity parameter for the dust settling was found to be
$\log_{10}\alpha\!=\!-2.9\pm 0.9$, the PAH abundance with respect to
interstellar standards $f_{\rm PAH}\!=\!0.005-0.8$, and the mean dust
opacity $\kappa_{850\,\mu{\rm m}}^{\rm
  abs}\!=\!6.3^{+3.5}_{-2.3}\rm\,cm^2/g(dust)$.  We found that some
discs can be so cold in the distant settled midplane, which contains
most of the dust mass, that the dust barely emits at 850\,$\mu$m and
1.3\,mm, which causes the standard method to derive disc masses from
millimetre fluxes to fail. We found indeed considerably larger gas
masses as compared to the values derived from this standard method.

\clearpage
\section{\bff Outlook}
\markboth{OUTLOOK}
         {4.\ \ OUTLOOK}
         
\begin{figure}
  \centering \includegraphics[width=150mm]{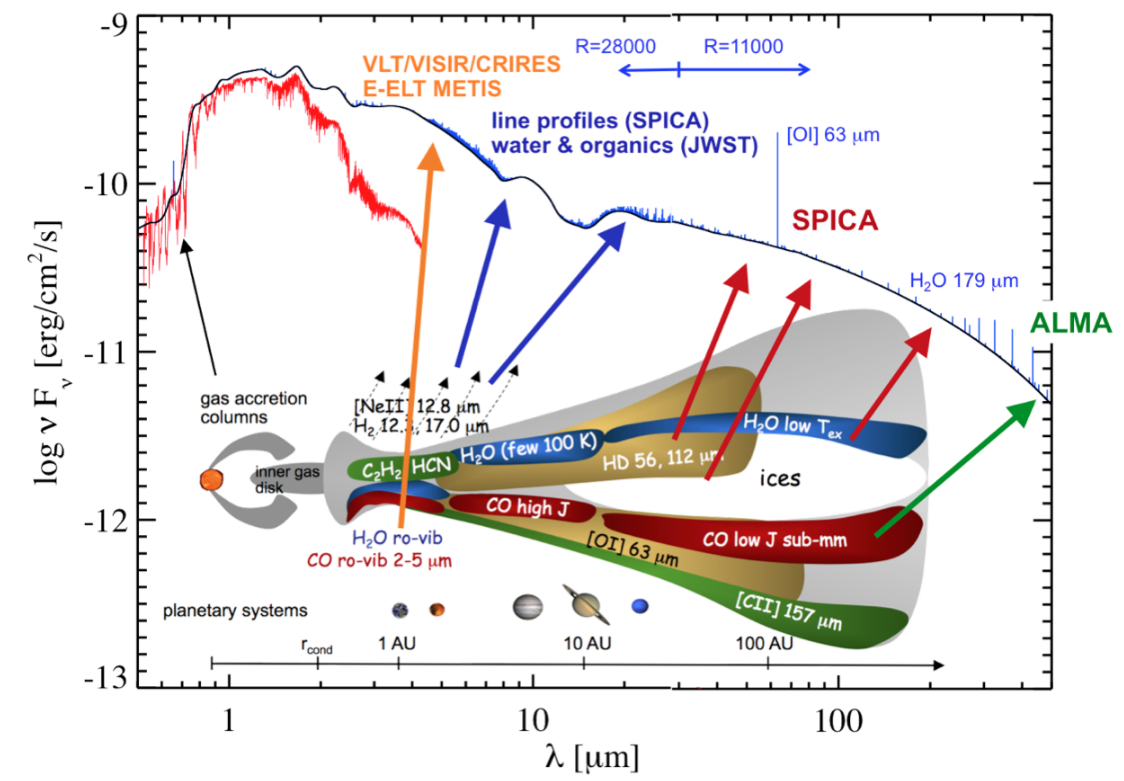}
  \resizebox{\textwidth}{!}{\parbox{18.4cm}{\caption{Multi-wavelength
        observations of protoplanetary discs -- which spectral tracers
        probe which disc locations, disc physics and chemistry?
        Figure re-printed from \citet{Kamp2020}.}
  \label{fig:outlook}}}
\end{figure}

The \ProDiMo\ disc models have been proven to be a very useful tool to
analyse and explain various disc observations, from optical emission
lines probing disc winds and outflows to continuum observations at
different wavelengths, and from mid-IR molecular emission lines
probing the inner disc to millimetre line observations and channel
maps probing the outer disc.  A summary of line emission mechanisms,
line emitting disc regions, and corresponding wavelength domains is
shown in Fig.~\ref{fig:outlook}.  The figure includes current
observational platforms and instruments, such as ALMA and JWST, as well as
future instruments like E-ELT/METIS and SPICA.  Unfortunately for disc
research, the European Space Agency ESA decided in 2020 to no longer
consider the SPace Infrared telescope for Cosmology and Astrophysics
(SPICA) as candidate M5 mission. However, hope is not entirely lost,
as space missions with similar mid-IR and far-IR instrument
capabilities have been proposed, for example LIFE, GREX-PLUS, and
SALTUS.

In the near future, the focus of the research of my group will be on
ALMA and JWST.  {\bff ALMA} observations are the prime choice to probe
the outer disc, the chemistry and ice formation, its time-dependence,
dust evolution, dust dynamics and the gas energy balance.  With the
high-spatial and high frequency resolution of ALMA, we can directly
(model-free) locate the line emitting regions and emission
temperatures of certain molecules from the observations and can
compare these data against the predictions from \ProDiMo\ models.

With {\bff JWST}, the ro-vibrational lines of the molecules in the
mid-infrared spectral region can be studied, emitted from the
upper layers of the inner disc, where these molecules
have temperatures between a few hundred to a few thousand Kelvin.
These studies have just begun, but first results
\citep[e.g.][]{Woitke2023} look very promising, because we can
interpret and understand the main characteristics of these molecular
line emissions (e.g.\ column densities and emission temperatures) in
terms of physically consistent 2D models, in similar ways as
astronomers have learnt to interpret stellar spectra to conclude about
the element composition of stars in the last century.  Among the
current hot JWST topics with regard to protoplanetary
discs are the shape of the inner disc, transport processes and oxygen
depletion, and the dynamic structure of the underlying dust particles.
Models are also required here to connect these measurements to the
chemical composition of the midplane in the planet-forming disc
regions, as these regions cannot be observed directly.

In this current spirit of optimism in the community, however, one
should not forget about the idea of a pan-chromatic
(i.e.\,multi-wavelength) modelling approach.  Certain observations can
only tell us about certain layers of the disc in certain radial
domains.  Only if we succeed in combining the observational data from
different instruments and wavelength regions, a holistic understanding
of planet-forming discs can be achieved.  Are here, ground-based
facilities such as {\bff VLT/CRIRES} and {\bff ELT/METIS} can provide
additional constraints which probe the dynamics of the gas emitting in
the infrared, thereby searching for signatures of planets in the
making.

Further future topics of research include our understanding of disc
and dust evolution over millions of years, where we need
hydrodynamical disc models that can predict the dynamical behaviour of
dust and gas, and the instabilities that can occur, which eventually
trigger the process of dust growth and planet formation. Here,
\ProDiMo\ models can be used to predict the spectral appearance of the
dust and gas in snapshots of these ``patterns of planet formation'',
to help interpreting the complex molecular structures that we are
bound to see quite soon.

Finally, the formation of refractory materials, phyllosilicates and
ices during the early disc evolution is key to understand the current
material composition of primitive bodies in the solar system, which
can be probed directly by solar system exploration missions.
\ProDiMo\ disc models can potentially make an impact here as well, in
combination with the {\sf GGchem} phase equilibrium models
\citep{Woitke2018}. For example, \citep{Steinmeyer2023} presented a
study about the sublimation of refractory minerals during pebble
accretion.  Once our disc chemistry is equipped with more
isotopologues, we will be able to predict the material composition and
isotopic ratios of pebbles, comets, and planetesimals at different
radii -- the building blocks for planet formation, including planet
Earth.

\markboth{ }{ }
\addtocounter{section}{1}
\addcontentsline{toc}{section}{\thesection\bff\ \ \,References}
\begin{small}
\setlength{\bibsep}{0pt plus 0.3ex}
\bibliographystyle{aa_Birnstiel}
\bibliography{references}
\end{small}

\cleardoublepage

\pagestyle{plain}
\addtocounter{section}{1}
\addcontentsline{toc}{section}{\thesection\bff\ \ \,Submitted Publications}
\subsection*{\bff\underline{Publication 1:}}
\addcontentsline{toc}{subsection}{\bff Publication 1}
\label{PUB1}

\vspace*{3mm}
\begin{spacing}{1.35}
  \noindent{\Large\bff Radiation thermo-chemical models of protoplanetary
  disks -- I. Hydrostatic disk structure and inner rim}\\[-3mm]
\end{spacing}

\noindent{\sf {\bff Woitke, P.}; Kamp, I.; Thi, W.-F.}\\[-2mm]

\noindent{\sl Astronomy \& Astrophysics, 2009, Volume 501, Issue 1, 2009, pp.\,383--406}

\begin{spacing}{0.85}\paragraph{Abstract:}{\footnotesize
Emission lines from protoplanetary disks originate mainly in the
irradiated surface layers, where the gas is generally warmer than the
dust. Therefore, interpreting emission lines requires detailed
thermo-chemical models, which are essential to converting line
observations into understanding disk physics.
We aim at hydrostatic disk models that are valid from 0.1\,AU to
1000\,AU to interpret gas emission lines from UV to sub-mm. In
particular, our interest lies in interpreting far IR gas emission
lines, such as will be observed by the Herschel observatory, related
to the GASPS open time key program. This paper introduces a new disk
code called ProDiMo.
We combine frequency-dependent 2D dust continuum radiative transfer,
kinetic gas-phase and UV photo-chemistry, ice formation, and detailed
non-LTE heating \& cooling with the consistent calculation of the
hydrostatic disk structure. We include Fe\,II and CO ro-vibrational
line heating/cooling relevant to the high-density gas close to the
star, and apply a modified escape-probability treatment. The models
are characterised by a high degree of consistency between the various
physical, chemical, and radiative processes, where the mutual
feedbacks are solved iteratively.
In application to a T Tauri disk extending from 0.5\,AU to 500\,AU, the
models show that the dense, shielded and cold midplane ($z/r\la 0.1$,
$\Tg\approx\Td$) is surrounded by a layer of hot ($\Tg\approx
5000\,$K) and thin ($\nH\approx 10^7$ to $10^8\rm\,cm^{-3}$) atomic gas
that extends radially to about 10\,AU and vertically up to $z/r\approx
0.5$.  This layer is predominantly heated by the stellar UV
(e.g.\ PAH-heating) and cools via Fe\,II semi-forbidden and O\,I
630\,nm optical line emission. The dust grains in this ``halo''
scatter the starlight back onto the disk, which affects the
photochemistry. The more distant regions are characterised by a cooler
flaring structure. Beyond $r\ga 100\,$AU, $\Tg$ decouples from $\Td$
even in the midplane and reaches values of about $\Tg\approx 2\,\Td$.
Our models show that the gas energy balance is the key to
understanding the vertical disk structure. Models calculated with the
assumption $\Tg=\Td$ show a much flatter disk structure. The
conditions in the close regions ($<\!10\,$AU) with densities $\nH\approx
10^8$ to $10^{15}\rm\,cm^{-3}$ resemble those of cool stellar
atmospheres and, thus, the heating and cooling is more like in stellar
atmospheres. The application of heating and cooling rates known from
PDR and interstellar cloud research alone can be misleading here, so
more work needs to be invested to identify the leading heating and
cooling processes.}
\end{spacing}
    
\vspace*{-1mm}
\paragraph{Own contribution:} This paper established ProDiMo as a new
modelling code for protoplanetary discs and is my best cited paper
until the present day (317 citations on November 2023).  It required
more than 2 years of programming work from scratch done by myself as a
post-doc at the ATC in Edinburgh, UK. The overall disc modelling
approach was based on Inga Kamp's COSTAR-program, but was considerably
enhanced and generalised by us in the code re-writing process. Some
important non-LTE atomic and molecular, and other physical and
chemical data was contributed by Wing-Fai Thi, including some related
code parts.  All paper text was written by me, and all figures were
created by myself.

\vspace*{-1mm}
\paragraph{Link:}
\url{https://ui.adsabs.harvard.edu/abs/2009A%26A...501..383W}

\clearpage
\subsection*{\bff \underline{Publication 2:}}
\addcontentsline{toc}{subsection}{\bff Publication 2}  
\label{PUB2}

\vspace*{3mm}
\begin{spacing}{1.35}
  \noindent{\Large\bff Continuum and line modelling of discs around young
  stars -- 300000 disc models for HERSCHEL/GASPS}\\[-3mm]
\end{spacing}
    
\noindent{\sf {\bff Woitke, P.}; Pinte, C.; Tilling, I.; M\'enard, F.; Kamp, I.; Thi, W.-F.; Duch\^ene, G.; Augereau, J.-C.}\\[-2mm]

\noindent{\sl Monthly Notices of the Royal Astronomical Society,
  2010, Letters, Vol.\,405, Issue 1, pp.\,L26--L30}

\begin{spacing}{0.85}\paragraph{Abstract:}{\footnotesize
We have combined the thermo-chemical disc code ProDiMo with the Monte
Carlo radiative transfer code MCFOST to calculate a grid of $\sim$300000
circumstellar disc models, systematically varying 11 stellar, disc and
dust parameters including the total disc mass, several disc shape
parameters and the dust-to-gas ratio. For each model, dust continuum
and line radiative transfer calculations are carried out for 29
far-infrared, sub-mm and mm lines of [OI], [CII], $^{12}$CO and o/p-H$_2$O
under five inclinations. The grid allows us to study the influence of
the input parameters on the observables, to make statistical
predictions for different types of circumstellar discs and to find
systematic trends and correlations between the parameters, the
continuum fluxes and the line fluxes. The model grid, comprising the
calculated disc temperature and chemical structures, the computed
spectral energy distributions, line fluxes and profiles, will be used
in particular for the data interpretation of the HERSCHEL open
time-key program GASPS. The calculated line fluxes show a strong
dependence on the assumed ultraviolet excess of the central star and
on the disc flaring. The fraction of models predicting [OI] and [CII]
fine-structure lines fluxes above HERSCHEL/PACS and SPICA/SAFARI
detection limits is calculated as a function of disc mass. The
possibility of deriving the disc gas mass from line observations is
discussed.}
\end{spacing}
    
\paragraph{Own contribution:} This paper marked the starting point of
a long-lasting, fruitful collaboration with disc researchers from the
Laboratoire d'Astrophysique de Grenoble (later called IPAG), in
particular Christophe Pinte and Francois M\'enard.  For the production
of 300000 disc models, it was necessary to use the fast MCFOST Monte
Carlo radiative transfer code, developed by C.~Pinte, to calculate the
dust temperature structure and UV radiation field and to pass these
results to our thermo-chemical disc code.  C.~Pinte modified MCFOST
accordingly, and I modified {\sc ProDiMo} to skip the radiative
transfer parts and use the MCFOST results instead.  A lot of code
testing and double-checking was done by myself.  My former PhD student
I.~Tilling helped to create an idl-tool to visualise the grid results,
but most of that work I actually did myself, too. The other co-authors
mostly had advisory functions.  All paper text was written by myself
and all figures and tables were created by myself. The modelling grid
was later used in subsequent papers, in particular by
Kamp et al.\,(2011). This grid of thermo-chemical models is still
available to the community via
\href{https://prodimo.iwf.oeaw.ac.at/models}{the IWF webpage}.

\vspace*{-1mm}
\paragraph{Link:}
\url{https://ui.adsabs.harvard.edu/abs/2010MNRAS.405L..26W}

\clearpage
\section*{\bff\underline{Publication 3:}}
\addcontentsline{toc}{subsection}{\bff Publication 3}  
\label{PUB3}

\vspace*{1mm}
\begin{spacing}{1.35}
  \noindent{\Large\bff The unusual protoplanetary disk around the
    T Tauri star ET Chamaeleontis}\\[-4mm]
\end{spacing}
    
\noindent{\sf {\bff Woitke, P.}; Riaz, B.; Duch\^ene, G.; Pascucci,
  I.; Lyo, A.-R.; Dent, W.R.F.; Phillips, N.; Thi, W.-F.; M\'enard, F.; Herczeg, G.J.; Bergin, E.; Brown, A.; Mora, A.; Kamp, I.; Aresu, G.; Brittain, S.; de Gregorio-Monsalvo, I.; Sandell}\\[-2mm]

\noindent{\sl Astronomy \& Astrophysics, 2011, Volume 534, id.\ A44, 27, pp.\,1-27}

\vspace*{-1mm}
\begin{spacing}{0.84}
\paragraph{Abstract:}{\footnotesize
We present new continuum and line observations, along with modelling,
of the faint (6-8)\,Myr old T Tauri star ET Cha belonging to the
$\eta$ Chamaeleontis cluster. We have acquired Herschel/PACS
photometric fluxes at 70\,$\mu$m and 160\,$\mu$m, as well as a
detection of the [OI] 63\,$\mu$m fine-structure line in emission, and
derived upper limits for some other far-IR OI, CII, CO and o-H$_2$O
lines. These observations were carried out in the frame of the open
time key programme GASPS, where ET\,Cha was selected as one of the
science demonstration phase targets. The Herschel data is complemented
by new simultaneous ANDICAM B--K photometry, new HST/COS and HST/STIS
UV-observations, a non-detection of CO $J\!=\!3\to2$ with APEX,
re-analysis of a UCLES high-resolution optical spectrum showing
forbidden emission lines like [OI] 6300\,\AA, [SII] 6731\,\AA\ and
6716\,\AA, and [NII] 6583\,\AA, and a compilation of existing
broad-band photometric data. We used the thermo-chemical disk code
ProDiMo and the Monte-Carlo radiative transfer code MCFOST to model
the protoplanetary disk around ET\,Cha. The paper also introduces a
number of physical improvements to the ProDiMo disk modelling code
concerning the treatment of PAH ionisation balance and heating, the
heating by exothermic chemical reactions, and several non-thermal
pumping mechanisms for selected gas emission lines. By applying an
evolutionary strategy to minimise the deviations between model
predictions and observations, we find a variety of united gas and dust
models that simultaneously fit all observed line and continuum fluxes
about equally well. Based on these models we can determine the disk
dust mass with confidence, $M_{\rm
  dust}\approx(2-5)\times10^{-8}\rm\,M_\odot$ whereas the total disk
gas mass is found to be only little constrained, $M_{\rm
  gas}\approx(5\times 10^{-5} - 3\times 10^{-3})\rm\,M_\odot$. Both
mass estimates are substantially lower than previously reported. In
the models, the disk extends from 0.022\,AU (just outside of the
co-rotation radius) to only about 10\,AU, remarkably small for single
stars, whereas larger disks are found to be inconsistent with the CO
$J\!=\!3\to2$ non-detection. The low velocity component of the [OI]
6300\,\AA\ emission line is centred on the stellar systematic
velocity, and is consistent with being emitted from the inner
disk. The model is also consistent with the line flux of $\rm H_2\ 
v\!=\!1\to 0\ S(1)$ at 2.122\,$\mu$m and with the [OI]\,63\,$\mu$m line
as seen with Herschel/PACS. An additional high-velocity component of
the [OI] 6300\,\AA\ emission line, however, points to the existence of
an additional jet/outflow of low velocity 40-65 km/s with mass loss
rate $\approx 10^{-9}\rm\,M_\odot/yr$. In relation to our low
estimations of the disk mass, such a mass loss rate suggests a disk
lifetime of only $\sim 0.05-3\rm\,Myr$, substantially shorter than the cluster
age. If a generic gas/dust ratio of 100 was assumed, the disk lifetime
would be even shorter, only $\sim$3000 yrs. The evolutionary state of this
unusual protoplanetary disk is discussed.}
\end{spacing}

\vspace*{-1mm}
\paragraph{Own contribution:} I was leading this paper on one of the
GASPS targets.  The long author list is because (a) the GASPS data
were reduced by a number of team members, (b) Riaz, Duch\^ene and
Pascucci provided auxiliary observational data and (c) because of the
GASPS consortium publication rules (PI Dent).  Except for the sections
detailling these observations and data reductions (sections 3.1, 3.2,
3.3, 3.5, 3.6) I wrote all the paper text and created all figures. I
did most of the basic code development on ProDiMo, with some
contributions from Wing-Fai Thi.  The paper enabled us to successfully
propose cycle~0 ALMA observational time, which confirmed that the disk
of ET Chamaeleontis has indeed an outer radius of only about 5\,AU
\citep{Woitke2013}.

\vspace*{-1mm}
\paragraph{Link:}
\url{https://ui.adsabs.harvard.edu/abs/2011A%26A...534A..44W}

\clearpage
\section*{\bff\underline{Publication 4:}}
\addcontentsline{toc}{subsection}{\bff Publication 4}  
\label{PUB4}

\vspace*{3mm}
\begin{spacing}{1.35}
  \noindent{\Large\bff Consistent dust and gas models for
    protoplanetary disks. I. Disk shape, dust settling,
    opacities, and PAHs}\\[-4mm]
\end{spacing}
  
\noindent{\sf {\bff Woitke, P.};  Min, M.; Pinte, C.; Thi, W.-F.; Kamp, I.; Rab, C.; Anthonioz, F.; Antonellini, S.; Baldovin-Saavedra, C.; Carmona, A.; Dominik, C.; Dionatos, O.; Greaves, J.; G\"udel, M.; Ilee, J.D.; Liebhart, A.; M\'enard, F.; Rigon, L.; Waters, L.B.F.M.; Aresu, G.; Meijerink, R.; Spaans, M.}\\[-2mm]

\noindent{\sl Astronomy \& Astrophysics, 2016, Volume 586, id.\,A103, pp.\,1--35}

\begin{spacing}{0.85}\paragraph{Abstract:}{\footnotesize
We propose a set of standard assumptions for the modelling of Class II
and III protoplanetary disks, which includes detailed continuum
radiative transfer, thermo-chemical modelling of gas and ice, and line
radiative transfer from optical to cm wavelengths. The first paper of
this series focuses on the assumptions about the shape of the disk,
the dust opacities, dust settling, and polycyclic aromatic
hydrocarbons (PAHs). In particular, we propose new standard dust
opacities for disk models, we present a simplified treatment of PAHs
in radiative equilibrium which is sufficient to reproduce the PAH
emission features, and we suggest using a simple yet physically
justified treatment of dust settling. We roughly adjust parameters to
obtain a model that predicts continuum and line observations that
resemble typical multi-wavelength continuum and line observations of
Class II T Tauri stars. We systematically study the impact of each
model parameter (disk mass, disk extension and shape, dust settling,
dust size and opacity, gas/dust ratio, etc.) on all mainstream
continuum and line observables, in particular on the SED, mm-slope,
continuum visibilities, and emission lines including [OI]\,63\,$\mu$m,
high-J CO lines, (sub-)mm CO isotopologue lines, and CO fundamental
ro-vibrational lines. We find that evolved dust properties, I.e. large
grains, often needed to fit the SED, have important consequences for
disk chemistry and heating/cooling balance, leading to stronger near-
to far-IR emission lines in general. Strong dust settling and missing
disk flaring have similar effects on continuum observations, but
opposite effects on far-IR gas emission lines. PAH molecules can
efficiently shield the gas from stellar UV radiation because of their
strong absorption and negligible scattering opacities in comparison to
evolved dust. The observable millimetre-slope of the SED can become
significantly more gentle in the case of cold disk midplanes, which we
find regularly in our T Tauri models. We propose to use line
observations of robust chemical tracers of the gas, such as O, CO, and
H$_2$, as additional constraints to determine a number of key properties
of the disks, such as disk shape and mass, opacities, and the dust/gas
ratio, by simultaneously fitting continuum and line observations.}
\end{spacing}
    
\vspace*{-1mm}
\paragraph{Own contribution:} This paper was the first common
publication that emerged from my FP7-SPACE project DIANA (PI
Woitke). It has set new standards for the physical and chemical modelling
of protoplanetary discs including stellar irradiation, dust opacities,
PAHs and dust settling. All consortium members were involved in the
definition of these standards, yet the numerical implementation of
these equations was mostly done by myself concerning ProDiMo, by Christophe
Pinte concerning MCFOST, and by Michiel Min concerning MCMax.  I did a
lot of cross-checking between the three modelling codes. The full
consortium appears in the author list, yet I have written all paper
text, and all figures were created by myself, with the exception of
figure B.1 which was provided by Christophe Pinte, and section A.3
which was provided by Armin Liebhart and Manuel G\"udel.

\vspace*{-1mm}
\paragraph{Link:}
\url{https://ui.adsabs.harvard.edu/abs/2016A%26A...586A.103W}

\clearpage
\section*{\bff\underline{Publication 5:}}
\addcontentsline{toc}{subsection}{\bff Publication 5}  
\label{PUB5}

\vspace*{1mm}
\begin{spacing}{1.35}
  \noindent{\Large\bff Modelling mid-infrared molecular emission lines from T~Tauri stars}\\[-3mm]
\end{spacing}
    
\noindent{\sf {\bff Woitke, P.}; Min, M.; Thi, W.-F.; Roberts, C.; Carmona, A.; Kamp, I.; M\'enard, F.; Pinte, C.}\\[-2mm]

\noindent{\sl Astronomy \& Astrophysics, 2018, Volume 618, id.\,A57, pp.\,1--18}

\vspace*{-1mm}
\begin{spacing}{0.84}
\paragraph{Abstract:}{\footnotesize
We introduce a new modelling framework including the Fast Line Tracer
(FLiTs) to simulate infrared line emission spectra from protoplanetary
discs. This paper focusses on the mid-IR spectral region between 9.7
and 40\,$\mu$m for T\,Tauri stars. The generated spectra contain
several tens of thousands of molecular emission lines of H$_2$O, OH,
CO, CO$_2$, HCN, C$_2$H$_2$, H$_2$, and a few other molecules, as well
as the forbidden atomic emission lines of S\,I, S\,II, S\,III, Si\,II,
Fe\,II, Ne\,II, Ne\,III, Ar\,II, and Ar\,III. In contrast to
previously published works, we do not treat the abundances of the
molecules nor the temperature in the disc as free parameters, but use
the complex results of detailed 2D ProDiMo disc models concerning gas
and dust temperature structure, and molecular concentrations. FLiTs
computes the line emission spectra by ray tracing in an efficient,
fast, and reliable way. The results are broadly consistent with
$R\!=\!600$ Spitzer/IRS observational data of T\,Tauri stars
concerning line strengths, colour, and line ratios. In order to
achieve that agreement, however, we need to assume either a high
gas/dust mass ratio of order 1000, or the presence of illuminated disc
walls at distances of a few au, for example, due to disc-planet
interactions. These walls are irradiated and heated by the star which
causes the molecules to emit strongly in the mid-IR. The molecules in
the walls cannot be photodissociated easily by UV because of the large
densities in the walls favouring their re-formation. Most observable
molecular emission lines are found to be optically thick. An abundance
analysis is hence not straightforward, and the results of simple slab
models concerning molecular column densities can be misleading. We
find that the difference between gas and dust temperatures in the disc
surface is important for the line formation. The mid-IR emission
features of different molecules probe the gas temperature at different
depths in the disc, along the following sequence: OH (highest) -- CO
-- H$_2$O -- CO$_2$ -- HCN -- C$_2$H$_2$ (deepest), just where these
molecules start to become abundant. We briefly discuss the effects of
C/O ratio and choice of chemical rate network on these results. Our
analysis offers new ways to infer the chemical and temperature
structure of T Tauri discs from future James Webb Space Telescope
(JWST)/MIRI observations, and to possibly detect secondary illuminated
disc walls based on their specific mid-IR molecular signature.}
\end{spacing}

\vspace*{-1mm}
\paragraph{Own contribution:} The paper presents the results of a part
project in the FP7-SPACE project DIANA (PI Woitke), in which Michiel
Min developed the Fast Line Tracer (FLiTs). FLiTs uses the molecular
abundances, dust opacity, temperatures, and non-LTE populations from
ProDiMo, and I coded the interface from ProDiMo to FLiTs. I also
performed all combined ProDiMo $\to$ FLiTs runs, created all figures
and tables, and wrote the main text of the paper and scientific
discussions, except for sections 4.1 and 4.2, which were provided by
Michiel Min. Wing-Fai Thi provided some of the atomic and molecular
data used, and my former master student Clayton Roberts helped with
post-processing the Spitzer spectroscopic data and identification of
molecular features in figure 4.  The other co-authors had advisory
functions.

\vspace*{-1mm}
\paragraph{Link:}
\url{https://ui.adsabs.harvard.edu/abs/2018A%26A...618A..57W}

\clearpage
\section*{\bff \underline{Publication 6:}}
\addcontentsline{toc}{subsection}{\bff Publication 6}  
\label{PUB6}

\vspace*{1mm}
\begin{spacing}{1.35}
\noindent{\Large\bff Consistent dust and gas models for protoplanetary
  disks. III. Models for Selected Objects from the FP7 DIANA Project}\\[-3mm]
\end{spacing}

\noindent{\sf {\bff Woitke, P.};  Kamp, I.; Antonellini, S.; Anthonioz, F.; Baldovin-Saveedra, C.; Carmona, A.; Dionatos, O.; Dominik, C.; Greaves, J.; G\"udel, M.; Ilee, J.D.; Liebhardt, A.; Menard, F.; Min, M.; Pinte, C.; Rab, C.; Rigon, L.; Thi, W.-F.; Thureau, N.; Waters, L.B.F.M.}\\[-2mm]

\noindent{\sl Publications of the Astronomical Society of the Pacific,
  2019, Volume 131, pp.\,064301, 71 pages.}

\vspace*{-1mm}
\begin{spacing}{0.85}\paragraph{Abstract:}{\footnotesize
The European FP7 project DIANA has performed a coherent analysis of a
large set of observational data of protoplanetary disks by means of
thermo-chemical disk models. The collected data include
extinction-corrected stellar UV and X-ray input spectra (as seen by
the disk), photometric fluxes, low and high resolution spectra,
interferometric data, emission line fluxes, line velocity profiles and
line maps, which probe the dust, polycyclic aromatic hydrocarbons
(PAHs) and the gas in these objects. We define and apply a
standardized modeling procedure to fit these data by state- of-the-art
modeling codes (ProDiMo, MCFOST, MCMax), solving continuum and line
radiative transfer (RT), disk chemistry, and the heating and cooling
balance for both the gas and the dust. 3D diagnostic RT tools (e.g.,
FLiTs) are eventually used to predict all available observations from
the same disk model, the DIANA-standard model. Our aim is to determine
the physical parameters of the disks, such as total gas and dust
masses, the dust properties, the disk shape, and the chemical
structure in these disks. We allow for up to two radial disk zones to
obtain our best-fitting models that have about 20 free
parameters. This approach is novel and unique in its completeness and
level of consistency. It allows us to break some of the degeneracies
arising from pure Spectral Energy Distribution (SED) modeling. In this
paper, we present the results from pure SED fitting for 27 objects and
from the all inclusive DIANA-standard models for 14 objects. Our
analysis shows a number of Herbig Ae and T Tauri stars with very cold
and massive outer disks which are situated at least partly in the
shadow of a tall and gas-rich inner disk. The disk masses derived are
often in excess to previously published values, since these disks are
partially optically thick even at millimeter wavelength and so cold
that they emit less than in the Rayleigh-Jeans limit. We fit most
infrared to millimeter emission line fluxes within a factor better
than 3, simultaneously with SED, PAH features and radial brightness
profiles extracted from images at various wavelengths. However, some
line fluxes may deviate by a larger factor, and sometimes we find
puzzling data which the models cannot reproduce. Some of these issues
are probably caused by foreground cloud absorption or object
variability. Our data collection, the fitted physical disk parameters
as well as the full model output are available to the community
through an \href{http://www.univie.ac.at/diana}{online database.}}
\end{spacing}

\vspace*{-1mm}
\paragraph{Own contribution:} This paper summarised the modelling
results of the FP7-SPACE project DIANA (PI Woitke). All authors
contributed to either the multi-wavelength data collection or the
various modelling efforts, but as the work package leader for the disk
modelling, it was my task to bring the various modelling results from
the different co-authors to common standards, re-calculating all final
models to create homogeneous figures.  I have writen most of the paper
text, and edited some text passages that were provided by my team
members.

\vspace*{-1mm}
\paragraph{Link:}
\url{https://ui.adsabs.harvard.edu/abs/2019PASP..131f4301W}

\clearpage
\section*{\bff\underline{Publication 7:}}
\addcontentsline{toc}{subsection}{\bff Publication 7}  
\label{PUB7}

\vspace*{1mm}
\begin{spacing}{1.35}
  \noindent{\Large\bff Consistent dust and gas models for
    protoplanetary disks. IV. A panchromatic view of protoplanetary
    disks}\\[-3mm]
\end{spacing}
    
\noindent{\sf Dionatos, O.; {\bff Woitke, P.}; G\"udel, M.; Degroote,
  P.; Liebhart, A.; Anthonioz, F.; Antonellini, S.; Baldovin-Saavedra,
  C.; Carmona, A.; Dominik, C.; Greaves, J.; Ilee, J. D.; Kamp, I.;
  M\'enard, F.; Min, M.; Pinte, C.; Rab, C.; Rigon, L.; Thi, W.-F.;
  Waters, L.B.F.M.}\\[-3mm]

\noindent{\sl Astronomy \& Astrophysics, 2019, Volume 625, id.\,A66, pp.\,1--32}

\vspace*{-1mm}
\begin{spacing}{0.84}
\paragraph{Abstract:}{\footnotesize
Consistent modeling of protoplanetary disks requires the simultaneous
solution of both continuum and line radiative transfer, heating and
cooling balance between dust and gas and, of course, chemistry. Such
models depend on panchromatic observations that can provide a complete
description of the physical and chemical properties and energy balance
of protoplanetary systems. Along these lines, we present a
homogeneous, panchromatic collection of data on a sample of 85 T Tauri
and Herbig Ae objects for which data cover a range from X-rays to
centimeter wavelengths. Datasets consist of photometric measurements,
spectra, along with results from the data analysis such as line fluxes
from atomic and molecular transitions. Additional properties resulting
from modeling of the sources such as disk mass and shape parameters,
dust size, and polycyclic aromatic hydrocarbon (PAH) properties are
also provided for completeness. The purpose of this data collection is
to provide a solid base that can enable consistent modeling of the
properties of protoplanetary disks. To this end, we performed an
unbiased collection of publicly available data that were combined to
homogeneous datasets adopting consistent criteria. Targets were
selected based on both their properties and the availability of data.
Data from more than 50 different telescopes and facilities were
retrieved and combined in homogeneous datasets directly from public
data archives or after being extracted from more than 100 published
articles. X-ray data for a subset of 56 sources represent an exception
as they were reduced from scratch and are presented here for the first
time.  Compiled datasets, along with a subset of continuum and
emission-line models are stored in a dedicated database and
distributed through a publicly accessible online system. All datasets
contain metadata descriptors that allow us to track them back to their
original resources. The graphical user interface of the online system
allows the user to visually inspect individual objects but also
compare between datasets and models. It also offers to the user the
possibility to download any of the stored data and metadata for
further processing.}
\end{spacing}

\vspace*{-1mm}
\paragraph{Own contribution:} This paper published the data collection
done for the FP7-SPACE project DIANA (PI Woitke). Many team members
were responsible for various types of observational data, and Odysseas
Dionatos was leading the corresponding work package. I provided the
archival Spitzer/IRS, the Herschel/SPIRE and the UV data collections,
I created and hosted the computer account were team members could
initially upload their data, to oversee the data collection process. I
created figures 1, 2, 3, C.1, C.2, and wrote the appendix B. Some of
the tables, for example tables 4 and C.1, resulted from spreadsheets I
updated regularly. I would estimate my contribution to this paper to
be about 35\%.

\vspace*{-1mm}
\paragraph{Link:}
\url{https://ui.adsabs.harvard.edu/abs/2019A%26A...625A..66D}

\cleardoublepage

\pagestyle{headings}
\markboth{ }{ }
\section{\bff International Research Activities}

\begin{enumerate}
\item European Marie-Curie Training network H2020-MSCA-ITN-2019
  project 860470 entitled ``Virtual Laboratories for Exoplanets and
  Planet-Forming Disks'' (or short {\sc Chameleon}), funded with a
  total European contribution of 4.100.000 from 06/2020 to 05/2024,
  see \url{https://chameleon.iwf.oeaw.ac.at/}. {\sc Chameleon}
  combines 6 European universities (Graz, Antwerp, Copenhagen,
  Edinburgh, Groningen, Leuven) and 5 non-university institutions,
  which includes the LUCA School of Arts (Antwerp), the Netherlands
  Institute for Space Research (SRON), MPIA Heidelberg, Copenhagen
  Games Lab, and the Scottish Parliament Information Centre. This
  MC ITN EJD network involves 15 staff members and educates
  15 PhD students. I am leading the work package~1
  entitled ``Modelling planet-forming disks''.
  
\item European FP7-SPACE collaboration project 284405 ``DiscAnalysis''
  (or short, {\sl DIANA}), funded with a total European contribution
  of \Euro 2.000.000 from 01/2012 to 03/2016, see
  \url{https://diana.iwf.oeaw.ac.at/}. I was the PI and scientific
  coordinator. Besides my former host University of St Andrews, there
  were four other European institutions involved: Astronomical
  Institute Anton Pannekoek (Amsterdam, NL), Universit{\'e} Joseph
  Fourier (UJF Grenoble, France), Kapteyn Astronomical Institute
  (Groningen, NL) and Universit{\"a}t Wien (Vienna, Austria). Tasks
  included the management of the consortium, organisation of team
  meetings, reporting, recruitment and sub-contracting, funds for
  outreach and conference organisation.

\item FAPESP/BEPE: In November 2016, a joint application
  with my Brazilian collaborator Prof.~S.~Pilling to the S\~ao Paulo
  Research Foundation (FAPESP) for a Research Internship entitled
  ``Formation of complex molecules on astrophysical ices in star
  forming regions triggered by UV and X-rays by employing the ProDiMo
  code and laboratory data'' was successful. It had a
  financial volume of about \textsterling 50.000. Dr.~Will Rocha worked as a
  post-doc at St Andrews University under my supervision for 12 months
  in 2017, fully funded by Brazil.

\item St Andrews astronomy group-application for STFC consolidation
  grant 04/2012 (Co-PI), about  \textsterling 1.400.000 total, and
  in 04/2015 (Co-PI), about \textsterling 1.600.000 total, 3 PDRA
  positions.

\item FWF National Research Program ``Pathways to Habitability'' (PI
  Manuel G{\"u}del) for period 03/2012 to 02/2020, the largest
  Austrian key project in Astronomy with a total of about \Euro
  7.000.000, organised in 7 sub-projects (SPs), see
  \url{http://path.univie.ac.at}. My contribution was to co-edit the
  application for SP\,2 ``Disks''.

\item Nederlandse Onderzoekschool Voor Astronomie (NOVA)-2 project
  {\em Pulsations, Winds and Dust Formation in AGB Stars},
  PhD position for 2.8 years (2005 -- 2008), in cooperation with
  A.~Quirrenbach (Leiden Observatory), volume \Euro 134.400.

\item Netherlands National Computing Facilities Foundation (NCF)
  project June~2004 -- May~2005 (renewal August~2005 -- July~2006),
  {\em AstroHydro3D}\ \ group application for supercomputing time on {\sc
    Teras/Aster/Lisa} parallel machines, 200.000 CPU hours per year.

\end{enumerate}

\section{\bff List of Teaching Courses}

\paragraph{Selected Topics in Space Physics and Aeronomy} {\sl (Summer
  2022)}:\ \ I contributed two lectures to the course PHM.024UB led by
Philippe-Andr{\'e} in the Joined Physics Teaching Programme of the TU
Graz and the University of Graz for advanced students. My lectures
were entitled {\sl ``The chemical diversity of planetary atmospheres''} and
{\sl ``Chemical Rates and Networks''}.

\paragraph{Physics 1A} {\sl (Autumn 2016 -- 2020)} at the
  University of St Andrews. I have coordinated this most important
  physics module for first-semester students of St Andrews University
  since 2016, and have lectured the mechanics part (about 1/3 of all
  lectures). 30 credits. The module was run with an increasing number
  of students (from 146 students in 2016 to 181 students in 2020)
  including about 35--50 students from other schools and about 10--15
  Gateway students.  I have coordinated the work of 3 lecturers,
  11--16 tutors, and 5 lab demonstrators. We have been running 4
  lectures and 1 workshop each week, as well as 20 tutorial groups to
  keep the size of the tutorial groups at an optimum of about 6--8
  students.
  classtest and exam correction work and set the final grades which
  have then been discussed with the external examiners. I have
  re-structured the Mechanics course, have invented a number of new
  procedures, and have sailed the PH1011-ship through the difficult
  COVID times.
  
\paragraph{Computational Astrophysics} {\sl (Spring 2012 --
    2021)} at the University of St Andrews. 3$^{\rm rd}$-yr students
  get hands-on experience in developing their own FORTRAN-90 programs
  under LINUX, study and test numerical algorithms, plot their results
  with PYTHON, and develop skills for systematic work and scientific
  report writing. 15 credits. Numerical methods and astrophysical
  topics include numerical integration (initial mass function),
  solving systems of coupled ordinary differential equations (stellar
  orbits) and Bayesian analysis (planet transits). 15 credits, about
  15\,--\,35 students, initial 6 lectures, 2x3.5 hours lab contact
  time during 13 weeks, weekly, full continuous assessment (4 assessed
  exercises). I was the module coordinator, the course was partly
  shared with other lecturers.

\paragraph{Stars and Nebulae II} {\sl (Spring 2015 -- 2021)} at
  the University of St Andrews about stellar interiors and
  atmospheres, radiative transfer, interaction of radiation with
  matter, line formation and emergent spectrum. 15 credits, 14
  lectures, about 15 4$^{\rm th}$-yr students.  Computer-aided
  tutorial exercises, assessed homework, written exams. The course is
  mostly shared with Prof.\ Andrew Cameron, but in 2020 I taught all
  parts with 28 lectures.

\paragraph{Summer School} {\sl (June 2014)}:\ \ I have co-organised a
  summer school entitled {\sl ``Protoplanetary discs: theory and
    modelling meet observations''}, Ameland, NL, where 57 participants
  (mostly PhD-students) learnt about the basic theories of
  protoplanetary discs, their formation and evolution, chemistry,
  radiative transfer, and the diversity of observational data such as
  SEDs, images, line emission and interferometry. I contributed two
  lectures about SED-fitting and gas heating \& cooling. The summer
  school aimed at introducing this exciting field to the next
  generation of scientists. From these lectures, we compiled a book
  published in 2015
  (\url{https://diana.iwf.oeaw.ac.at/summer-school}).  I have
  co-edited this volume.


\paragraph{Astrophysics I} {\sl (Winter 2011/12)}:\ \ compulsory lecture for
  4$^{\rm th}$-yr astronomy students at Vienna university on theoretical
  astrophysics (hydrostatics, stars, planets, white dwarfs,
  Bonnor-Ebert spheres, stellar structure, fragmentation, stability,
  virial theorem, shock waves, radiative transfer, radiative
  processes, MHD, waves, dissipation), 2x2 hours weekly, about 40
  students, assessed tutorials, oral examinations.

\paragraph{Astrophysics II} {\sl (Spring 2011)}:\ \ compulsory lecture for 
  4$^{\rm th}$-yr astronomy students at Vienna university on theoretical
  astrophysics (interstellar medium, heating \& cooling, shock waves,
  ionisation fronts, HII regions, stellar winds, supernovae, galactic
  astrophysics, gravitation, motion in mean gravitational potentials,
  spiral structure), 2x2 hours weekly, about 40 students, assessed
  tutorials, oral examinations.

\section{\bff List of Publications}

\begin{tabular}{lcll}
1993 -- 2023 & 200 (120 refereed) & $>5500$ citations & $h=45$\\
2013 -- 2023 & 101 (68 refereed)  & $>2200$ citations & $h=27$\\
\end{tabular}\\[1mm]
\hspace*{1mm} citation metrics from ADS
\url{https://ui.adsabs.harvard.edu}, October 2023:\\[3mm]
\hspace*{0mm}\begin{tabular}{cc}
{\bff citations per year} & {\bff citation metrics}\\[2mm]
\includegraphics[width=7.6cm,height=5.5cm,trim=3 1 1 4,clip]{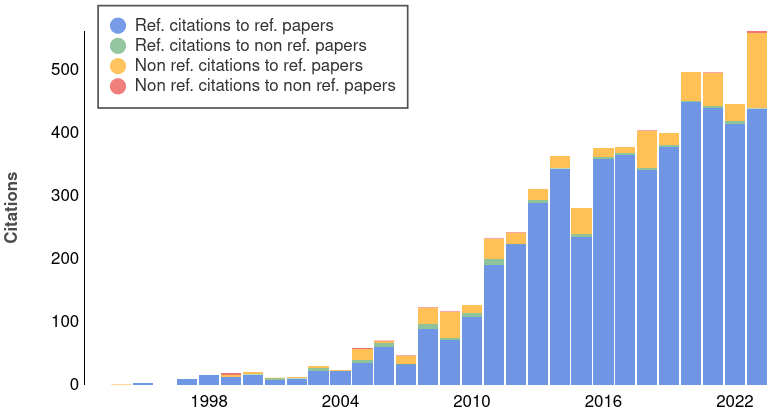}
&
\includegraphics[width=7.6cm,height=5.5cm,trim=3 1 1 4,clip]{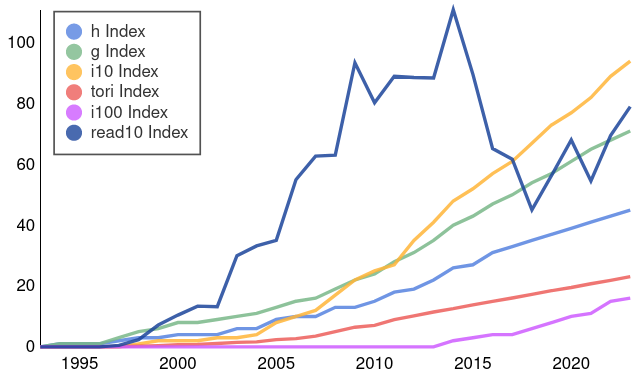}
\end{tabular}\\[0mm]
\begin{itemize*}
\item My refereed publications since 2013 have attracted a total of
  2237 citations (about 32 times per paper) and have been read
  about 41000 times (about 600 times per paper).
\item Five of my ten best-cited papers are first-author papers.
\end{itemize*}
\medskip

\noindent In the following, I have listed these publications
separately in categories {\it Books}, {\it Refereed Articles}, and
{\sl Conference Proceedings} (only those listed by ADS are
included).

\subsection*{Books}

\begin{enumerate}

\item{\sc {Kamp}, I., {\bf {Woitke}, P.}, {Ilee}, J.} (2015, September).
\newblock {Summer School ``Protoplanetary Disks: Theory and Modelling
  Meet Observations''}, {\em Preface}.
\newblock In {European Physical Journal Web of Conferences}, Band
102, pp.~1. 

\item{\sc {\bf {Woitke}, P.}} (2015b, September).  \newblock {Summer School
  ``Protoplanetary Disks: Theory and Modelling Meet Observations''},
  {\em Modelling and interpretation of SEDs}.  \newblock In {European
  Physical Journal Web of Conferences}, Band 102, chapter~7.

\item{\sc {\bf {Woitke}, P.}} (2015a, September).  \newblock {Summer School
  ``Protoplanetary Disks: Theory and Modelling Meet Observations'',}
  {\em Heating and cooling processes in disks}.  \newblock In {European
  Physical Journal Web of Conferences}, Band 102, chapter~11.

\item{\sc {Simis}, Y., {\bf {Woitke}, P.}} (2003).  \newblock {Dynamics and
  Instabilities in Dusty Winds}.  \newblock In H.~J. {Habing} \&
  H.~{Olofsson} (eds.), {\em Asymptotic giant branch stars}, 
    Astronomy and astrophysics library,
    New York, Berlin: Springer, 2003, pp.\ 291.

\item{\sc {\bf {Woitke}, P.}} (1997).  \newblock {\em Heating and Cooling in
  Circumstellar Envelopes}.  \newblock Dissertation, Thesis,
  Tech.~Univ.~Berlin, (1997).

\end{enumerate}

\subsection*{Refereed Articles}

\begin{enumerate}

\item{\sc {Marigo}, {\bf P., {Woitke}, P.}, {Tognelli}, E., {Girardi}, L.,
  {Aringer}, B., {Bressan}, A.} (2023, October).  \newblock {AESOPUS
  2.0: Low-Temperature Opacities with Solid Grains}.  \newblock {\em
  arXiv e-prints\/}, arXiv:2310.14588.

\item{\sc {Janssen}, L.~J., {\bf {Woitke}, P.}, {Herbort}, O., {Min}, M.,
  {Chubb}, K.~L., {helling}, C., {Carone}, L.} (2023, October).
  \newblock {The sulphur species in hot rocky exoplanet atmospheres}.
  \newblock {\em arXiv e-prints\/}, arXiv:2310.07701.

\item{\sc {Kanwar}, J., {Kamp}, I., {\bf {Woitke}, P.}, {Rab}, C., {Thi},
  W.-F., {Min}, M.} (2023, October).  \newblock {Hydrocarbon chemistry
  in inner regions of planet forming disks}.  \newblock {\em arXiv
  e-prints\/}, arXiv:2310.04505.

\item{\sc {Steinmeyer}, M.-L., {\bf {Woitke}, P.}, {Johansen}, A.} (2023,
  September).  \newblock {Sublimation of refractory minerals in the
    gas envelopes of accreting rocky planets}.  \newblock {\em
    \aap\/}~{\bf 677}, A181.

\item{\sc {Balduin}, T., {\bf {Woitke}, P.}, {Thi}, W.-F., {Gr{\r{a}}e
    J{\o}rgensen}, U., {Narita}, Y.} (2023, August).  \newblock
  {Size-dependent charging of dust particles in protoplanetary disks
    Can turbulence cause charge separation and lightning?}  \newblock
  {\em arXiv e-prints\/}, arXiv:2308.04335.

\item{\sc {Rocha}, W.~R.~M., {\bf {Woitke}, P.}, {Pilling}, S., {Thi},
  W.~F., {J{\o}rgensen}, J.~K., {Kristensen}, L.~E., {Perotti}, G.,
  {Kamp}, I.} (2023, May).  \newblock {Simulation of CH$_{3}$OH ice UV
  photolysis under laboratory conditions}.  \newblock {\em
  \aap\/}~{\bf 673}, A70.

\item{\sc {Kaeufer}, T., {\bf {Woitke}, P.}, {Min}, M., {Kamp}, I., {Pinte},
  C.} (2023, April).  \newblock {Analysing the SEDs of protoplanetary
  disks with machine learning}.  \newblock {\em \aap\/}~{\bf 672},
  A30.

\item{\sc {Helling}, C., {Samra}, D., {Lewis}, D., {Calder}, R.,
  {Hirst}, G., {\bf {Woitke}, P.}, {Baeyens}, R., {Carone}, L., {Herbort},
  O., {Chubb}, K.~L.}  (2023, March).  \newblock {Exoplanet weather
  and climate regimes with clouds and thermal ionospheres. A model
  grid study in support of large-scale observational campaigns}.
  \newblock {\em \aap\/}~{\bf 671}, A122.

\item{\sc {Rensen}, F., {Miguel}, Y., {Zilinskas}, M., {Louca}, A.,
  {\bf {Woitke}, P.}, {Helling}, C., {Herbort}, O.} (2023, February).
  \newblock {The Deep Atmospheric Composition of Jupiter from
    Thermochemical Calculations Based on Galileo and Juno Data}.
  \newblock {\em Remote Sensing\/}~{\bf 15\/}(3), 841.

\item{\sc {Erkaev}, N.~V., {Scherf}, M., {Herbort}, O., {Lammer}, H.,
  {Odert}, P., {Kubyshkina}, D., {Leitzinger}, M., {\bf {Woitke}, P.},
  {O'Neill}, C.} (2023, January).  \newblock {Modification of the
  radioactive heat budget of Earth-like exoplanets by the loss of
  primordial atmospheres}.  \newblock {\em \mnras\/}~{\bf 518\/}(3),
  3703--3721.

\item{\sc {\bf {Woitke}, P.}, {Arabhavi}, A.~M., {Kamp}, I., {Thi},
  W.~F.}  (2022, December).  \newblock {Mixing and diffusion in
  protoplanetary disc chemistry}.  \newblock {\em \aap\/}~{\bf 668},
  A164.

\item{\sc {Rab}, C., {Weber}, M., {Grassi}, T., {Ercolano}, B.,
  {Picogna}, G., {Caselli}, P., {Thi}, W.~F., {Kamp}, I., {\bf {Woitke},
  P.}} (2022, December).  \newblock {Interpreting molecular hydrogen
  and atomic oxygen line emission of T Tauri disks with
  photoevaporative disk-wind models}.  \newblock {\em \aap\/}~{\bf
  668}, A154.

\item{\sc {Oberg}, N., {Kamp}, I., {Cazaux}, S., {\bf {Woitke}, P.},
  {Thi}, W.~F.} (2022, November).  \newblock {Circumplanetary disk
  ices. I. Ice formation vs. viscous evolution and grain drift}.
  \newblock {\em \aap\/}~{\bf 667}, A95.

\item{\sc {Arabhavi}, A.~M., {\bf {Woitke}, P.}, {Cazaux}, S.~M., {Kamp},
  I., {Rab}, C., {Thi}, W.-F.} (2022, October).  \newblock {Ices in
  planet-forming disks: Self-consistent ice opacities in disk models}.
  \newblock {\em \aap\/}~{\bf 666}, A139.

\item{\sc {Herbort}, O., {\bf {Woitke}, P.}, {Helling}, C., {Zerkle},
  A.~L.} (2022, February).  \newblock {The atmospheres of rocky
  exoplanets. II. Influence of surface composition on the diversity of
  cloud condensates}.  \newblock {\em A\&A}~{\bf 658}, A180.
  
\item{\sc {Khorshid}, N., {Min}, M., {D{\'e}sert}, J.~M., {\bf
    {Woitke}, P.}, {Dominik}, C.} (2022, November).  \newblock {SimAb:
  A simple, fast, and flexible model to assess the effects of planet
  formation on the atmospheric composition of gas giants}.  \newblock
  {\em \aap\/}~{\bf 667}, A147.

\item{\sc {Rimmer}, P.~B., {Jordan}, S., {Constantinou}, T., {\bf
    {Woitke}, P.}, {Shorttle}, O., {Hobbs}, R., {Paschodimas}, A.}
  (2021, August).  \newblock {Hydroxide Salts in the Clouds of Venus:
  Their Effect on the Sulfur Cycle and Cloud Droplet pH}.  \newblock
  {\em \psj\/}~{\bf 2\/}(4), 133.

\item{\sc {\bf {Woitke}, P.} and {Herbort}, O. and {Helling}, Ch. and
  {St{\"u}eken}, E. and {Dominik}, M. and {Barth}, P. and {Samra},
  D.}, (2021, February).  \newblock {Coexistence of CH$_{4}$,
  CO$_{2}$, and H$_{2}$O in exoplanet atmospheres},
  \newblock {\em A\&A} {\bf 646}, 43.

\item{\sc {Rab}, Ch. and {Kamp}, I. and {Dominik}, C. and {Ginski},
  C. and {Muro-Arena}, G.~A. and {Thi}, W. -F. and {Waters},
  L.~B.~F.~M. and {\bf {Woitke}, P.}} (2020, October).  \newblock
  {Interpreting high spatial resolution line observations of
    planet-forming disks with gaps and rings: the case of HD 163296},
  \newblock{\em A\&A} {\bf 642}, 165.

\item{\sc {Antonellini}, S. and {Banzatti}, A. and {Kamp}, I. and
  {Thi}, W. -F. and {\bf {Woitke}, P.}} (2020, May).  \newblock {Model
  exploration of near-IR ro-vibrational CO emission as a tracer of
  inner cavities in protoplanetary disks},
  \newblock{\em A\&A} {\bf 637}, 29.

\item{\sc {Herbort}, O. and {\bf {Woitke}, P.} and {Helling}, Ch. and
  {Zerkle}, A.} (2020, April).  \newblock {The atmospheres of rocky
  exoplanets. I. Outgassing of common rock and the stability of liquid
  water},
  \newblock{\em A\&A} {\bf 636}, 71.

\item{\sc {Molaverdikhani}, K. and {Helling}, Ch. and {Lew},
  B.~W.~P. and {MacDonald}, R.~J. and {Samra}, D. and {Iro}, N. and
  {\bf {Woitke}, P.} and {Parmentier}, V.} (2020, March).  \newblock
  {Understanding the atmospheric properties and chemical composition
    of the ultra-hot Jupiter HAT-P-7b. II. Mapping the effects of gas
    kinetics},
  \newblock{\em A\&A} {\bf 635}, 31.

\item{\sc {Thi}, W.~F. and {Hocuk}, S. and {Kamp}, I. and {\bf {Woitke},
  P.} and {Rab}, Ch. and {Cazaux}, S. and {Caselli}, P. and {D'Angelo},
  M.} (2020, March).  \newblock {Warm dust surface chemistry in
  protoplanetary disks. Formation of phyllosilicates}.
  \newblock {\em A\&A} {\bf 635}, 16.
  
\item{\sc {Thi}, W.~F. and {Hocuk}, S. and {Kamp}, I. and {\bf {Woitke},
  P.} and {Rab}, Ch. and {Cazaux}, S. and {Caselli}, P.} (2020,
  February).   \newblock {Warm dust surface chemistry. H$_{2}$ and HD
    formation}.
  \newblock{\em A\&A} {\bf 634}, 42.

\item{\sc {\bf {Woitke}, Peter} and {Helling}, Christiane and {Gunn},
  Ophelia} (2020, February).  \newblock {Dust in brown dwarfs and
  extra-solar planets. VII. Cloud formation in diffusive atmospheres},
  \newblock{\em A\&A} {\bf 634}, 23.

\item{\sc {Thi}, W.~F. and {Lesur}, G. and {\bf {Woitke}, P.} and {Kamp},
  I. and {Rab}, Ch. and {Carmona}, A.} (2019, December).
  \newblock {Radiation thermo-chemical models of protoplanetary
    disks. Grain and polycyclic aromatic hydrocarbon charging},
  \newblock{{\em A\&A} {\bf {\bf 632}}, 44}.

\item{\sc {Greenwood}, A.~J. and {Kamp}, I. and {Waters},
  L.~B.~F.~M. and {\bf {Woitke}, P.} and {Thi}, W. -F.} (2019,
  November). \newblock {The infrared line-emitting regions of T Tauri
    protoplanetary disks},
  \newblock{{\em A\&A} {\bf 631}, 81}.

\item{\sc {Helling}, Ch. and {Iro}, N. and {Corrales}, L. and {Samra},
  D. and {Ohno}, K. and {Alam}, M.~K. and {Steinrueck}, M. and {Lew},
  B. and {Molaverdikhani}, K. and {MacDonald}, R.~J. and {Herbort},
  O. and {\bf {Woitke}, P.} and {Parmentier}, V.} (2019, November).
  \newblock {Understanding the atmospheric properties and chemical
    composition of the ultra-hot Jupiter HAT-P-7b. I. Cloud and
    chemistry mapping},
  \newblock{{\em A\&A} {\bf 631}, 79}

\item{\sc {\bf {Woitke}, P.} and {Kamp}, I. and {Antonellini}, S. and
  {Anthonioz}, F. and {Baldovin-Saveedra}, C. and {Carmona}, A. and
  {Dionatos}, O. and {Dominik}, C. and {Greaves}, J. and {G{\"u}del},
  M. and {Ilee}, J.~D. and {Liebhardt}, A. and {Menard}, F. and {Min},
  M. and {Pinte}, C. and {Rab}, C. and {Rigon}, L. and {Thi},
  W.~F. and {Thureau}, N. and {Waters}, L.~B.~F.~M.} (2019, June).
  \newblock {Consistent Dust and Gas Models for Protoplanetary
    Disks. III. Models for Selected Objects from the FP7 DIANA
    Project},
  \newblock{{\em PASP} {\bf 131}, 64301}

\item{\sc {Helling}, Ch. and {Gourbin}, P. and {\bf {Woitke}, P.} and
  {Parmentier}, V.} (2019, June).  \newblock {Sparkling nights and
  very hot days on WASP-18b: the formation of clouds and the emergence
  of an ionosphere},
  \newblock{{\em A\&A} {\bf 626}, 133}

\item{\sc {Greenwood}, A.~J. and {Kamp}, I. and {Waters},
  L.~B.~F.~M. and {\bf {Woitke}, P.} and {Thi}, W. -F.} (2019, June).
  \newblock {Effects of dust evolution on protoplanetary disks in the
    mid-infrared},
  \newblock{{\em A\&A} {\bf 626}, 6}

\item{\sc {Dionatos}, O. and {\bf {Woitke}, P.} and {G{\"u}del}, M. and
  {Degroote}, P. and {Liebhart}, A. and {Anthonioz}, F. and
  {Antonellini}, S. and {Baldovin-Saavedra}, C. and {Carmona}, A. and
  {Dominik}, C. and {Greaves}, J. and {Ilee}, J.~D. and {Kamp}, I. and
  {M{\'e}nard}, F. and {Min}, M. and {Pinte}, C. and {Rab}, C. and
  {Rigon}, L. and {Thi}, W.~F. and {Waters}, L.~B.~F.~M.} (2019, May).
  \newblock {Consistent dust and gas models for protoplanetary
    disks. IV. A panchromatic view of protoplanetary disks},
  \newblock{{\em A\&A} {\bf 625}, 66}.

\item{\sc {Rab}, Ch. and {Kamp}, I. and {Ginski}, C. and {Oberg},
  N. and {Muro-Arena}, G.~A. and {Dominik}, C. and {Waters},
  L.~B.~F.~M. and {Thi}, W. -F. and {\bf {Woitke}, P.}} (2019, April).
  \newblock {Observing the gas component of circumplanetary disks
    around wide-orbit planet-mass companions in the (sub)mm
    regime},
  \newblock{{\em A\&A} {\bf 624}, 16}.

\item{\sc {D'Angelo}, M. and {Cazaux}, S. and {Kamp}, I. and {Thi},
  W. -F. and {\bf {Woitke}, P.}} (2019, February).  \newblock {Water
  delivery in the inner solar nebula. Monte Carlo simulations of
  forsterite hydration},
  \newblock{{\em A\&A} {\bf 622}, 208}.

\item{\sc {\bf {Woitke}, P.} and {Min}, M. and {Thi}, W. -F. and {Roberts},
  C. and {Carmona}, A. and {Kamp}, I. and {M{\'e}nard}, F. and
  {Pinte}, C.} (2018, October).  \newblock {Modelling mid-infrared
  molecular emission lines from T Tauri stars},
  \newblock{{\em A\&A} {\bf 618}, 57}.

\item{\sc {\bf {Woitke}, P.} and {Helling}, Ch. and {Hunter}, G.~H. and
  {Millard}, J.~D. and {Turner}, G.~E. and {Worters}, M. and {Blecic},
  J. and {Stock}, J.~W.} (2018, June).  \newblock {Equilibrium
  chemistry down to 100 K. Impact of silicates and phyllosilicates on
  the carbon to oxygen ratio},
  \newblock{{\em A\&A} {\bf 614}, 1}.

\item{\sc {Rab}, Ch. and {G{\"u}del}, M. and {\bf {Woitke}, P.} and {Kamp},
  I. and {Thi}, W. -F. and {Min}, M. and {Aresu}, G. and {Meijerink},
  R.} (2018, January).  \newblock {X-ray radiative transfer in
  protoplanetary disks. The role of dust and X-ray background fields},
  \newblock{{\em A\&A} {\bf 609}, 91}.

\item{\sc {Pinte}, C. and {M{\'e}nard}, F. and {Duch{\^e}ne}, G. and
  {Hill}, T. and {Dent}, W.~R.~F. and {\bf {Woitke}, P.} and {Maret}, S. and
  {van der Plas}, G. and {Hales}, A. and {Kamp}, I. and {Thi},
  W.~F. and {de Gregorio-Monsalvo}, I. and {Rab}, C. and {Quanz},
  S.~P. and {Avenhaus}, H. and {Carmona}, A. and {Casassus}, S.}
  (2018, January).  \newblock {Direct mapping of the temperature and
  velocity gradients in discs. Imaging the vertical CO snow line
  around IM Lupi},
  \newblock{{\em A\&A} {\bf 609}, 47}.

\item{\sc {Kamp}, I. and {Thi}, W. -F. and {\bf {Woitke}, P.} and {Rab},
  C. and {Bouma}, S. and {M{\'e}nard}, F.} (2017, November).
  \newblock {Consistent dust and gas models for protoplanetary
    disks. II. Chemical networks and rates},
  \newblock{{\em A\&A} {\bf 607}, 41}.

\item{\sc {Rab}, Ch. and {Elbakyan}, V. and {Vorobyov}, E. and
  {G{\"u}del}, M. and {Dionatos}, O. and {Audard}, M. and {Kamp},
  I. and {Thi}, W. -F. and {\bf {Woitke}, P.} and {Postel}, A.} (2017,
  July).  \newblock {The chemistry of episodic accretion in embedded
    objects. 2D radiation thermo-chemical models of the post-burst
    phase},
  \newblock{{\em A\&A} {\bf 604}, 15}.

\item{\sc {Helling}, Ch. and {Tootill}, D. and {\bf {Woitke}, P.} and {Lee},
  G.} (2017, July).  \newblock {Dust in brown dwarfs and extrasolar
  planets. V. Cloud formation in carbon- and oxygen-rich
  environments},
  \newblock{{\em A\&A} {\bf 603}, 123}.

\item{\sc {Rab}, Ch. and {G{\"u}del}, M. and {Padovani}, M. and
  {Kamp}, I. and {Thi}, W. -F. and {\bf {Woitke}, P.} and {Aresu}, G.}
  (2017, July).  \newblock {Stellar energetic particle ionization in
  protoplanetary disks around T Tauri stars},
  \newblock{{\em A\&A} {\bf 603}, 96}.

\item{\sc {Greenwood}, A.~J. and {Kamp}, I. and {Waters},
  L.~B.~F.~M. and {\bf {Woitke}, P.} and {Thi}, W. -F. and {Rab}, Ch. and
  {Aresu}, G. and {Spaans}, M.} (2017, May).  \newblock
  {Thermochemical modelling of brown dwarf discs},
  \newblock{{\em A\&A} {\bf 601}, 44}.

\item{\sc {Antonellini}, S. and {Bremer}, J. and {Kamp}, I. and
  {Riviere-Marichalar}, P. and {Lahuis}, F. and {Thi}, W. -F. and
  {\bf {Woitke}, P.} and {Meijerink}, R. and {Aresu}, G. and {Spaans}, M.}
  (2017, January).  \newblock {Mid-IR water and silicate relation in
  protoplanetary disks},
  \newblock{{\em A\&A} {\bf 597}, 72}.

\item{\sc {Drabek-Maunder}, E. and {Mohanty}, S. and {Greaves}, J. and
  {Kamp}, I. and {Meijerink}, R. and {Spaans}, M. and {Thi},
  W. -F. and {\bf {Woitke}, P.}} (2016, December).  \newblock {HCO+
  Detection of Dust-depleted Gas in the Inner Hole of the LkCa 15
  Pre-transitional Disk},
  \newblock{{\em ApJ} {\bf 833}, 260}

\item{\sc {Haworth}, Thomas J. and {Ilee}, John D. and {Forgan},
  Duncan H. and {Facchini}, Stefano and {Price}, Daniel J. and
  {Boneberg}, Dominika M. and {Booth}, Richard A. and {Clarke}, Cathie
  J. and {Gonzalez}, Jean-Fran{\c{c}}ois and {Hutchison}, Mark A. and
  {Kamp}, Inga and {Laibe}, Guillaume and {Lyra}, Wladimir and {Meru},
  Farzana and {Mohanty}, Subhanjoy and {Pani{\'c}}, Olja and {Rice},
  Ken and {Suzuki}, Takeru and {Teague}, Richard and {Walsh},
  Catherine and {\bf {Woitke}, Peter} and {Community authors}} (2016,
  October).  \newblock {Grand Challenges in Protoplanetary Disc
    Modelling},
  \newblock{{\em PASA} {\bf 33}, e053}

\item{\sc {Greaves}, J.~S. and {Holland}, W.~S. and {Matthews},
  B.~C. and {Marshall}, J.~P. and {Dent}, W.~R.~F. and {\bf {Woitke},
  P.} and {Wyatt}, M.~C. and {Matr{\`a}}, L. and {Jackson}, A.} (2016,
  October).  \newblock {Gas and dust around A-type stars at tens of
    Myr: signatures of cometary breakup},
  \newblock{{\em MNRAS} {\bf 461}, 3910}.

\item{\sc {Lee}, G. and {Dobbs-Dixon}, I. and {Helling}, Ch. and
  {Bognar}, K. and {\bf {Woitke}, P.}} (2016, October).  \newblock
  {Dynamic mineral clouds on HD 189733b. I. 3D RHD with kinetic,
    non-equilibrium cloud formation},
  \newblock{{\em A\&A} {\bf 594}, 48}.

\item{\sc {Min}, M. and {Bouwman}, J. and {Dominik}, C. and {Waters},
  L.~B.~F.~M. and {Pontoppidan}, K.~M. and {Hony}, S. and {Mulders},
  G.~D. and {Henning}, Th. and {van Dishoeck}, E.~F. and {\bf {Woitke},
    P.} and {Evans}, Neal J., II and {Digit Team}} (2016, August).
  \newblock {The abundance and thermal history of water ice in the
    disk surrounding HD 142527 from the DIGIT Herschel Key Program},
  \newblock{{\em A\&A} {\bf 593}, 11}.

\item{\sc {Hein Bertelsen}, R.~P., {Kamp}, I., {van der Plas}, G.,
  {van den Ancker}, M.~E., {Waters}, L.~B.~F.~M., {Thi}, W.-F.,
  {\bf {Woitke}, P.}} (2016, May).  \newblock {Variability in the CO
  ro-vibrational lines from HD163296}.
  \newblock {\em \mnras\/}~{\bf 458}, 1466--1477.

\item{\sc {Hein Bertelsen}, R.~P., {Kamp}, I., {van der Plas}, G.,
  {van den Ancker}, M.~E., {Waters}, L.~B.~F.~M., {Thi}, W.-F.,
  {\bf {Woitke}, P.}} (2016, May).  \newblock {A proposed new diagnostic
  for Herbig disc geometry. FWHM versus J of CO ro-vibrational lines}.
  \newblock {\em \aap\/}~{\bf 590}, A98.

\item{\sc {\bf {Woitke}, P.}, {Min}, M., {Pinte}, C., {Thi}, W.-F.,
  {Kamp}, I., {Rab}, C., 
  {Anthonioz}, F., {Antonellini}, S., {Baldovin-Saavedra}, C., {Carmona}, A.,
  {Dominik}, C., {Dionatos}, O., {Greaves}, J., {G{\"u}del}, M., {Ilee}, J.~D.,
  {Liebhart}, A., {M{\'e}nard}, F., {Rigon}, L., {Waters}, L.~B.~F.~M.,
  {Aresu}, G., {Meijerink}, R., {Spaans}, M.} (2016, February).
\newblock {Consistent dust and gas models for protoplanetary disks. I. Disk
  shape, dust settling, opacities, and PAHs}.
\newblock {\em \aap\/}~{\bf 586}, A103.

\item{\sc {Antonellini}, S., {Kamp}, I., {Lahuis}, F., {\bf {Woitke}, P.},
  {Thi}, W.-F., {Meijerink}, R., {Aresu}, G., {Spaans}, M.,
  {G{\"u}del}, M., {Liebhart}, A.}  (2016, January).  \newblock
  {Mid-IR spectra of pre-main sequence Herbig stars: An explanation
    for the non-detections of water lines}.  \newblock {\em
    \aap\/}~{\bf 585}, A61.

\item{\sc {Min}, M., {Rab}, C., {\bf {Woitke}, P.}, {Dominik}, C., {M{\'e}nard}, F.}
  (2016, January).
\newblock {Multiwavelength optical properties of compact dust aggregates in
  protoplanetary disks}.
\newblock {\em \aap\/}~{\bf 585}, A13.

\item{\sc {Antonellini}, S., {Kamp}, I., {Riviere-Marichalar}, P.,
  {Meijerink}, R., {\bf {Woitke}, P.}, {Thi}, W.-F., {Spaans}, M., {Aresu},
  G., {Lee}, G.} (2015, October).  \newblock {Understanding the water
  emission in the mid- and far-IR from protoplanetary disks around T
  Tauri stars}.  \newblock {\em \aap\/}~{\bf 582}, A105.

\item{\sc {van der Wiel}, M.~H.~D., {Naylor}, D.~A., {Kamp}, I.,
  {M{\'e}nard}, F., {Thi}, W.-F., {\bf {Woitke}, P.}, {Olofsson}, G.,
  {Pontoppidan}, K.~M., {Di Francesco}, J., {Glauser}, A.~M.,
  {Greaves}, J.~S., {Ivison}, R.~J.} (2014, November).  \newblock
  {Signatures of warm carbon monoxide in protoplanetary discs observed
    with Herschel SPIRE}.  \newblock {\em \mnras\/}~{\bf 444},
  3911--3925.  

\item{\sc {Garufi}, A., {Podio}, L., {Kamp}, I., {M{\'e}nard}, F.,
  {Brittain}, S., {Eiroa}, C., {Montesinos}, B.,
  {Alonso-Mart{\'{\i}}nez}, M., {Thi}, W.~F., {\bf {Woitke}, P.}} (2014,
  July).  \newblock {The protoplanetary disk of FT Tauri:
    multiwavelength data analysis and modelling}.  \newblock {\em
    \aap\/}~{\bf 567}, A141.  

\item{\sc {Carmona}, A., {Pinte}, C., {Thi}, W.~F., {Benisty}, M.,
  {M{\'e}nard}, F., {Grady}, C., {Kamp}, I., {\bf {Woitke}, P.}, {Olofsson},
  J., {Roberge}, A., {Brittain}, S., {Duch{\^e}ne}, G., {Meeus}, G.,
  {Martin-Za{\"i}di}, C., {Dent}, B., {Le Bouquin}, J.~B., {Berger},
  J.~P.} (2014, July).  \newblock {Constraining the structure of the
  transition disk HD 135344B (SAO 206462) by simultaneous modelling of
  multiwavelength gas and dust observations}.  \newblock {\em
  \aap\/}~{\bf 567}, A51.  

\item{\sc {Keane}, J.~T., {Pascucci}, I., {Espaillat}, C., {\bf {Woitke},
  P.}, {Andrews}, S., {Kamp}, I., {Thi}, W.-F., {Meeus}, G., {Dent},
  W.~R.~F.} (2014, June).  \newblock {Herschel Evidence for Disk
  Flattening or Gas Depletion in Transitional Disks}.  \newblock {\em
  \apj\/}~{\bf 787}, 153.  

\item{\sc {Aresu}, G., {Kamp}, I., {Meijerink}, R., {Spaans}, M.,
  {Vicente}, S., {Podio}, L., {\bf {Woitke}, P.}, {Menard}, F., {Thi},
  W.-F., {G{\"u}del}, M., {Liebhart}, A.} (2014, June).  \newblock {[O
    I] disk emission in the Taurus star-forming region}.  \newblock
  {\em \aap\/}~{\bf 566}, A14.  

\item{\sc {Helling}, C., {\bf {Woitke}, P.}, {Rimmer}, P.~B., {Kamp}, I.,
  {Thi}, W.-F., {Meijerink}, R.} (2014, April).  \newblock {Disk
  Evolution, Element Abundances and Cloud Properties of Young Gas
  Giant Planets}.  \newblock {\em Life\/}~{\bf 4}, 142--173.  

\item{\sc {Podio}, L., {Kamp}, I., {Codella}, C., {Nisini}, B.,
  {Aresu}, G., {Brittain}, S., {Cabrit}, S., {Dougados}, C., {Grady},
  C., {Meijerink}, R., {Sandell}, G., {Spaans}, M., {Thi}, W.-F.,
  {White}, G.~J., {\bf {Woitke}, P.}}  (2014, March).  \newblock {Probing
  the Gaseous Disk of T Tau N with CN 5-4 Lines}.  \newblock {\em
  \apjl\/}~{\bf 783}, L26.  

\item{\sc {Hein Bertelsen}, R.~P., {Kamp}, I., {Goto}, M., {van der
    Plas}, G., {Thi}, W.-F., {Waters}, L.~B.~F.~M., {van den Ancker},
  M.~E., {\bf {Woitke}, P.}}  (2014, January).  \newblock {CO ro-vibrational
  lines in HD 100546. A search for disc asymmetries and the role of
  fluorescence}.  \newblock {\em \aap\/}~{\bf 561}, A102.  

\item{\sc {Thi}, W.-F., {Pinte}, C., {Pantin}, E., {Augereau}, J.~C.,
  {Meeus}, G., {M{\'e}nard}, F., {Martin-Za{\"i}di}, C., {\bf {Woitke}, P.},
  {Riviere-Marichalar}, P., {Kamp}, I., {Carmona}, A., {Sandell}, G.,
  {Eiroa}, C., {Dent}, W., {Montesinos}, B., {Aresu}, G., {Meijerink},
  R., {Spaans}, M., {White}, G., {Ardila}, D., {Lebreton}, J.,
  {Mendigut{\'{\i}}a}, I., {Brittain}, S.} (2014, January).  \newblock
  {Gas lines from the 5-Myr old optically thin disk around HD 141569A.  
  Herschel observations and modelling}.  \newblock {\em
    \aap\/}~{\bf 561}, A50.  

\item{\sc {Kamp}, I., {Thi}, W.-F., {Meeus}, G., {\bf {Woitke}, P.},
  {Pinte}, C., {Meijerink}, R., {Spaans}, M., {Pascucci}, I., {Aresu},
  G., {Dent}, W.~R.~F.}  (2013, November).  \newblock {Uncertainties
  in water chemistry in disks: An application to TW Hydrae}.
  \newblock {\em \aap\/}~{\bf 559}, A24.  

\item{\sc {Meijerink}, R., {Spaans}, M., {Kamp}, I., {Aresu}, G.,
  {Thi}, W.-F., {\bf {Woitke}, P.}} (2013, October).  \newblock {Tracing the
  Physical Conditions in Active Galactic Nuclei with Time-Dependent
  Chemistry}.  \newblock {\em Journal of Physical Chemistry A\/}~{\bf
  117}, 9593--9604.  

\item{\sc {Thi}, W.~F., {M{\'e}nard}, F., {Meeus}, G., {Carmona}, A.,
  {Riviere-Marichalar}, P., {Augereau}, J.-C., {Kamp}, I., {\bf {Woitke},
  P.}, {Pinte}, C., {Mendigut{\'{\i}}a}, I., {Eiroa}, C., {Montesinos},
  B., {Britain}, S., {Dent}, W.} (2013a, September).  \newblock
  {Nature of the gas and dust around 51 Ophiuchi. Modelling continuum
    and Herschel line observations}.  \newblock {\em \aap\/}~{\bf
    557}, A111. 

\item{\sc {Roberge}, A., {Kamp}, I., {Montesinos}, B., {Dent},
  W.~R.~F., {Meeus}, G., {Donaldson}, J.~K., {Olofsson}, J.,
  {Mo{\'o}r}, A., {Augereau}, J.-C., {Howard}, C., {Eiroa}, C., {Thi},
  W.-F., {Ardila}, D.~R., {Sandell}, G., {\bf {Woitke}, P.}} (2013, July).
  \newblock {Herschel Observations of Gas and Dust in the Unusual 49
    Ceti Debris Disk}.  \newblock {\em \apj\/}~{\bf 771}, 69.  

\item{\sc {Dent}, W.~R.~F., {Thi}, W.~F., {Kamp}, I., {Williams},
  J.~P., {Menard}, F., {Andrews}, S., {Ardila}, D., {Aresu}, G.,
  {Augereau}, J.-C., {Barrado y Navascues}, D., {Brittain}, S.,
  {Carmona}, A., {Ciardi}, D., {Danchi}, W., {Donaldson}, J.,
  {Duchene}, G., {Eiroa}, C., {Fedele}, D., {Grady}, C., {de
    Gregorio-Molsalvo}, I., {Howard}, C., {Hu{\'e}lamo}, N., {Krivov},
  A., {Lebreton}, J., {Liseau}, R., {Martin-Zaidi}, C., {Mathews}, G.,
  {Meeus}, G., {Mendigut{\'{\i}}a}, I., {Montesinos}, B.,
  {Morales-Calderon}, M., {Mora}, A., {Nomura}, H., {Pantin}, E.,
  {Pascucci}, I., {Phillips}, N., {Pinte}, C., {Podio}, L., {Ramsay},
  S.~K., {Riaz}, B., {Riviere-Marichalar}, P., {Roberge}, A.,
  {Sandell}, G., {Solano}, E., {Tilling}, I., {Torrelles}, J.~M.,
  {Vandenbusche}, B., {Vicente}, S., {White}, G.~J., {\bf {Woitke}, P.}}
  (2013, May).  \newblock {GASPS - A Herschel Survey of Gas and Dust
  in Protoplanetary Disks: Summary and Initial Statistics}.  \newblock
  {\em \pasp\/}~{\bf 125}, 477--505.  

\item{\sc {Podio}, L., {Kamp}, I., {Codella}, C., {Cabrit}, S.,
  {Nisini}, B., {Dougados}, C., {Sandell}, G., {Williams}, J.~P.,
  {Testi}, L., {Thi}, W.-F., {\bf {Woitke}, P.}, {Meijerink}, R., {Spaans},
  M., {Aresu}, G., {M{\'e}nard}, F., {Pinte}, C.} (2013b, March).
  \newblock {Water Vapor in the Protoplanetary Disk of DG Tau}.
  \newblock {\em \apjl\/}~{\bf 766}, L5.  

\item{\sc {Thi}, W.~F., {Kamp}, I., {\bf {Woitke}, P.}, {van der Plas}, G.,
  {Bertelsen}, R., {Wiesenfeld}, L.} (2013, March).  \newblock
  {Radiation thermo-chemical models of protoplanetary discs. IV.
    Modelling CO ro-vibrational emission from Herbig Ae discs}.
  \newblock {\em \aap\/}~{\bf 551}, A49.  

\item{\sc {Aresu}, G., {Meijerink}, R., {Kamp}, I., {Spaans}, M.,
  {Thi}, W.-F., {\bf {Woitke}, P.}} (2012, November).  \newblock
  {Far-ultraviolet and X-ray irradiated protoplanetary disks: a grid
    of models. II. Gas diagnostic line emission}.  \newblock {\em
    \aap\/}~{\bf 547}, A69.  

\item{\sc {Meijerink}, R., {Aresu}, G., {Kamp}, I., {Spaans}, M.,
  {Thi}, W.-F., {\bf {Woitke}, P.}} (2012, November).  \newblock
  {Far-ultraviolet and X-ray irradiated protoplanetary disks: a grid
    of models. I. The disk structure}.  \newblock {\em \aap\/}~{\bf
    547}, A68.  

\item{\sc {Lebreton}, J., {Augereau}, J.-C., {Thi}, W.-F., {Roberge},
  A., {Donaldson}, J., {Schneider}, G., {Maddison}, S.~T.,
  {M{\'e}nard}, F., {Riviere-Marichalar}, P., {Mathews}, G.~S.,
  {Kamp}, I., {Pinte}, C., {Dent}, W.~R.~F., {Barrado}, D.,
  {Duch{\^e}ne}, G., {Gonzalez}, J.-F., {Grady}, C.~A., {Meeus}, G.,
  {Pantin}, E., {Williams}, J.~P., {\bf {Woitke}, P.}} (2012, March).
  \newblock {An icy Kuiper belt around the young solar-type star HD
    181327}.  \newblock {\em \aap\/}~{\bf 539}, A17.  

\item{\sc {Riviere-Marichalar}, P., {M{\'e}nard}, F., {Thi}, W.~F.,
  {Kamp}, I., {Montesinos}, B., {Meeus}, G., {\bf {Woitke}, P.}, {Howard},
  C., {Sandell}, G., {Podio}, L., {Dent}, W.~R.~F.,
  {Mendigut{\'{\i}}a}, I., {Pinte}, C., {White}, G.~J., {Barrado}, D.}
  (2012, February).  \newblock {Detection of warm water vapour in
  Taurus protoplanetary discs by Herschel}.  \newblock {\em
  \aap\/}~{\bf 538}, L3.   

\item{\sc {Tilling}, I., {\bf {Woitke}, P.}, {Meeus}, G., {Mora}, A.,
  {Montesinos}, B., {Riviere-Marichalar}, P., {Eiroa}, C., {Thi},
  W.-F., {Isella}, A., {Roberge}, A., {Martin-Zaidi}, C., {Kamp}, I.,
  {Pinte}, C., {Sandell}, G., {Vacca}, W.~D., {M{\'e}nard}, F.,
  {Mendigut{\'{\i}}a}, I., {Duch{\^e}ne}, G., {Dent}, W.~R.~F.,
  {Aresu}, G., {Meijerink}, R., {Spaans}, M.} (2012, February).
  \newblock {Gas modelling in the disc of HD 163296}.  \newblock {\em
    \aap\/}~{\bf 538}, A20.   

\item{\sc {\bf {Woitke}, P.}, {Riaz}, B., {Duch{\^e}ne}, G., {Pascucci}, I.,
  {Lyo}, A.-R., {Dent}, W.~R.~F., {Phillips}, N., {Thi}, W.-F.,
  {M{\'e}nard}, F., {Herczeg}, G.~J., {Bergin}, E., {Brown}, A.,
  {Mora}, A., {Kamp}, I., {Aresu}, G., {Brittain}, S., {de
    Gregorio-Monsalvo}, I., {Sandell}, G.} (2011, October).  \newblock
  {The unusual protoplanetary disk around the T Tauri star ET
    Chamaeleontis}.  \newblock {\em \aap\/}~{\bf 534}, A44.   

\item{\sc {Kamp}, I., {Tilling}, I., {\bf {Woitke}, P.}, {Thi}, W.-F.,
  {Hogerheijde}, M.}  (2010, January).  \newblock {Radiation
  thermo-chemical models of protoplanetary disks. II. Line
  diagnostics}.  \newblock {\em \aap\/}~{\bf 510}, A18. 

\item{\sc {de Kok}, R.~J., {Helling}, C., {Stam}, D.~M., {\bf {Woitke}, P.},
  {Witte}, S.}  (2011, July).  \newblock {The influence of
  non-isotropic scattering of thermal radiation on spectra of brown
  dwarfs and hot exoplanets}.  \newblock {\em \aap\/}~{\bf 531}, A67.

\item{\sc {Thi}, W.-F., {M{\'e}nard}, F., {Meeus}, G.,
  {Martin-Za{\"i}di}, C., {\bf {Woitke}, P.}, {Tatulli}, E., {Benisty}, M.,
  {Kamp}, I., {Pascucci}, I., {Pinte}, C., {Grady}, C.~A., {Brittain},
  S., {White}, G.~J., {Howard}, C.~D., {Sandell}, G., {Eiroa}, C.}
  (2011, June).  \newblock {Detection of CH$^{+}$ emission from the
  disc around HD 100546}.  \newblock {\em \aap\/}~{\bf 530}, L2.

\item{\sc {Fedele}, D., {Pascucci}, I., {Brittain}, S., {Kamp}, I.,
  {\bf {Woitke}, P.}, {Williams}, J.~P., {Dent}, W.~R.~F., {Thi}, W.-F.}
  (2011, May).  \newblock {Water Depletion in the Disk Atmosphere of
  Herbig AeBe Stars}.  \newblock {\em \apj\/}~{\bf 732}, 106.

\item{\sc {Thi}, W.-F., {\bf {Woitke}, P.}, {Kamp}, I.} (2011, April).
  \newblock {Radiation thermo-chemical models of protoplanetary discs.
    III. Impact of inner rims on spectral energy distributions}.
  \newblock {\em \mnras\/}~{\bf 412}, 711--726.

\item{\sc {Aresu}, G., {Kamp}, I., {Meijerink}, R., {\bf {Woitke}, P.},
  {Thi}, W.-F., {Spaans}, M.} (2011a, February).  \newblock {X-ray
  impact on the protoplanetary disks around T Tauri stars}.  \newblock
  {\em \aap\/}~{\bf 526}, A163.

\item{\sc {Thi}, W.-F., {\bf {Woitke}, P.}, {Kamp}, I.} (2010, September).
  \newblock {Warm non-equilibrium gas phase chemistry as a possible
    origin of high HDO/H$_{2}$O ratios in hot and dense gases:
    application to inner protoplanetary discs}.  \newblock {\em
    \mnras\/}~{\bf 407}, 232--246.

\item{\sc {Mathews}, G.~S., {Dent}, W.~R.~F., {Williams}, J.~P.,
  {Howard}, C.~D., {Meeus}, G., {Riaz}, B., {Roberge}, A., {Sandell},
  G., {Vandenbussche}, B., {Duch{\^e}ne}, G., {Kamp}, I.,
  {M{\'e}nard}, F., {Montesinos}, B., {Pinte}, C., {Thi}, W.~F.,
  {\bf {Woitke}, P.}, {Alacid}, J.~M., {Andrews}, S.~M., {Ardila}, D.~R.,
  {Aresu}, G., {Augereau}, J.~C., {Barrado}, D., {Brittain}, S.,
  {Ciardi}, D.~R., {Danchi}, W., {Eiroa}, C., {Fedele}, D., {Grady},
  C.~A., {de Gregorio-Monsalvo}, I., {Heras}, A., {Huelamo}, N.,
  {Krivov}, A., {Lebreton}, J., {Liseau}, R., {Martin-Zaidi}, C.,
  {Mendigut{\'{\i}}a}, I., {Mora}, A., {Morales-Calderon}, M.,
  {Nomura}, H., {Pantin}, E., {Pascucci}, I., {Phillips}, N., {Podio},
  L., {Poelman}, D.~R., {Ramsay}, S., {Rice}, K.,
  {Riviere-Marichalar}, P., {Solano}, E., {Tilling}, I., {Walker}, H.,
  {White}, G.~J., {Wright}, G.} (2010, July).  \newblock {GAS in
  Protoplanetary Systems (GASPS). I. First results}.  \newblock {\em
  \aap\/}~{\bf 518}, L127.

\item{\sc {Pinte}, C., {\bf {Woitke}, P.}, {M{\'e}nard}, F., {Duch{\^e}ne},
  G., {Kamp}, I., {Meeus}, G., {Mathews}, G., {Howard}, C.~D.,
  {Grady}, C.~A., {Thi}, W.-F., {Tilling}, I., {Augereau}, J.-C.,
  {Dent}, W.~R.~F., {Alacid}, J.~M., {Andrews}, S., {Ardila}, D.~R.,
  {Aresu}, G., {Barrado}, D., {Brittain}, S., {Ciardi}, D.~R.,
  {Danchi}, W., {Eiroa}, C., {Fedele}, D., {de Gregorio-Monsalvo}, I.,
  {Heras}, A., {Huelamo}, N., {Krivov}, A., {Lebreton}, J., {Liseau},
  R., {Martin-Za{\"i}di}, C., {Mendigut{\'{\i}}a}, I., {Montesinos},
  B., {Mora}, A., {Morales-Calderon}, M., {Nomura}, H., {Pantin}, E.,
  {Pascucci}, I., {Phillips}, N., {Podio}, L., {Poelman}, D.~R.,
  {Ramsay}, S., {Riaz}, B., {Rice}, K., {Riviere-Marichalar}, P.,
  {Roberge}, A., {Sandell}, G., {Solano}, E., {Vandenbussche}, B.,
  {Walker}, H., {Williams}, J.~P., {White}, G.~J., {Wright}, G.}
  (2010, July).  \newblock {The Herschel view of GAS in Protoplanetary
  Systems (GASPS). First comparisons with a large grid of models}.
  \newblock {\em \aap\/}~{\bf 518}, L126.

\item{\sc {Thi}, W.-F., {Mathews}, G., {M{\'e}nard}, F., {\bf {Woitke}, P.},
  {Meeus}, G., {Riviere-Marichalar}, P., {Pinte}, C., {Howard}, C.~D.,
  {Roberge}, A., {Sandell}, G., {Pascucci}, I., {Riaz}, B., {Grady},
  C.~A., {Dent}, W.~R.~F., {Kamp}, I., {Duch{\^e}ne}, G., {Augereau},
  J.-C., {Pantin}, E., {Vandenbussche}, B., {Tilling}, I., {Williams},
  J.~P., {Eiroa}, C., {Barrado}, D., {Alacid}, J.~M., {Andrews}, S.,
  {Ardila}, D.~R., {Aresu}, G., {Brittain}, S., {Ciardi}, D.~R.,
  {Danchi}, W., {Fedele}, D., {de Gregorio-Monsalvo}, I., {Heras}, A.,
  {Huelamo}, N., {Krivov}, A., {Lebreton}, J., {Liseau}, R.,
  {Martin-Zaidi}, C., {Mendigut{\'{\i}}a}, I., {Montesinos}, B.,
  {Mora}, A., {Morales-Calderon}, M., {Nomura}, H., {Phillips}, N.,
  {Podio}, L., {Poelman}, D.~R., {Ramsay}, S., {Rice}, K., {Solano},
  E., {Walker}, H., {White}, G.~J., {Wright}, G.} (2010, July).
  \newblock {Herschel-PACS observation of the 10 Myr old T Tauri disk
    TW Hya.  Constraining the disk gas mass}.  \newblock {\em
    \aap\/}~{\bf 518}, L125.

\item{\sc {Meeus}, G., {Pinte}, C., {\bf {Woitke}, P.}, {Montesinos}, B.,
  {Mendigut{\'{\i}}a}, I., {Riviere-Marichalar}, P., {Eiroa}, C.,
  {Mathews}, G.~S., {Vandenbussche}, B., {Howard}, C.~D., {Roberge},
  A., {Sandell}, G., {Duch{\^e}ne}, G., {M{\'e}nard}, F., {Grady},
  C.~A., {Dent}, W.~R.~F., {Kamp}, I., {Augereau}, J.~C., {Thi},
  W.~F., {Tilling}, I., {Alacid}, J.~M., {Andrews}, S., {Ardila},
  D.~R., {Aresu}, G., {Barrado}, D., {Brittain}, S., {Ciardi}, D.~R.,
  {Danchi}, W., {Fedele}, D., {de Gregorio-Monsalvo}, I., {Heras}, A.,
  {Huelamo}, N., {Krivov}, A., {Lebreton}, J., {Liseau}, R.,
  {Martin-Zaidi}, C., {Mora}, A., {Morales-Calderon}, M., {Nomura},
  H., {Pantin}, E., {Pascucci}, I., {Phillips}, N., {Podio}, L.,
  {Poelman}, D.~R., {Ramsay}, S., {Riaz}, B., {Rice}, K., {Solano},
  E., {Walker}, H., {White}, G.~J., {Williams}, J.~P., {Wright}, G.}
  (2010, July).  \newblock {Gas in the protoplanetary disc of HD
  169142: Herschel's view}.  \newblock {\em \aap\/}~{\bf 518}, L124.

\item{\sc {\bf {Woitke}, P.}, {Pinte}, C., {Tilling}, I., {M{\'e}nard}, F.,
  {Kamp}, I., {Thi}, W.-F., {Duch{\^e}ne}, G., {Augereau}, J.-C.}
  (2010, June).  \newblock {Continuum and line modelling of discs
  around young stars. I. 300000 disc models for HERSCHEL/GASPS}.
  \newblock {\em \mnras\/}~{\bf 405}, L26--L30.

\item{\sc {Kamp}, I., {\bf {Woitke}, P.}, {Pinte}, C., {Tilling}, I., {Thi},
  W.-F., {Menard}, F., {Duchene}, G., {Augereau}, J.-C.} (2011,
  August).  \newblock {Continuum and line modelling of discs around
    young stars. II. Line diagnostics for GASPS from the DENT grid}.
  \newblock {\em \aap\/}~{\bf 532}, A85.

\item{\sc {\bf {Woitke}, P.}, {Thi}, W.-F., {Kamp}, I., {Hogerheijde},
  M.~R.} (2009, July).  \newblock {Hot and cool water in Herbig Ae
  protoplanetary disks. A challenge for Herschel}.  \newblock {\em
  \aap\/}~{\bf 501}, L5--L8.

\item{\sc {\bf {Woitke}, P.}, {Kamp}, I., {Thi}, W.-F.} (2009, July).
  \newblock {Radiation thermo-chemical models of protoplanetary
    disks. I.  Hydrostatic disk structure and inner rim}.  \newblock
            {\em \aap\/}~{\bf 501}, 383--406.

\item{\sc {Pinte}, C., {Harries}, T.~J., {Min}, M., {Watson}, A.~M.,
  {Dullemond}, C.~P., {\bf {Woitke}, P.}, {M{\'e}nard}, F.,
  {Dur{\'a}n-Rojas}, M.~C.} (2009, May).  \newblock {Benchmark
  problems for continuum radiative transfer. High optical depths,
  anisotropic scattering, and polarisation}.  \newblock {\em
  \aap\/}~{\bf 498}, 967--980.

\item{\sc {Helling}, C., {Ackerman}, A., {Allard}, F., {Dehn}, M.,
  {Hauschildt}, P., {Homeier}, D., {Lodders}, K., {Marley}, M.,
  {Rietmeijer}, F., {Tsuji}, T., {\bf {Woitke}, P.}} (2008a, December).
  \newblock {A comparison of chemistry and dust cloud formation in
    ultracool dwarf model atmospheres}.  \newblock {\em \mnras\/}~{\bf
    391}, 1854--1873.

\item{\sc {Helling}, C., {\bf {Woitke}, P.}, {Thi}, W.-F.} (2008, July).
  \newblock {Dust in brown dwarfs and extra-solar planets. I. Chemical
    composition and spectral appearance of quasi-static cloud layers}.
  \newblock {\em \aap\/}~{\bf 485}, 547--560.

\item{\sc {Helling}, C., {Dehn}, M., {\bf {Woitke}, P.}, {Hauschildt},
  P.~H.} (2008b, April).  \newblock {Erratum: ``Consistent Simulations
  of Substellar Atmospheres and Nonequilibrium Dust Cloud Formation''}.
  \newblock {\em \apjl\/}~{\bf 677}, L157--L157.

\item{\sc {Johnas}, C.~M.~S., {Helling}, C., {Dehn}, M., {\bf {Woitke}, P.},
  {Hauschildt}, P.~H.} (2008, March).  \newblock {The influence of
  dust formation modelling on Na I and K I line profiles in substellar
  atmospheres}.  \newblock {\em \mnras\/}~{\bf 385}, L120--L124.

\item{\sc {Helling}, C., {Dehn}, M., {\bf {Woitke}, P.}, {Hauschildt},
  P.~H.} (2008a, March).  \newblock {Consistent Simulations of
  Substellar Atmospheres and Nonequilibrium Dust Cloud Formation}.
  \newblock {\em \apjl\/}~{\bf 675}, L105--L108.

\item{\sc {Helling}, C., {Ackerman}, A.~S., {Allard}, F., {Dehen}, M.,
  {Hauschildt}, P., {Hubeny}, I., {Homeier}, D., {Lodders}, K.,
  {Marley}, M., {Tsuji}, T., {\bf {Woitke}, P.}} (2007, September).
  \newblock {Comparative study of dust cloud modelling for substellar
    atmospheres}.  \newblock {\em Astronomische Nachrichten\/}~{\bf
    328}, 655.

\item{\sc {\bf {Woitke}, P.}} (2006b, December).  \newblock {Too little
  radiation pressure on dust in the winds of oxygen-rich AGB stars}.
  \newblock {\em \aap\/}~{\bf 460}, L9--L12.

\item{\sc {Helling}, C., {\bf {Woitke}, P.}} (2006a, August).  \newblock
  {Dust in brown dwarfs. V. Growth and evaporation of dirty dust
    grains}.  \newblock {\em \aap\/}~{\bf 455}, 325--338.

\item{\sc {\bf {Woitke}, P.}} (2006a, June).  \newblock {2D models for
  dust-driven AGB star winds}.  \newblock {\em \aap\/}~{\bf 452},
  537--549.

\item{\sc {Helling}, C., {Thi}, W.-F., {\bf {Woitke}, P.}, {Fridlund}, M.}
  (2006, May).  \newblock {Detectability of dirty dust grains in brown
  dwarf atmospheres}.  \newblock {\em \aap\/}~{\bf 451}, L9--L12.

\item{\sc {\bf {Woitke}, P.}, {Niccolini}, G.} (2005, April).  \newblock
  {Dust cloud formation in stellar environments. II. Two-dimensional
    models for structure formation around AGB stars}.  \newblock {\em
    \aap\/}~{\bf 433}, 1101--1115.

\item{\sc {Helling}, C., {Klein}, R., {\bf {Woitke}, P.}, {Nowak}, U.,
  {Sedlmayr}, E.}  (2004, August).  \newblock {Dust in brown
  dwarfs. IV. Dust formation and driven turbulence on mesoscopic
  scales}.  \newblock {\em \aap\/}~{\bf 423}, 657--675.

\item{\sc {Pascucci}, I., {Wolf}, S., {Steinacker}, J., {Dullemond},
  C.~P., {Henning}, T., {Niccolini}, G., {\bf {Woitke}, P.}, {Lopez}, B.}
  (2004, April).  \newblock {The 2D continuum radiative transfer
  problem. Benchmark results for disk configurations}.  \newblock {\em
  \aap\/}~{\bf 417}, 793--805.

\item{\sc {\bf {Woitke}, P.}, {Helling}, C.} (2004b, January).  \newblock
  {Dust in brown dwarfs. III. Formation and structure of quasi-static
    cloud layers}.  \newblock {\em \aap\/}~{\bf 414}, 335--350.

\item{\sc {Schirrmacher}, V., {\bf {Woitke}, P.}, {Sedlmayr}, E.} (2003,
  June).  \newblock {On the gas temperature in the shocked
    circumstellar envelopes of pulsating stars. III. Dynamical models
    for AGB star winds including time-dependent dust formation and
    non-LTE cooling}.  \newblock {\em \aap\/}~{\bf 404}, 267--282.

\item{\sc {Richter}, H., {Wood}, P.~R., {\bf {Woitke}, P.}, {Bolick}, U.,
  {Sedlmayr}, E.}  (2003, March).  \newblock {On the shock-induced
  variability of emission lines in M-type Mira variables. II. Fe II
  and [Fe II] emission lines as a diagnostic tool}.  \newblock {\em
  \aap\/}~{\bf 400}, 319--328.

\item{\sc {Niccolini}, G., {\bf {Woitke}, P.}, {Lopez}, B.} (2003b,
  February).  \newblock {High precision Monte Carlo radiative transfer
    in dusty media}.  \newblock {\em \aap\/}~{\bf 399}, 703--716.

\item{\sc {\bf {Woitke}, P.}, {Helling}, C.} (2003a, February).  \newblock
  {Dust in brown dwarfs. II. The coupled problem of dust formation and
    sedimentation}.  \newblock {\em \aap\/}~{\bf 399}, 297--313.

\item{\sc {\bf {Woitke}, P.}} (2001a).  \newblock {Dust Induced Structure
  Formation}.  \newblock In R.~E. {Schielicke} (Hrsg.), {\em Reviews
  in Modern Astronomy}, Band~14 of {\em Reviews in Modern Astronomy},
  pp.\ 185.

\item{\sc {\bf {Woitke}, P.}, {Sedlmayr}, E., {Lopez}, B.} (2000, June).
  \newblock {Dust cloud formation in stellar environments. I. A
    radiative/thermal instability of dust forming gases}.  \newblock
            {\em \aap\/}~{\bf 358}, 665--670.

\item{\sc {\bf {Woitke}, P.}, {Helling}, C., {Winters}, J.~M., {Jeong},
  K.~S.} (1999, August).  \newblock {On the formation of warm
  molecular layers}.  \newblock {\em \aap\/}~{\bf 348}, L17--L20.

\item{\sc {\bf {Woitke}, P.}, {Sedlmayr}, E.} (1999, July).  \newblock
  {Heating and cooling by iron in cool star winds}.  \newblock {\em
    \aap\/}~{\bf 347}, 617--629.

\item{\sc {Clayton}, G.~C., {Ayres}, T.~R., {Lawson}, W.~A.,
  {Drilling}, J.~S., {\bf {Woitke}, P.}, {Asplund}, M.} (1999, April).
  \newblock {First Observations of an R Coronae Borealis Star with the
    Space Telescope Imaging Spectrograph: RY Sagittarii near Maximum
    Light}.  \newblock {\em \apj\/}~{\bf 515}, 351--355.

\item{\sc {\bf {Woitke}, P.}, {Goeres}, A., {Sedlmayr}, E.} (1996a,
  September).  \newblock {On the gas temperature in the shocked
    circumstellar envelopes of pulsating stars. II. Shock induced
    condensation around R Coronae Borealis stars.}  \newblock {\em
    \aap\/}~{\bf 313}, 217--228.

\item{\sc {\bf {Woitke}, P.}, {Krueger}, D., {Sedlmayr}, E.} (1996, July).
  \newblock {On the gas temperature in the shocked circumstellar
    envelopes of pulsating stars. I. Radiative heating and cooling
    rates.}  \newblock {\em \aap\/}~{\bf 311}, 927--944.

\item{\sc {Krueger}, D., {\bf {Woitke}, P.}, {Sedlmayr}, E.} (1995,
  November).  \newblock {A general multi-component method for the
    description of dust grain processing.}  \newblock {\em
    \aaps\/}~{\bf 113}, 593.

\item{\sc {\bf {Woitke}, P.}, {Dominik}, C., {Sedlmayr}, E.} (1993, July).
  \newblock {Dust destruction in the transition region between stellar
    wind and interstellar medium}.  \newblock {\em \aap\/}~{\bf 274},
  451.

\item{\sc {\bf {Woitke}, P.}, {Dominik}, C., {Winters}, J.~M., {Sedlmayr},
  E.} (1993).  \newblock {Dust formation in the Ejecta of SN 1987A}.
  \newblock In B.~{Baschek}, G.~{Klare}, \& J.~{Lequeux} (Hrsg.), {\em
    New Aspects of Magellanic Cloud Research}, Band 416 of {\em
    Lecture Notes in Physics, Berlin Springer Verlag}, pp.\ 224--225.

\end{enumerate}

\subsection*{Conference Proceedings}

\begin{enumerate}

\item{\sc {Scherf}, M., {Benedikt}, M., {Erkaev}, N., {Lammer}, H.,
  {Herbort}, O., {Marcq}, E., {\bf {Woitke}, P.}, {Odert}, P., {O'Neill},
  C., {Kubyshkina}, D., {Leitzinger}, M.} (2022, September).
  \newblock {Escape of moderately volatile elements from protoplanets
    and its potential effect on habitability}.  \newblock In {\em
    European Planetary Science Congress}, pp.\ EPSC2022--999.

\item{\sc {Jorge}, D., {Kamp}, I., {Waters}, R., {\bf {Woitke}, P.},
  {Spaargaren}, R.}  (2021, September).  \newblock {Forming planets
  around stars with non-solar composition}.  \newblock In {\em
  European Planetary Science Congress}, pp.\ EPSC2021--853.

\item{\sc {Khorshid}, N., {Min}, M., {Desert}, J.~M., {Dominik}, C.,
  {\bf {Woitke}, P.}}  (2021, September).  \newblock {Planet formation
  and its effect on planetara atmosphere composition}.  \newblock In
  {\em European Planetary Science Congress}, pp.\ EPSC2021--660.

\item{\sc {\bf {Woitke}, P.}, {Helling}, C.} (2021, April).
\newblock {GGchem: Fast thermo-chemical equilibrium code}.
\newblock {\em Bibcode: 2021ascl.soft04018W}.

\item{\sc {Herbort}, O., {\bf {Woitke}, P.}, {Helling}, C., {Zerkle},
  A.} (2021, March).  \newblock {From clouds to crust {\textemdash}
  Cloud diversity and surface conditions in atmospheres of rocky
  exoplanets}.  \newblock In {\em Bulletin of the American
  Astronomical Society}, Band~53, pp.\ 1134.

\item{\sc {Waters}, L.~B.~F.~M., {Van Hoolst}, T., {\bf {Woitke}, P.},
  {Helling}, C., {Kamp}, I., {Vazan}, A., {Van der Vleut}, B.,
  {Spaargaren}, R., {Noack}, L., {Loes Ten Kate}, I., {Min}, M.,
  {Ormel}, C.} (2021, January).  \newblock {The impact of planet
  formation scenarios on the composition of rocky exoplanets}.
  \newblock In {\em 43rd COSPAR Scientific Assembly. Held 28 January -
    4 February}, Band~43, pp.\ 520.

\item{\sc {Nikolaou}, Athanasia and {Mugnai}, Lorenzo and {Herbort},
  Oliver and {Pascale}, Enzo and {\bf {Woitke}, Peter}}, (2020,
  September).  \newblock {Characteristics of an hybrid atmosphere with
    disk-captured and degassing contributions over a rocky planet's
    magma ocean. A modeling approach},
  in \newblock {European Planetary Science Congress, EPSC2020-937}. 

\item{\sc {Rab}, C. and {Elbakyan}, V. and {Vorobyov}, E. and
  {Postel}, A. and {G{\"u}del}, M. and {Dionatos}, O. and {Audard},
  M. and {Kamp}, I. and {Thi}, W. -F. and {\bf {Woitke}, P.}} (2020,
  January).  \newblock {The chemistry of episodic accretion},
  in \newblock{\em Laboratory Astrophysics: From Observations to
    Interpretation, p.440-442},
  
\item{\sc {Rab}, C. and {Muro-Arena}, G.~A. and {Kamp}, I. and
  {Dominik}, C. and {Waters}, L.~B.~F.~M. and {Thi}, W. -F. and
  {\bf {Woitke}, P.}} (2020, January).  \newblock {The gas structure
  of the HD 163296 planet-forming disk - gas gaps or not?},
  in \newblock{\em Laboratory Astrophysics: From Observations to
    Interpretation, p.445-447}  

\item{\sc {Rab}, Ch. and {Muro-Arena}, G.~A. and {Kamp}, I. and
  {Dominik}, C. and {Waters}, L.~B.~F.~M. and {Thi}, W. -F. and
  {\bf {Woitke}, P.}} (2020, January).  \newblock {Searching for
  chemical signatures of planet formation},
  in \newblock{\em Origins: From the Protosun to the First Steps of Life,
  p. 362-364}.

\item{\sc {Rab}, Ch. and {Padovani}, M. and {G{\"u}del}, M. and
  {Kamp}, I. and {Thi}, W. -F. and {\bf {Woitke}, P.}} (2020,
  January).  \newblock {Constraining the stellar energetic particle
    flux in young solar-like stars},
  in \newblock{\em Origins: From the Protosun to the First Steps of
    Life, p. 310-311}.
    
\item{\sc {\bf {Woitke}, P.}, {Dent}, W.~R.~F., {Thi}, W.-F., {Menard}, F.,
  {Pinte}, C., {Duchene}, G., {Sandell}, G., {Lawson}, W., {Kamp}, I.}
  (2013, July).  \newblock {ET Cha - a single T Tauri star with a disk
  of radius $\sim$\,5 AU ?}  \newblock In {\em Protostars and Planets
  VI Posters}, pp.\ ~13.

\item{\sc {Thi}, W.-F., {Pinte}, C., {Pantin}, E., {Augereau}, J.-C.,
  {Meeus}, G., {M{\'e}nard}, F., {Martin-Zaidi}, C., {\bf {Woitke}, P.},
  {Riviere-Marichalar}, P., {Kamp}, I., {Carmona}, A., {Sandell}, G.,
  {Eiroa}, C., {Dent}, W., {Montesinos}, B., {Aresu}, G., {Meijerink},
  R., {Spaans}, M., {White}, G., {Ardila}, D., {Lebreton}, J.,
  {Mendigutia}, I., {Brittain}, S.} (2013, July).  \newblock
  {Herschel-PACS observation of gas lines from the disc around
    HD141569A}.  \newblock In {\em Protostars and Planets VI Posters},
  pp.\ ~21.

\item{\sc {Rab}, C., {\bf {Woitke}, P.}, {G{\"u}del}, M., {Min}, M., {Diana
    Team}} (2013, July).  \newblock {X-ray Radiative Transfer in
  Protoplanetary Disks with ProDiMo}.  \newblock In {\em Protostars
  and Planets VI Posters}, pp.\ ~38.

\item{\sc {Min}, M., {Rab}, C., {Dominik}, C., {\bf {Woitke}, P.}} (2013,
  July).  \newblock {The appearance of large aggregates in
    protoplanetary disks}.  \newblock In {\em Protostars and Planets
    VI Posters}, pp.\ ~44.

\item{\sc {Antonellini}, S., {Kamp}, I., {\bf {Woitke}, P.}, {Thi}, W.-F.}
  (2013, July).  \newblock {H2O in Protoplanetary Disks: the Snow Line
  and the Planets' Nest}.  \newblock In {\em Protostars and Planets VI
  Posters}, pp.\ ~35.

\item{\sc {Hein Bertelsen}, R., {Kamp}, I., {Goto}, M., {van der
    Plas}, G., {Thi}, W.-F., {Waters}, R., {\bf {Woitke}, P.}, {van den
    Ancker}, M.} (2013, July).  \newblock {Co Ro-Vib Observations of
  HD100546: a Symmetric Disk or Not?}  \newblock In {\em Protostars
  and Planets VI Posters}, pp.\ ~11.

\item{\sc {Lahuis}, F., {Kamp}, I., {Thi}, W.-F., {van Dishoeck}, E.,
  {Harsono}, D., {\bf {Woitke}, P.}} (2013, July).  \newblock {Time
  variation in the molecular infrared spectrum of IRS 46}.  \newblock
  In {\em Protostars and Planets VI Posters}, pp.\ ~25.

\item{\sc {Podio}, L., {Kamp}, I., {Codella}, C., {Cabrit}, S.,
  {Nisini}, B., {Dougados}, C., {Sandell}, G., {Williams}, J.~P.,
  {Testi}, L., {Thi}, W.-F., {\bf {Woitke}, P.}, {Meijerink}, R., {Spaans},
  M., {Aresu}, G., {M{\'e}nard}, F., {Pinte}, C.} (2013a, July).
  \newblock {Water Vapor in the Protoplanetary Disk of DG Tau}.
  \newblock In {\em Protostars and Planets VI Posters}, pp.\ ~28.

\item{\sc {Dent}, W.~R.~F., {Thi}, W.~F., {Kamp}, I., {Williams},
  J.~P., {Menard}, F., {Andrews}, S., {Ardila}, D., {Aresu}, G.,
  {Augereau}, J.-C., {Barrado Y Navascues}, D., {Brittain}, S.,
  {Carmona}, A., {Ciardi}, D., {Danchi}, W., {Donaldson}, J.,
  {Duchene}, G., {Eiroa}, C., {Fedele}, D., {Grady}, C., {de
    Gregorio-Monsalvo}, I., {Howard}, C., {Huelamo}, N., {Krivov}, A.,
  {Lebreton}, J., {Liseau}, R., {Martin-Zaidi}, C., {Mathews}, G.,
  {Meeus}, G., {Mendigutia}, I., {Montesinos}, B., {Morales-Calderon},
  M., {Mora}, A., {Nomura}, H., {Pantin}, E., {Pascucci}, I.,
  {Phillips}, N., {Pinte}, C., {Podio}, L., {Ramsay}, M.~K., {Riaz},
  B., {Riviere-Marichalar}, P., {Roberge}, A., {Sandell}, G.,
  {Solano}, E., {Tilling}, I., {Torrelles}, J.~M., {Vandenbussche},
  B., {Vicente}, S., {White}, G.~J., {\bf {Woitke}, P.}}  (2013, August).
  \newblock {VizieR Online Data Catalog: Gas Survey of Protoplanetary
    Systems. I.  (Dent+, 2013)}.  \newblock {\em VizieR Online Data
    Catalog\/}~{\bf 612}, 50477.

\item{\sc {Aresu}, G., {Kamp}, I., {Meijerink}, R., {\bf {Woitke}, P.},
  {Thi}, W.~F., {Spaans}, M.} (2011b, May).  \newblock {X-rays in
  protoplanetary disks: their impact on the thermal and chemical
  structure, a grid of models.}  \newblock In J.~{Cernicharo} \&
  R.~{Bachiller} (Hrsg.), {\em IAU Symposium}, Band 280 of {\em IAU
    Symposium}, pp.\ 87P.

\item{\sc {Helling}, C., {Ackerman}, A., {Allard}, F., {Dehn}, M.,
  {Hauschildt}, P., {Homeier}, D., {Lodders}, K., {Marley}, M.,
  {Rietmeijer}, F., {Tsuji}, T., {\bf {Woitke}, P.}} (2008b, May).
  \newblock {Comparison of cloud models for Brown Dwarfs}.  \newblock
  In Y.-S. {Sun}, S.~{Ferraz-Mello}, \& J.-L. {Zhou} (Hrsg.), {\em IAU
    Symposium}, Band 249 of {\em IAU Symposium}, pp.\ 173--177.

\item{\sc {Dehn}, M., {Helling}, C., {\bf {Woitke}, P.}, {Hauschildt}, P.}
  (2007, May).  \newblock {The influence of convective energy
  transport on dust formation in brown dwarf atmospheres}.  \newblock
  In F.~{Kupka}, I.~{Roxburgh}, \& K.~L. {Chan} (Hrsg.), {\em IAU
    Symposium}, Band 239 of {\em IAU Symposium}, pp.\ 227--229.

\item{\sc {Helling}, C., {\bf {Woitke}, P.}} (2005, March).  \newblock {Rain
  and clouds in brown dwarf atmospheres: a coupled problem from small
  to large}.  \newblock In F.~{Favata}, G.~A.~J. {Hussain}, \&
  B.~{Battrick} (Hrsg.), {\em 13th Cambridge Workshop on Cool Stars,
    Stellar Systems and the Sun}, Band 560 of {\em ESA Special
    Publication}, pp.\ 249.

\item{\sc {Dehn}, M., {Helling}, C., {\bf {Woitke}, P.}, {Hauschildt}, P.}
  (2005).  \newblock {First Steps Towards Modelling a Brown Dwarf
  Atmosphere Including the Formation of Dust}.  \newblock In {\em
  Protostars and Planets V Posters}, pp.\ 8158.

\item{\sc {Helling}, C., {\bf {Woitke}, P.}} (2004, December).  \newblock
  {Theory of Precipitating Dust Formation in Substellar Atmospheres}.
  \newblock In J.~{Beaulieu}, A.~{Lecavelier Des Etangs}, \&
  C.~{Terquem} (Hrsg.), {\em Extrasolar Planets: Today and Tomorrow},
  Band 321 of {\em Astronomical Society of the Pacific Conference
    Series}, pp.\ 199.

\item{\sc {Helling}, C., {Klein}, R., {\bf {Woitke}, P.}, {Nowak}, U.,
  {Seldmayr}, E.}  (2003, July).  \newblock {On the Sub-grid Modelling
  of Turbulent Dust Formation in Substellar Atmospheres}.  \newblock
  {\em Astronomische Nachrichten Supplement\/}~{\bf 324}, 2.

\item{\sc {Helling}, C., {Klein}, R., {\bf {Woitke}, P.}, {Sedlmayr}, E.}
  (2003).  \newblock {Dust Formation in Brown Dwarf Atmospheres Under
  Conditions of Driven Turbulence}.  \newblock In N.~{Piskunov},
  W.~W. {Weiss}, \& D.~F. {Gray} (Hrsg.), {\em Modelling of Stellar
    Atmospheres}, Band 210 of {\em IAU Symposium}, pp.\ 12P.

\item{\sc {Helling}, C., {\bf {Woitke}, P.}} (2006b, August).  \newblock
  {Time-dependent modelling of oxygen-rich dust formation}.  \newblock
  In {\em IAU Joint Discussion}, Band~11 of {\em IAU Joint
    Discussion}, pp.\ ~13.

\item{\sc {Helling}, C., {\bf {Woitke}, P.}, {Klein}, R., {Sedlmayr}, E.}
  (2005).  \newblock {Dust Formation in Substellar Atmospheres: A
  Multi-Scale Problem}.  \newblock In H.~U. {K{\"a}ufl},
  R.~{Siebenmorgen}, \& A.~{Moorwood} (Hrsg.), {\em High Resolution
    Infrared Spectroscopy in Astronomy}, pp.\ 503--509.

\item{\sc {Helling}, C., {\bf {Woitke}, P.}, {Winters}, J.~M., {Sedlmayr},
  E.} (1998).  \newblock {Influence of molecular opacities on the
  generation of cool star winds}.  \newblock In {\em IAU Symposium},
  Band 191 of {\em IAU Symposium}, pp.\ 209P.

\item{\sc {Johnas}, C.~M.~S., {Helling}, C., {Witte}, S., {Dehn}, M.,
  {\bf {Woitke}, P.}, {Hauschildt}, P.~H.} (2008).  \newblock {The
  Consistent Modelling of Alkali Lines and Dust Formation in Extreme
  Exo--Planets}.  \newblock In D.~{Fischer}, F.~A. {Rasio},
  S.~E. {Thorsett}, \& A.~{Wolszczan} (Hrsg.), {\em Extreme Solar
    Systems}, Band 398 of {\em Astronomical Society of the Pacific
    Conference Series}, pp.\ 393.

\item{\sc {Kamp}, I., {Aresu}, G., {Chaparro}, G., {\bf {Woitke}, P.},
  {Thi}, W.~F.}  (2011, May).  \newblock {The structure and appearance
  of irradiated protoplanetary disks: the role of chemistry}.
  \newblock In J.~{Cernicharo} \& R.~{Bachiller} (Hrsg.), {\em IAU
    Symposium}, Band 280 of {\em IAU Symposium}, pp.\ 213P.

\item{\sc {Keane}, J.~T., {Pascucci}, I., {Andrews}, S.~M., {Dent},
  B., {Espaillat}, C., {Meeus}, G., {Thi}, W., {\bf {Woitke}, P.}} (2013,
  January).  \newblock {From Classical Disks to Transition Disks: An
    Increasing Dust-to-Gas Ratio?}  \newblock In {\em American
    Astronomical Society Meeting Abstracts \#221}, Band 221 of {\em
    American Astronomical Society Meeting Abstracts}, pp.\ \#220.02.

\item{\sc {Lahuis}, F., {Kamp}, I., {Thi}, W.~F., {van Dishoeck},
  E.~F., {\bf {Woitke}, P.}} (2011, May).  \newblock {Epic changes in the
  IRS46 mid-infrared spectrum; an inner disk chemistry study}.
  \newblock In J.~{Cernicharo} \& R.~{Bachiller} (Hrsg.), {\em IAU
    Symposium}, Band 280 of {\em IAU Symposium}, pp.\ 223P.

\item{\sc {L{\"u}ttke}, M., {Helling}, C., {John}, M., {Jeong}, K.~S.,
  {\bf {Woitke}, P.}, {Sedlmayr}, E.} (2000).  \newblock {Dust Formation in
  Brown Dwarfs}.  \newblock In R.~E. {Schielicke} (Hrsg.), {\em
  Astronomische Gesellschaft Meeting Abstracts}, Band~17 of {\em
  Astronomische Gesellschaft Meeting Abstracts}, pp.\ ~7.

\item{\sc {Meier}, S., {Patzer}, A.~B.~C., {L{\"u}ttke}, M., {\bf {Woitke},
  P.}, {Sedlmayr}, E.} (2000).  \newblock {Circumstellar Dust Shells of
  Pulsating Red Giants as Dynamical Systems}.  \newblock In
  R.~E. {Schielicke} (Hrsg.), {\em Astronomische Gesellschaft Meeting
    Abstracts}, Band~17 of {\em Astronomische Gesellschaft Meeting
    Abstracts}, pp.\ ~14.

\item{\sc {Niccolini}, G., {Lopez}, B., {\bf {Woitke}, P.}} (2003).
  \newblock {Radiative Transfer and Dust formation in the Envelopes of
    Evolved Stars}.  \newblock In F.~{Combes}, D.~{Barret},
  T.~{Contini}, \& L.~{Pagani} (Hrsg.), {\em SF2A-2003: Semaine de
    l'Astrophysique Francaise}, pp.\ 555.

\item{\sc {Pervan}, {\v S}., {Helling}, C., {\bf {Woitke}, P.}, {Sedlmayr},
  E.} (2003, July).  \newblock {Study on Synthetic Spectra of
  Substellar Objects}.  \newblock {\em Astronomische Nachrichten
  Supplement\/}~{\bf 324}, 127.

\item{\sc {Richter}, H., {\bf {Woitke}, P.}, {Bolick}, U., {Sedlmayr}, E.,
  {Wood}, P.~R.}  (2003, July).  \newblock {Fe II and [Fe II] Emission
  Lines as a Diagnostic Tool to Probe the Shocked Atmospheres of
  M-type Miras}.  \newblock {\em Astronomische Nachrichten
  Supplement\/}~{\bf 324}, 17.

\item{\sc {Richter}, H., {\bf {Woitke}, P.}, {Sedlmayr}, E., {Wood}, P.~R.}
  (2000).  \newblock {The Variability of Emission Lines in Shocked M
  Mira Atmospheres}.  \newblock In R.~E. {Schielicke} (Hrsg.), {\em
  Astronomische Gesellschaft Meeting Abstracts}, Band~17 of {\em
  Astronomische Gesellschaft Meeting Abstracts}, pp.\ ~24.

\item{\sc {Richter}, H., {Wood}, P.~R., {Bolick}, U., {\bf {Woitke}, P.},
  {Sedlmayr}, E.}  (2003, January).  \newblock {NLTE Modelling of FeII
  and [FeII] Lines in the Shocked Atmospheres of M-type Miras}.
  \newblock In I.~{Hubeny}, D.~{Mihalas}, \& K.~{Werner} (Hrsg.), {\em
    Stellar Atmosphere Modelling}, Band 288 of {\em Astronomical
    Society of the Pacific Conference Series}, pp.\ 344.

\item{\sc {Schirrmacher}, V., {\bf {Woitke}, P.}, {Sedlmayr}, E.} (2000).
  \newblock {Dynamical Model Calculations of AGB Star Winds Including
    Time Dependent Dust Formation and Non-LTE Radiative Cooling}.
  \newblock In R.~E. {Schielicke} (Hrsg.), {\em Astronomische
    Gesellschaft Meeting Abstracts}, Band~17 of {\em Astronomische
    Gesellschaft Meeting Abstracts}, pp.\ ~31.

\item{\sc {Schirrmacher}, V., {\bf {Woitke}, P.}, {Sedlmayr}, E.} (2001).
  \newblock {A Radiative Instability in Post-shock-cooling
    Circumstellar Gas}.  \newblock In E.~R. {Schielicke} (Hrsg.), {\em
    Astronomische Gesellschaft Meeting Abstracts}, Band~18 of {\em
    Astronomische Gesellschaft Meeting Abstracts}, pp.\ ~65.

\item{\sc {Schnabel}, K., {Helling}, C., {\bf {Woitke}, P.}, {Sedlmayr}, E.}
  (2001).  \newblock {Radiation Transfer Through an Extended Planetary
  Atmosphere}.  \newblock In E.~R. {Schielicke} (Hrsg.), {\em
  Astronomische Gesellschaft Meeting Abstracts}, Band~18 of {\em
  Astronomische Gesellschaft Meeting Abstracts}, pp.\ ~44.

\item{\sc {Thi}, W.~F., {Menard}, F., {Meeus}, G., {Martin-Zaidi}, C.,
  {\bf {Woitke}, P.}, {Tatulli}, E., {Benisty}, M., {Kamp}, I., {Pascucci},
  I., {Pinte}, C., {Grady}, C.~A., {Brittain}, S., {White}, G.~J.,
  {Howard}, C.~D., {Sandell}, G., {Eiroa}, C.} (2011, May).  \newblock
  {Modelling CH\^{}+ in the protoplanetary disk HD100546}.  \newblock
  In J.~{Cernicharo} \& R.~{Bachiller} (Hrsg.), {\em IAU Symposium},
  Band 280 of {\em IAU Symposium}, pp.\ 355P.

\item{\sc {\bf {Woitke}, P.}} (1998).  \newblock {Time-Dependent Behavior of
  Cool-Star Winds}.  \newblock In L.~{Kaper} \& A.~W. {Fullerton}
  (Hrsg.), {\em Cyclical Variability in Stellar Winds}, pp.\ 278.

\item{\sc {\bf {Woitke}, P.}} (1999).  \newblock {Dust formation in
  Radioactive Environments}.  \newblock In R.~{Diehl} \& D.~{Hartmann}
  (Hrsg.), {\em Astronomy with Radioactivities}, pp.\ 163.

\item{\sc {\bf {Woitke}, P.}} (2001b).  \newblock {Self-organization of Dust
  Forming Media}.  \newblock In E.~R. {Schielicke} (Hrsg.), {\em
  Astronomische Gesellschaft Meeting Abstracts}, Band~18 of {\em
  Astronomische Gesellschaft Meeting Abstracts}, pp.\ 809.

\item{\sc {\bf {Woitke}, P.}} (2003).  \newblock {Modelling the Mass Loss of
  Cool AGB Stars}.  \newblock In N.~{Piskunov}, W.~W. {Weiss}, \&
  D.~F. {Gray} (Hrsg.), {\em Modelling of Stellar Atmospheres}, Band
  210 of {\em IAU Symposium}, pp.\ 387.

\item{\sc {\bf {Woitke}, P.}} (2005, January).  \newblock {2D models for the
  winds of AGB stars}.  \newblock In A.~{Wilson} (Hrsg.), {\em ESA
  Special Publication}, Band 577 of {\em ESA Special Publication},
  pp.\ 461--462.

\item{\sc {\bf {Woitke}, P.}} (2007, November).  \newblock {What Drives the
  Mass Loss of Oxygen-Rich AGB Stars?}  \newblock In F.~{Kerschbaum},
  C.~{Charbonnel}, \& R.~F. {Wing} (Hrsg.), {\em Why Galaxies Care
    About AGB Stars: Their Importance as Actors and Probes}, Band 378
  of {\em Astronomical Society of the Pacific Conference Series},
  pp.\ 156.

\item{\sc {\bf {Woitke}, P.}} (2008a, October).  \newblock {Dust-driven
  Winds Beyond Spherical Symmetry}.  \newblock In L.~{Deng} \&
  K.~L. {Chan} (Hrsg.), {\em IAU Symposium}, Band 252 of {\em IAU
    Symposium}, pp.\ 229--234.

\item{\sc {\bf {Woitke}, P.}} (2008b).  \newblock {Monte Carlo Radiative
  Transfer in Multi-D Dynamical Models of Dust-Driven Winds}.
  \newblock In S.~{Wolf}, F.~{Allard}, \& P.~{Stee} (Hrsg.), {\em EAS
    Publications Series}, Band~28 of {\em EAS Publications Series},
  pp.\ 75--83.

\item{\sc {\bf {Woitke}, P.}} (2012, March).  \newblock {Modelling
  planet-forming circumstellar discs}.  \newblock In {\em From Atoms
  to Pebbles: Herschel's view of Star and Planet Formation}, pp.\ ~39.

\item{\sc {\bf {Woitke}, P.}, {Dent}, B., {Thi}, W.-F., {Sibthorpe}, B.,
  {Rice}, K., {Williams}, J., {Sicilia-Aguilar}, A., {Brown}, J.,
  {Kamp}, I., {Pascucci}, I., {Alexander}, R., {Roberge}, A.} (2009,
  February).  \newblock {Gas Evolution in Protoplanetary Disks}.
  \newblock In E.~{Stempels} (Hrsg.), {\em 15th Cambridge Workshop on
    Cool Stars, Stellar Systems, and the Sun}, Band 1094 of {\em
    American Institute of Physics Conference Series}, pp.\ 225--233.

\item{\sc {\bf {Woitke}, P.}, {GASPS Consortium}} (2010).  \newblock {Gas in
  Protoplanetary Systems}.  \newblock In {\em 38th COSPAR Scientific
  Assembly}, Band~38 of {\em COSPAR Meeting}, pp.\ 2474.

\item{\sc {\bf {Woitke}, P.}, {Goeres}, A., {Sedlmayr}, E.} (1996b).
  \newblock {Shock-induced condensation in R CrB stars}.  \newblock In
  C.~S. {Jeffery} \& U.~{Heber} (Hrsg.), {\em Hydrogen Deficient
    Stars}, Band~96 of {\em Astronomical Society of the Pacific
    Conference Series}, pp.\ ~83.

\item{\sc {\bf {Woitke}, P.}, {Helling}, C.} (2003b, July).  \newblock {The
  Structure of Quasi-static Cloud Layers in Brown Dwarf Atmospheres}.
  \newblock {\em Astronomische Nachrichten Supplement\/}~{\bf 324},
  126.

\item{\sc {\bf {Woitke}, P.}, {Helling}, C.} (2004a, August).  \newblock {2D
  Models for the Winds of AGB Stars}.  \newblock {\em Astronomische
  Nachrichten Supplement\/}~{\bf 325}, 95.

\item{\sc {\bf {Woitke}, P.}, {Helling}, C.} (2005, March).  \newblock {2D
  dynamical models for dust-driven winds of AGB stars}.  \newblock In
  F.~{Favata}, G.~A.~J. {Hussain}, \& B.~{Battrick} (Hrsg.), {\em 13th
    Cambridge Workshop on Cool Stars, Stellar Systems and the Sun},
  Band 560 of {\em ESA Special Publication}, pp.\ 1039.

\item{\sc {\bf {Woitke}, P.}, {Kr{\"u}ger}, D., {Goeres}, A., {Sedlmayr},
  E.} (1994).  \newblock {On the temperature in the shocked envelopes
  of R CrB stars.}  \newblock In G.~{Klare} (Hrsg.), {\em
  Astronomische Gesellschaft Abstract Series}, Band~10 of {\em
  Astronomische Gesellschaft Abstract Series}, pp.\ 160.

\item{\sc {\bf {Woitke}, P.}, {Quirrenbach}, A.} (2008).  \newblock {The
  Chaotic Winds of AGB Stars: Observation Meets Theory}.  \newblock In
  A.~{Richichi}, F.~{Delplancke}, F.~{Paresce}, \& A.~{Chelli}
  (Hrsg.), {\em The Power of Optical/IR Interferometry: Recent
    Scientific Results and 2nd Generation}, pp.\ 181.

\item{\sc {\bf {Woitke}, P.}, {Sedlmayr}, E.} (1998).  \newblock {Thermal
  Bifurcations in the Circumstellar Envelopes of C-Stars}.  \newblock
  In {\em IAU Symposium}, Band 191 of {\em IAU Symposium}, pp.\ 317P.

\item{\sc {Niccolini}, G., {\bf {Woitke}, P.}, {Lopez}, B.} (2003a, April).
  \newblock {Formation and evolution of dust clumps around cool
    stars}.  \newblock In Y.~{Nakada}, M.~{Honma}, \& M.~{Seki}
  (Hrsg.), {\em Mass-Losing Pulsating Stars and their Circumstellar
    Matter}, Band 283 of {\em Astrophysics and Space Science Library},
  pp.\ 245--246.

\end{enumerate}

\end{document}